%% file: lanthanide_metrology.tex
\documentclass[10pt]{iopart}

\usepackage{graphicx}				
\usepackage{enumitem}
\usepackage{color}
\definecolor{orange}{rgb}{1,0.5,0}
\definecolor{gray}{rgb}{0.4,0.4,0.4}
\definecolor{midnight}{rgb}{0.1,0.1,0.6}
\usepackage{sparklines}

\newcommand{\added}[1]{{\color{red}#1}}
\newcommand{\removed}[1]{{\color{midnight}\sout{#1}}}

\renewcommand{\added}[1]{#1}
\renewcommand{\removed}[1]{}

\usepackage{enumitem,amssymb}
\newlist{todolist}{itemize}{2}
\setlist[todolist]{label=$\square$}
\usepackage{pifont}
\usepackage[
style=numeric,
sorting=none,
giveninits=true,
maxbibnames=8,
date=year,
isbn=false,
doi=false,
url=false,
]{biblatex}
\DeclareNameAlias{author}{last-first}
\renewbibmacro{in:}{}
\AtEveryBibitem{\clearfield{number}}
\AtBeginBibliography{\small}
\DeclareFieldFormat[article]{volume}{\mkbibbold{#1}}

\DeclareFieldFormat[article]
{pages}{#1}
\DeclareFieldFormat[inproceedings, incollection, inbook]
{pages}{p. #1}

\begin{document}
\title[Absolute x-ray energies and line shapes of lanthanide metals]{Absolute energies and emission line shapes of the L x-ray transitions of lanthanide metals}

\newcommand{\affilNISTPML}{$^2$}
\newcommand{\affilCU}{$^1$}
\newcommand{\affilNISTPMLCU}{$^{1,2}$}
\newcommand{\affilNISTITL}{$^3$}
\newcommand{\affilNISTRadPhys}{$^4$}
\newcommand{\affilTheiss}{$^5$}
\author{J. W. Fowler\affilNISTPMLCU, 
%
G.~C. O'Neil\affilNISTPML, 
B.~K. Alpert\affilNISTITL, 
D.~A. Bennett\affilNISTPML, 
E.~V. Denison\affilNISTPML,
W.~B. Doriese\affilNISTPML, 
G.~C. Hilton\affilNISTPML, 
L.~T. Hudson\affilNISTRadPhys, 
Y.-I. Joe\affilNISTPMLCU, 
K.~M. Morgan\affilNISTPMLCU, 
D.~R. Schmidt\affilNISTPML, 
D.~S. Swetz\affilNISTPML, 
C.~I. Szabo\affilNISTRadPhys$^,$\affilTheiss, 
and
J.~N. Ullom\affilNISTPMLCU
}

\address{\affilCU Department of Physics, University of Colorado, Boulder, Colorado 80309, USA}
\address{\affilNISTPML Quantum Electromagnetics Division, National Institute of Standards and Technology, Boulder, Colorado 80305, USA}
\address{\affilNISTITL Applied \& Computational Mathematics Division, National Institute of Standards and Technology, Boulder, Colorado 80305, USA}
\address{\affilNISTRadPhys Radiation Physics Division, National Institute of Standards and Technology, Gaithersburg, Maryland 20899, USA}
\address{\added{\affilTheiss Theiss Research, 7411 Eads Ave, La Jolla, California 92037, USA}}

\date{\today}							

\begin{abstract}
We use an array of transition-edge sensors, cryogenic microcalorimeters with 4\,eV energy resolution, to measure L x-ray emission-line profiles of four elements of the lanthanide series: praseodymium, neodymium, terbium, and holmium. The spectrometer also surveys numerous x-ray standards in order to establish an absolute-energy calibration traceable to the International System of Units for the energy range 4\,keV to 10\,keV\@. The new results include emission line profiles for 97 lines, each expressed as a sum of one or more Voigt functions; improved absolute energy uncertainty on 71 of these lines relative to existing reference data; a median uncertainty on the peak energy of 0.24\,eV, four to ten times better than the median of prior work; and 6 lines that lack any measured values in existing reference tables. 
The 97 lines comprise nearly all of the most intense L lines from these elements under broad-band x-ray excitation. The work improves on previous measurements made with a similar cryogenic spectrometer by the use of sensors with better linearity in the absorbed energy and a gold x-ray absorbing layer that has a Gaussian energy-response function. It also employs a novel sample holder that enables rapid switching between science targets and calibration targets with excellent gain balancing. 
Most of the results for peak energy values shown here should be considered as replacements for the currently tabulated standard reference values, while the line shapes given here represent a significant expansion of the scope of available reference data.
\end{abstract}

\submitto{\MET}

\maketitle
\ioptwocol  

\section{Introduction}

Excited atoms emit x~rays at characteristic energies when outer-shell electrons fill inner-shell vacancies. The properties of elements including the energies and line shapes of emitted x~rays, fluorescence yields, mass-attenuation coefficients, and absorption edges of the elements are known collectively as \emph{x-ray fundamental parameters}. These parameters are of practical significance because they allow us to analyze the composition of complex materials. They also impose powerful constraints on atomic theory, because x-ray emission and absorption involve electronic transitions between quantized, element-specific orbitals. Fluorescence line energies in particular are used both directly to calibrate x-ray energy or wavelength scales, and indirectly to calibrate the lattice parameters of reference materials that in turn calibrate x-ray diffractometers~\cite{Mendenhall2017}.

Tabulations of x-ray line energies made in the 21st century include those of  Deslattes et al.~\cite{Deslattes:2003} and Zschornack~\cite{Zschornack:2007wu}. The Deslattes compilation is also published online as the NIST Standard Reference Database (SRD) 128, the official United States reference on K and L x-ray transition energies~\cite{NIST:SRD128}. Such tabulations necessarily favor completeness over accuracy. Unfortunately, relatively few of their values have been measured by modern instruments. For instance, fully 77\,\% of the 1263 L-line energies listed with measurements in SRD-128 cite only publications before 1970. The other 1402 L lines found in SRD-128 state only a theoretical energy value, outnumbering the entries that give measurements of any age. Most pre-1970 measurements are taken from a re-scaling of the 1967 compilation of Bearden~\cite{Bearden:1967tg}. Many of its entries were themselves decades old at the time of its publication. The 1978 compilation of Cauchois and S\'en\'emaud~\cite{Cauchois:1978,Jonnard:2011} adds to the picture. It contains some diagram lines absent from SRD-128, such as L-O shell transitions, as well as several non-diagram (satellite) L lines for each element; none were new measurements.  In short, most line energies in any collection of x-ray reference data are at least 50 years old, if indeed they have ever been reliably measured at all.

In the intervening decades, x-ray-optical interferometry and crystal lattice comparators have been developed, techniques that at last connect x-ray wavelengths directly to the Syst\`{e}me International d'Unit\'{e}s (SI) definition of the meter~\cite{Deslattes:1973}. The older measurements relied on x-ray emulsion films, where line asymmetries are nearly impossible to quantify, and data suffer from countless other systematic uncertainties that went unestimated in the original reports. Further, most existing reference data on fluorescence lines offer only a line energy without any sense of the critically important line profiles. Modern instrumentation offers the chance to improve on the older measurements~\cite{Mendenhall2017,Fowler:2017Metrology,Hudson:2020,Menesguen:2018,Menesguen:2019}; at the same time, new needs continue to arise for improved accuracy and for expanded compilations of fundamental parameters~\cite{Beckhoff:2012um}. The \emph{fundamental parameter method} of analysis depends on accurate and complete compilations, for instance~\cite{Beckhoff:2008,Wolff:2011,Sitko:2012}.

X-ray fundamental parameters are particularly needed for laboratory x-ray fluorescence (XRF) analysis. Numerous peaks overlap between x~rays from different shells of elements with very different Z values~\cite{Newbury:2015}. Efforts to extract elemental composition from overlapping features via peak fitting benefit from better knowledge of the underlying peak energies and shapes. Alpha-induced XRF instruments on the Mars Curiosity~\cite{Gellert:2015} and Perseverance~\cite{Allwood:2015} rovers are prominent examples of instruments that rely on reference data instead of reference samples.  In an example more specific to the x-ray features studied in this work, XRF analyses of rare-earth materials are often limited by the spectral overlap of emission lines from multiple analytes, given the large number of L lines per element~\cite{Kirsanov:2015fi}. The analysis of heavy elements typically uses L-line emissions, because the more intense K lines require excitation energies beyond the reach of most laboratory x-ray sources~\cite{Krishna2016}.  Also possible are overlaps in spectra of rare-earth ores with K lines of 3d transition metals, which are commonly present at much higher abundance than the rare earths~\cite{Smolinski2016,Silva2020}. To disentangle complicated XRF spectra with overlapping lines requires improved data on line energies and profiles for the rare-earth materials. 

Fundamental parameters are also necessary for astrophysical observations and for tests of atomic and nuclear theory. The \emph{Athena} x-ray observatory, under development, has an ambitious requirement to calibrate absolute energies to 0.4\,eV~\cite{AthenaAssessment:2011}. It will rely on knowledge of atomic and ionic emission energies to reach this goal. Highly charged ions provide stringent tests of atomic theory and quantum electrodynamics (QED)~\cite{Indelicato:2019,Machado:2020}. Exotic atoms---in which an unstable particle such as a pion~\cite{Anagnostopoulos:2003,Gotta:2015,Okada:2016}, kaon~\cite{Curceanu:2019,Hashimoto:2019}, muon~\cite{Okada:2020}, or antiproton~\cite{Gotta:2008} replaces an electron---probe both QED and the nuclear theory of quantum chromodynamics (QCD). Measurements of ionized or exotic atoms require instruments with excellent calibration, typically achieved through the use of x-ray lines of known energy.

This paper presents new measurements of x-ray fluorescence emission, enabled by improvements to an energy-dispersive spectrometer capable of measuring the absolute energies and line profiles of x-ray fluorescence lines. The instrument has been used previously for metrological measurements of lanthanide L lines~\cite{Fowler:2017Metrology} and for emission spectroscopy to distinguish spin states in iron compounds~\cite{Joe:2016}. The spectrometer makes nearly simultaneous measurements of many emission lines, including numerous lines previously measured with great precision and SI-traceable accuracy through the crystal diffraction technique. The known lines serve to calibrate the absolute energy scale and establish the energies of all other, unknown features. Such calibration linearizes the energy response of the spectrometer. The advantages of the energy-dispersive sensors include broad-band response, which allows spectra to be measured over a wide energy range at once, and a high efficiency suited to detection of even very faint features.

The spectrometer used in this work consists of 50 superconducting transition-edge sensors (TESs) selected from an array of 192 TESs, and a multiplexed readout based on superconducting quantum interference device (SQUID) amplifiers (section~\ref{sec:spectrometer}). The measurement technique features switching between calibration and science samples for optimal stability of the sensor gain between samples (section~\ref{sec:measurements}). Energy calibration (section~\ref{sec:calibration}) is anchored by the K lines of the 3d transition metals.

\begin{figure}
    \centering
    \includegraphics[width=\linewidth, keepaspectratio]{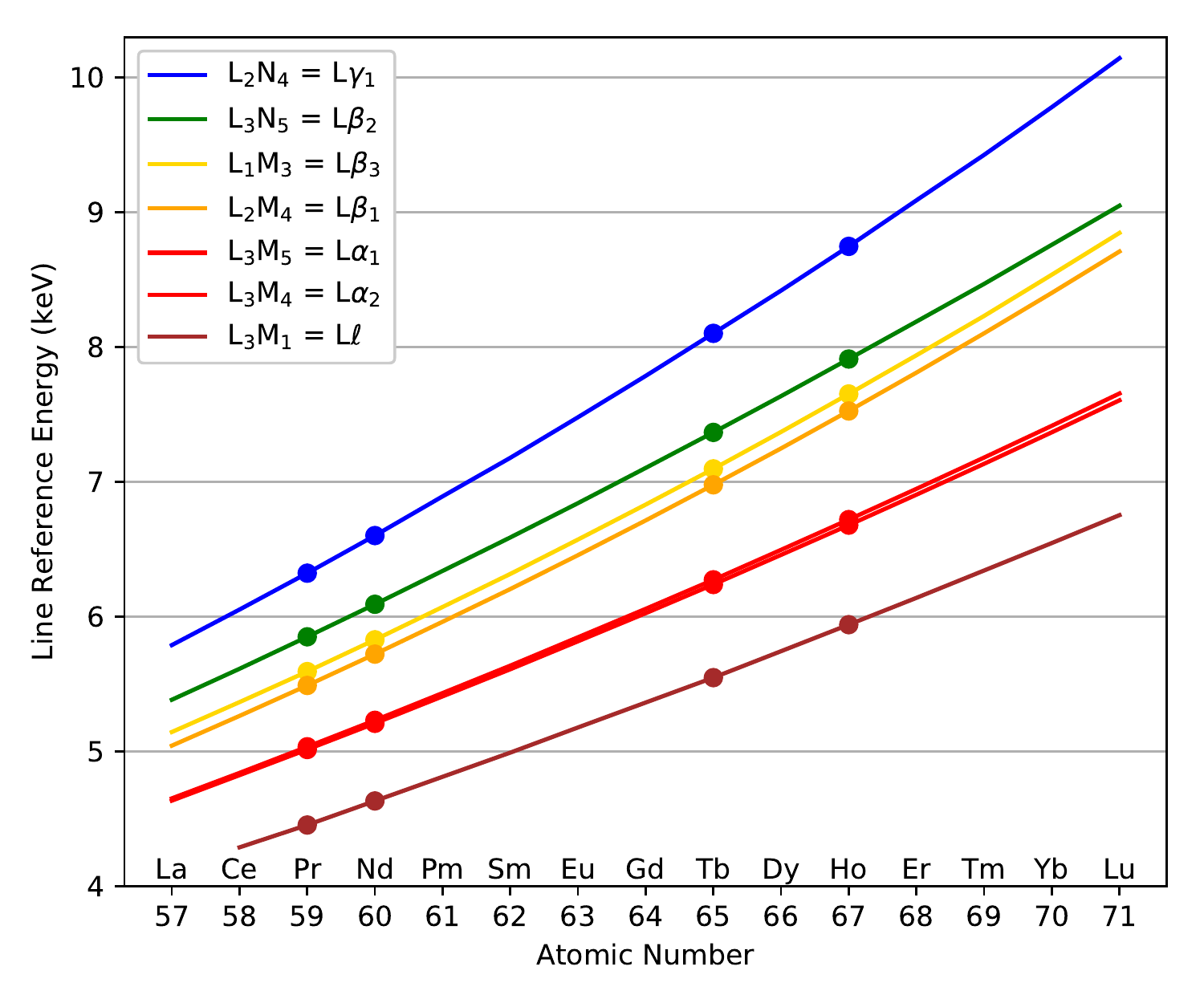}
    \caption{The most intense fluorescence L lines of lanthanide elements range in energy between 4\,keV and 10\,keV\@. IUPAC and Siegbahn notation are shown in the legend, from highest to lowest energy. The elements Pr, Nd, Tb, and Ho are studied here. Data are from Deslattes et al.~\cite{Deslattes:2003}.}
    \label{fig:lanthanide_L_summary}
\end{figure}

We have studied the L-line fluorescence emission of four lanthanide metals: praseodymium, neodymium, terbium, and holmium (Z = 59, 60, 65, 67). The L lines of the lanthanide-series elements span a range of hard x-ray energies in which numerous excellent calibrators are available. All L lines of the chosen elements, from the lowest-energy L$_3$M$_1$ up to the L$_1$ edge, lie in the well-calibrated range, 4.5\,keV to 10\,keV (figure~\ref{fig:lanthanide_L_summary}). Approximately two dozen L lines per element are found in the emission spectra. They are the L lines of highest intensity under photo-excitation by a broad-band source, making them the most relevant to typical users of x-ray fundamental parameters. The four elements offer a mix of lines both with and without relatively modern data in the literature. Modern measurements include eleven lines of Nd and Ho that were measured anew in Deslattes et al.~\cite{Deslattes:2003}, and the L$\ell$ lines and the L$\alpha_{1,2}$ doublet of each element studied by Mauron~\cite{Mauron:2003}. The elements Nd, Tb, and Ho were also measured with an earlier generation of our TES spectrometer~\cite{Fowler:2017Metrology}.

Many compilations of fluorescence line reference data, notably SRD-128, assign to each emission line a single ``peak energy'' and an uncertainty on it. While the peak energy is an appealing shorthand that facilitates quick comparisons between new work and earlier reports, it is next to impossible to define and measure peak energy in a robust fashion readily transferable between different instruments. Peak energies---local maxima in a spectrum---depend strongly on the exact energy response of an instrument and are also susceptible to instrument noise, at least in the absence of specific prior information about the spectral shape of an asymmetric emission line. 
Such terse, one-parameter summaries of what are often richly asymmetric line profiles undermine the very purpose of a reference collection. One cannot make use of a peak energy at the stated uncertainty level without knowledge of both the line shape and the detector response. For instance, we achieve an absolute energy-scale calibration better than 0.15\,eV uncertainty over much of the spectrum. Yet the peaks of many of the lines we measure differ by at least twice this much between the ideal spectrum and the one measured with the 4\,eV resolution of our instrument.

Users of reference data instead need a representation of complete line profiles. Such profiles allow one to make much more precise calibrations than a peak energy alone would permit; see, for example, ref.~\cite{Okada:2016}. The absolute calibration of the present work itself depends completely on reference data that consist of line profiles. Fortunately, this need is now widely recognized and generally accepted~\cite{Beckhoff:2012um,Chantler:2020}, and modern digital instrumentation and computing make it possible to meet the need.

Our primary goal, therefore, is to represent the full line profiles of the various lanthanide L lines that we measure. The result is an absolutely calibrated, transferable standard, complete with careful estimation of statistical and systematic uncertainties (section~\ref{sec:results}). As a secondary goal, we also estimate the peak energies of each line, for comparison to previously published reference values.


We believe that new x-ray measurements should supersede existing reference data when they improve on prior work in self-consistency, accuracy, and precision. We will show that our new results fulfill all three criteria. The results are more internally consistent than those of SRD-128, as they were measured by a single spectrometer calibrated by the same method over a two-week measurement period. By contrast, the SRD-128 data come from a broad diversity of sources, measured over many decades with a wide variety of instruments. Critical analysis choices are generally left unstated in the source documents, such as how the authors specified peak energies on asymmetric line profiles. We use a simple statistical metric to assess the consistency our new line energies with previous tabulations. Our results agree well with SRD-128 on the subset of lines for which the best reference data are available (if not with the reference data set as a whole). The self-consistency of the measurement implies that accuracy on the lines where stringent tests are possible extends to all of our measurements. Finally, the precision of our peak estimates is better than that of the reference data overall. Our estimates improve the median estimated uncertainty (including systematics) by at least a factor of four compared to published values and by a factor closer to ten after a reappraisal of past uncertainty values.

\section{The Transition-Edge-Sensor Spectrometer} \label{sec:spectrometer}

\begin{figure}
  \includegraphics[width=\linewidth]{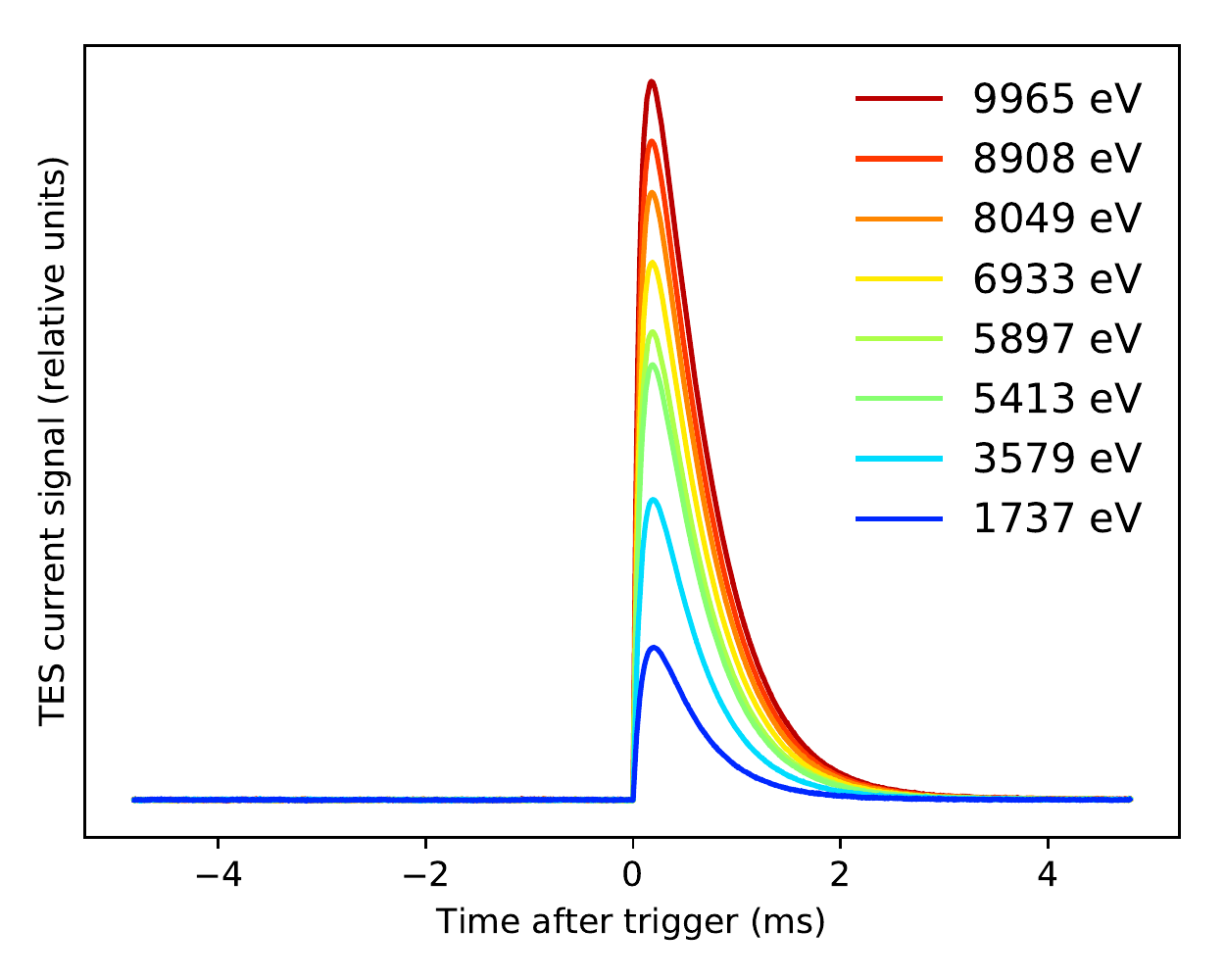}
  \caption{ \label{fig:raw_traces}
  Eight x-ray pulse records from a single TES that approximately span the fully calibrated energy range.  All pulses are nearly the same shape. The pulse size indicates the energy. The plotted records include detector noise, though it is too small to see at this scale. By convention, the transient \emph{decrease} in current is plotted as a positive quantity.}
\end{figure}

The spectrometer consists of an array of 192 superconducting transition-edge sensors, microcalorimeter detectors operated at an approximate transition temperature 111\,mK\@. All are fabricated on a single silicon wafer. The sensors act as parallel, independent x-ray spectrometers. Each consists of two components in close thermal contact: a thermalizing absorber and a thermometer. An absorbing layer of gold 1\,$\mu$m thick and 340\,$\mu$m square converts x-ray photons to thermal energy.  In thermal contact with and adjacent to the absorber is a molybdenum-copper bilayer (215\,nm Cu atop 65\,nm Mo) maintained in its narrow superconducting transition by Joule self-heating. This voltage-biased bilayer serves as a sensitive resistive thermometer. Its composition and dimensions are chosen to achieve the desired transition temperature and electrical resistance. Lateral separation of the two components into this ``sidecar'' arrangement permits use of a metallic absorber without its electrical properties changing those of the superconducting bilayer~\cite{Szypryt2019}.  Absorption of an x-ray photon causes a sudden rise in the bilayer temperature and electrical resistance and thus a rapid decrease in the bias current through it. Each sensor is in weak thermal contact with a colder heat bath at 70\,mK\@. A combination of reduced Joule heating in the sensor and thermal conduction removes the excess heat, and the TES returns to its quiescent state in a matter of milliseconds. The amplitude of the pulse indicates the amount of energy deposited. Figure~\ref{fig:raw_traces} shows example pulses at various energies. Further explanation of TES design and operation is given by Morgan~\cite{Morgan:2018}.
 
We employ a time-division multiplexed (TDM) readout system to reduce the wiring count and the attendant heat load onto the coldest stages of the cryogenic system. TDM reads out a set of up to 24 sensors in rapid succession through a single chain of two-stage SQUID amplifiers, so eight amplifier channels suffice to read out the entire TES array~\cite{Doriese:2016}. The multiplexing technique and the electronic and cryogenic engineering required to build an operational spectrometer are described elsewhere~\cite{Doriese:2017}. The review by Ullom and Bennett~\cite{Ullom:2015kp} surveys the design principles of TESs, both single sensors and arrays, as well as numerous applications of TES spectrometers.

\begin{figure}
    \centering
    \includegraphics[width=\linewidth, keepaspectratio]{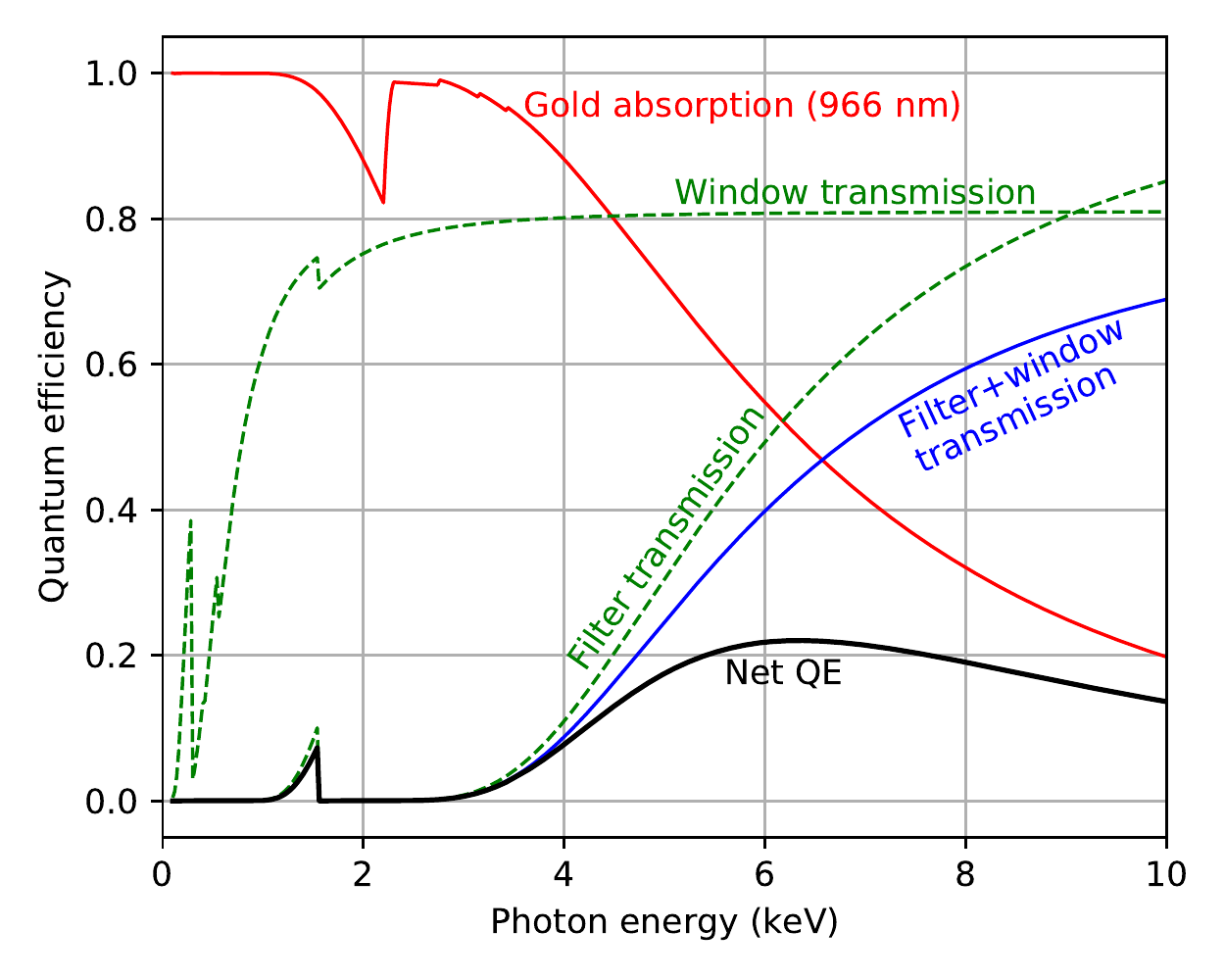}
    \caption{Quantum efficiency model. The three IR-blocking aluminum filters are 22.7\,$\mu$m thick in total. The aluminum filter transmission and that of the window are shown separately (green, dashed) and combined (blue, solid). The gold absorber is measured by a stylus profilometer to be 966\,nm thick, $\pm1\,\%$~\cite{Szypryt2019}; its absorption is shown. The product of the transmission and gold absorption is shown as \emph{Net QE}. 
}
    \label{fig:qe}
\end{figure}

The microcalorimeter array is located inside an adiabatic demagnetization refrigerator, separated from the x-ray emission source by a vacuum window and a series of thin aluminum filters that block infrared and visible radiation~\cite{Szypryt2019}. Figure~\ref{fig:qe} shows the x-ray transmission of the window and filters as a function of energy. The vacuum window transmission is computed to be quite flat, between 80\,\% and 81\,\% for the entire energy range of interest (4.5\,keV to 10\,keV), because it is made of a thin polymer filter backed by a thick stainless steel mesh with a 19\,\% areal fill factor. The absorption efficiency of the gold falls from 80\,\% at 5\,keV to 20\,\% at 10\,keV\@. The transmission of the combined 22.7\,$\mu$m of aluminum rises from 20\,\% to 85\,\% over the same range.  No discontinuities in the quantum efficiency as large as 0.05\,\% are present between gold's M edges (2\,keV to 3\,keV) and its L edges (11.9\,keV and higher). All efficiencies are estimated from standard mass attenuation values~\cite{Schoonjans:2011} and nominal or measured dimensions.  An estimated variation in the absorber thickness of less than 1\,\% across the array contributes negligible systematic uncertainties to the fluorescence line profiles that are the subject of this work. The absolute efficiency is not relevant to our results.

The spectrometer improves on that used for our previous metrological measurements with microcalorimeters~\cite{Fowler:2017Metrology} in a few important ways. A new sensor design allows us to achieve an energy resolution of approximately 4.0\,eV full-width at half-maximum (FWHM) for 7\,keV x~rays; the resolution was 5\,eV in the earlier work. The sensors are also being used farther from their saturation energy (typically 20\,keV), leading to an energy response that is more nearly linear by a factor of at least two, as detailed in section~\ref{sec:cal_curves}. Still, the energy resolution is an increasing function of energy, both because the TES resistance grows less than linearly with temperature as the sensor moves up the superconducting transition, and because the feedback loop in the TDM readout system is most challenged by the rising edges of pulses at higher energies.

The use of a gold layer as the photon absorber produces a nearly perfect Gaussian energy-response function. The evaporated bismuth used as an absorber in earlier spectrometers tends~\cite{Yan2017} instead to yield a Bortels function~\cite{Bortels:1987}: a Gaussian-like shape with a pronounced and unwelcome exponential tail to low energies. The tail is attributed to trapped energy in long-lived states at the  boundaries of the many tiny grains of evaporated bismuth layers~\cite{Yan2018,ONeil:2019}.

For the metrological measurements, we elected to use only 64 sensors out of 192 in the array to minimize cross-talk effects. Lead tape placed on the 50\,mK aluminum filter prevented x~rays from reaching the unused TESs. By reading out only one-third of the sensors, we reduced unwanted electrical cross-talk in the system, and by blocking x~rays from the others, we also eliminated any possible thermal effects that might affect the active detectors.

These changes in the spectrometer improve the results compared with our earlier measurements in several ways. Better energy resolution is directly helpful, because the lines being studied are not fully resolved. The reduction in crosstalk also makes the energy-response function more nearly Gaussian, as crosstalk tends to produce long and poorly understood tails in that function. More important to the energy-response function is the use of gold absorbers, which nearly eliminates the low-energy exponential tail. The improved linearity allows for better energy resolution at the higher energies and also reduces the systematic uncertainties inherent in the interpolation of energy-calibration curves. Finally, the use of a fast-switching sample holder (section~\ref{sec:6SSS}) allows us to employ calibration lines very near to energies of the lanthanide lines of interest.

\section{X-ray fluorescence line measurements} \label{sec:measurements}

A commercial x-ray source induces the fluorescence by first accelerating electrons from a cathode through 12.5\,kV toward a tungsten primary target.  Some x~rays produced in the tungsten (both bremsstrahlung and characteristic fluorescence emission) strike the nearby secondary target, consisting of one or more high-purity metal samples, the elements of interest. The TES array is located at a distance of 7\,cm from the secondary target. The enclosures of the tube source, the sample holder, and the TES array together shield the array from direct illumination by the x-ray tube. Only photons reflected, scattered, or absorbed and re-emitted by the target reach the sensors. Each sensor has an x-ray absorbing region defined by a single square aperture in an aperture array of gold-plated silicon. Its holes are each 280\,$\mu$m $\times$ 280\,$\mu$m, for a total of approximately 4\,mm$^2$ active area across the 50 best active sensors. The volume containing the x-ray output of the tube source, the secondary targets and the window of the spectrometer is partially evacuated to an absolute pressure of roughly 40\,kPa.

The secondary targets are pieces of metal or metal foils purchased from chemical-supply companies. For calibration, we used metallic titanium, vanadium, chromium, manganese, iron, cobalt, nickel, and copper (atomic number $Z$ from 22 to 29) with purity of at least 99.9\,\%. The K$\alpha$ and K$\beta$ emission-line energies and shapes of these transition metals are well established~\cite{Holzer:1997ts,Chantler:2006va,Chantler:2013wp}. The lanthanide metal samples are similarly 99.9\,\% pure, with the exception of praseodymium, which has a 0.4\,\% neodymium content according to the supplier's assay.

The mean photon rate was intentionally limited to a maximum of 7.0 photons per second per sensor. Although these TES detectors are capable of operation at 100 or more photons per second, the lower rate was chosen for this work in order to minimize the effects of detector-to-detector crosstalk and to optimize the energy resolution.

\subsection{The six-sided sample switcher} \label{sec:6SSS}

The absolute calibration of the energy scale demands measurements of numerous calibration metals, ideally spanning the widest possible range of energy and including well-known fluorescence lines quite close to the lanthanide lines of interest. Calibration lines and science lines that overlap in energy are nearly impossible to disentangle if measured at the same time. On the other hand, if they are not measured simultaneously, or nearly so, then small, slow drifts in the spectrometer gains of even one part in $10^4$ can render spectra incomparable and the calibration useless~\cite{Szypryt2019,Tatsuno:2020}. A six-sided sample holder, rotated under computer control, was designed to meet our need for measurement of closely placed lines in rapid succession. To reduce magnetic interference with TES electrical signals, we used a piezoelectric motor to turn the sample holder. This switching technique allows us to perform a near-simultaneous calibration; in every rotation we measure calibration lines and lanthanide L lines that are closely interspersed in energy.

\begin{figure}
    \centering
    \includegraphics[width=\linewidth, keepaspectratio]{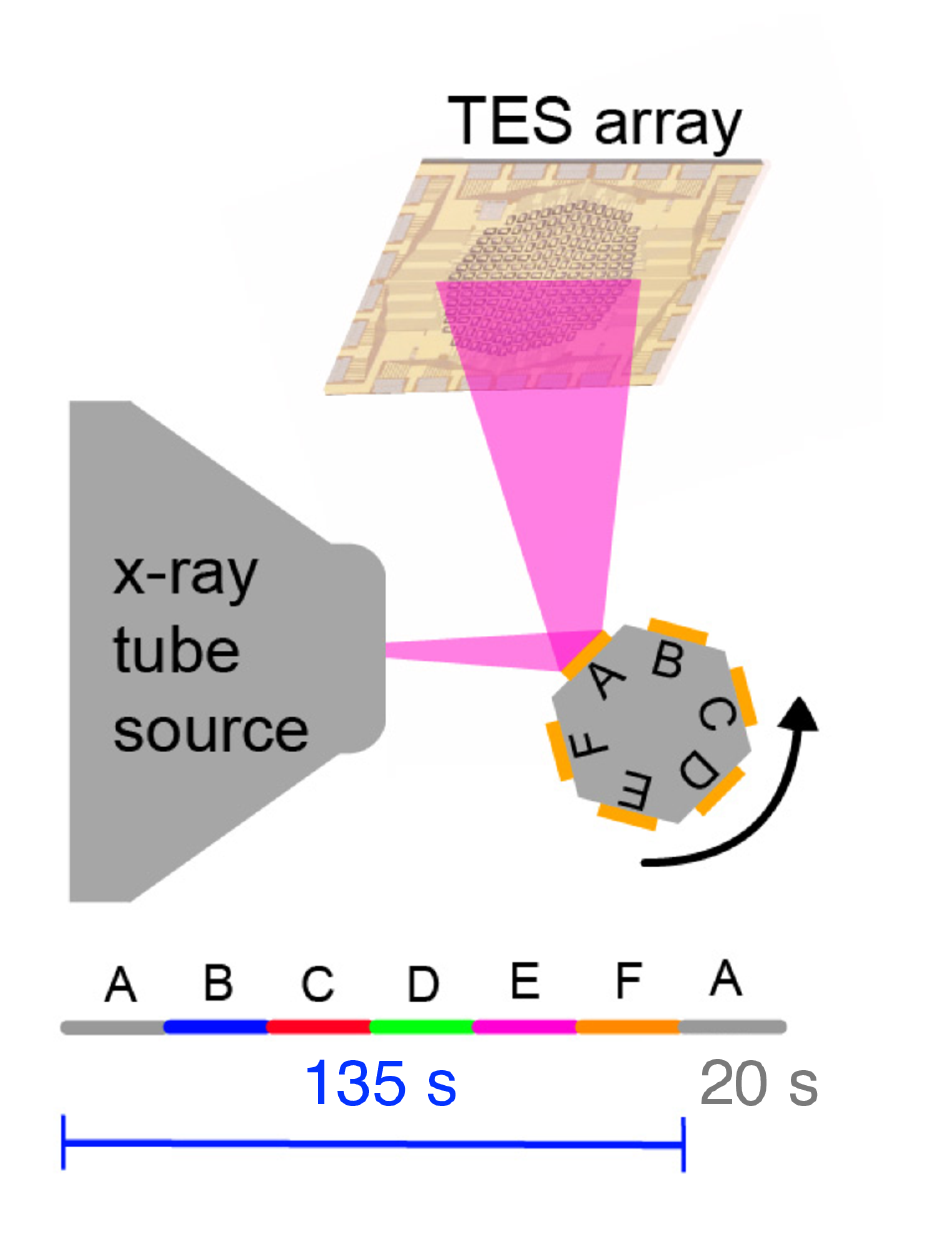}
    \caption{A diagram of the six-sided sample switcher. The sample targets are approximately {7\,mm} square, of which approximately {4\,mm} are illuminated by the tube source. The TES array is located 7\,cm from the samples. The hexagonal sample holder is enclosed in a copper structure (not shown) with entrance and exit apertures for x~rays.}
    \label{fig:sixSSS}
\end{figure}

The sample holder, illustrated in figure~\ref{fig:sixSSS}, is a hexagonal piece of aluminum onto each side of which a separate target is mounted. Targets consist of one, two, or three metal pieces affixed to the sample holder with rubber cement. The x-ray source illuminates an area of each target approximately 4\,mm in diameter. A  sample chamber of oxygen-free high-conductivity copper encloses the sample holder, with small apertures designed to ensure that only the desired target is struck by x~rays from the tube source. Additional effective targets beyond the six primary targets can be created by positioning the sample holder in between two of its standard orientations. We use six targets for the Tb and Ho measurement, seven for Pr and Nd. Each target is individually irradiated by the source for 20 seconds (Tb and Ho measurement) or 1 minute (Pr and Nd), after which the sample holder is rotated to the next target. X~rays emitted from the metals in a target are collected by our microcalorimeter array. The return to each target within two or seven minutes ensures that any changes in gains over longer time scales affect all measurements equally.

\begin{table}
%
\begin{tabular}{l|lrrr}
Side & Elements & Avg $E$ & Rate & Total x~rays \\ \hline
4A & Cr, Fe, Ni & 6899.1\,eV &  5.86 & 7\,399\,132 \\
4B & Mn, Co, Cu & 6843.9\,eV &  5.88 & 7\,418\,571 \\
4C & Fe, Ho     & 6906.0\,eV &  5.95 & 7\,512\,721 \\
4D & Ho         & 7027.7\,eV &  5.72 & 7\,227\,430 \\
4E & Tb         & 6664.7\,eV &  5.95 & 7\,509\,928 \\
4F & Cr, Co, Cu & 6752.7\,eV &  5.90 & 7\,445\,000 \\
\\
5A & Mn, Co, Ti & 5350.9\,eV &  6.37 & 6\,677\,900 \\
5B & Cr, Fe, Ti & 5330.7\,eV &  6.35 & 6\,658\,448 \\
5C & V, Mn, Co  & 5688.4\,eV &  6.41 & 6\,722\,120 \\
5D & Nd         & 5746.6\,eV &  6.40 & 6\,706\,974 \\
5E & Nd, Cr     & 5692.5\,eV &  6.39 & 6\,700\,528 \\
5F & Pr, Fe     & 5684.6\,eV &  6.42 & 6\,726\,801 \\
5G & Pr         & 5601.4\,eV &  6.51 & 6\,827\,905 \\
\end{tabular}

\caption{\label{tab:samples}
Elements in the sample switcher's targets. The first six entries (4A--4F) correspond to the April 2018 measurement of terbium and holmium; the last seven (5A--5G) to the May 2018 measurement of praseodymium and neodymium. The seventh position was created by turning the switcher to an intermediate angle. \emph{Avg E} gives the mean photon energy in eV\@. \emph{Rate} is the rate of valid pulse events (pulses per second) averaged over the approximately 45 to 50 highest-resolution§ TESs. We attempt to balance both the total array rate and (to the extent possible) the average energy across the sides. \emph{Total x~rays} is the number of valid pulse events per side, summed over all sensors and the approximately 56 and 51 hours of observation in April and May, respectively.}
\end{table}

To achieve the best possible control over systematic uncertainties, it was necessary to balance the spectra generated by the several targets as well as possible. We equalized both the photon count rate and the mean energy per detected photon. The goal was to ensure that the detectors were operating under the same conditions for all targets, in order to minimize any differences in response. A voltage of 12.5\,kV on the x-ray tube was used throughout. Samples with two or three targets were first adjusted by changing the placement or relative amounts of the target metals to approximately equalize the mean photon energy from all six or seven targets in the same measurement. Next, the x-ray tube's current was set to a different value for each target, in order to match the rates of detected photons. Data collected for 2.5 seconds after rotations of the sample holder are ignored, mainly to allow for settling of the tube current, but also to avoid any motor-induced transient signals and to ensure rotation is complete. The elements that are included on more than one target allow us to assess how well this balancing act equalized the gains across the six or seven targets, as discussed in section~\ref{sec:cal_consistency}.

Table~\ref{tab:samples} gives the contents of the several targets and the mean photon energy and mean rate for each. Sides 4D, 4E, 5D, and 5G each contain one lanthanide element and are the primary science targets. Sides 4C, 5E, and 5F contain both a lanthanide and a transition metal; they are intended for cross-checking results. The other six sides contain three transition metals and are used for the absolute calibration (section~\ref{sec:cal_transition_metals}). All figures in the table are computed after all pulse selection steps given in the following section, so they give the rates and numbers of photons used in the final energy spectra. The true rates and numbers of photons striking the sensors are approximately 9\,\% higher than these values when events dropped from the final analysis are included.

\subsection{Pulse selection and estimation of pulse heights} \label{sec:pulse_analysis}

The TES measurements consist of the sensor current, sampled regularly at intervals of 2.56\,$\mu$s. The data acquisition software scans the stream for the sudden changes in TES current that signify the onset of x-ray pulses. 
When a pulse is identified, a record of a fixed length (3750 samples, or 9.6\,ms) is recorded for later analysis, in equal parts before and after the trigger. Eight examples are shown in figure~\ref{fig:raw_traces}.

The millions of pulse records are analyzed by techniques described more thoroughly by Fowler et~al.~\cite{Fowler:2015is}, which we summarize here. ``Clean'' single-pulse events are selected by identification and removal of any records that are unusable: piled-up (multi-pulse) events, photons that arrive before the energy of the previous photon is fully dissipated, and any other transient problems. Approximately 93\,\% of pulse records remain valid after these cuts; the removed records contain either a second pulse or a substantial tail from a prior pulse in roughly equal measure.

Another cut applied to the data is one that identifies and removes a subtly different category of pulses. We estimate that approximately 1.5\,\% of all x~rays absorbed in a TES stop not in the gold absorber, as intended, but in the silicon nitride membrane on which the gold is deposited (thickness approximately 500\,nm) or the SiO$_2$ etch-stop layer (120\,nm). Such \emph{membrane hits} in the energy range of the present study are found to generate pulses $0.982\pm0.002$ times the size of pulses caused by non-membrane photons of equal photon energy. Membrane hits would therefore appear in a spectrum as echoes of the main peaks at a correspondingly lower energy. Fortunately, we can identify membrane hits on a pulse-by-pulse basis, because they show a pronounced lengthening of the decay tail at least $\sim 2$\,ms after the pulse peak. Removal of long-tailed pulses cuts an additional 3\,\% of records, including nearly all membrane hits and an estimated 2\,\% of all absorber hits at 5\,keV, falling to 0.5\,\% at 9\,keV\@. The energy-dependent effect of this selection is treated as a factor in the spectrometer quantum efficiency when line profiles are fit.

Linear optimal filtering reduces each clean record from 3750 samples to a single estimator of pulse height. A specific weighting of the samples in a record is chosen---based on training data---to yield the minimum-variance, unbiased estimator for the pulse height. This weighting is statistically optimal under several assumptions that are known to be only approximately true: that pulses are transient departures from a strictly constant baseline level; that the noise is an additive, stationary, multivariate Gaussian process independent of the signal level; and that all pulses at any energy are proportional to a single standard pulse shape. 

The filtered pulse height depends to a small degree, unfortunately, on variables such as the baseline level and the photon's exact arrival time at the sub-sample level. Slow variations in the baseline level track changes in the temperature of the cryogenic thermal bath, which change the sensor gain at the level of $10^{-4}$ to $10^{-3}$. We correct the gain by a factor linear in the baseline level; we  select the correction factor for each sensor that minimizes the information entropy of its resulting energy spectrum, because a minimum-entropy spectrum has the sharpest possible features~\cite{Fowler:2015is}.

A small bias in pulse heights that depends on the exact photon arrival time relative to the 2.56\,$\mu$s sampling clock also depends on the photon energy. 
We correct for the bias with a smooth function of arrival time and energy, again by minimization of spectral entropy. The adjustments for gain drift and sub-sample arrival time are typically equivalent to changes in the energy of 5\,eV and 1\,eV, respectively. After these corrections, each record has been summarized by a minimum-noise, bias-corrected estimate of its pulse height. Conversion of this quantity to the x-ray photon energy requires an extensive calibration procedure.

\section{Absolute Calibration of the Spectrometer} \label{sec:calibration}

Calibration of the absolute energy scale of a TES spectrometer requires measurement of multiple, narrow x-ray emission features whose energy and intrinsic line shapes are already well known, relative to the features under study. Although the gain and nonlinearity of a TES can be roughly estimated from a combination of first principles and measurements of sensor properties, the results will not attain anywhere near the required level of accuracy~\cite{Irwin-Hilton:2005,Pappas2018}. Such estimates might have 10\,\% relative uncertainties at best, while the spectrometer has intrinsic precision at least two orders of magnitude better than this.

The goal of the empirical, absolute-energy calibration is to determine a function $f_i$ for each sensor  $i$ such that when a pulse with optimally filtered and bias-corrected pulse height $P$ is measured in that sensor, $E=f_i(P)$ is the best possible energy estimate. We learn such functions directly from the pulse data by the deliberate inclusion of well established lines, such as the K lines of transition metals, among the lines being studied.  The procedure is similar to that described previously \cite{Fowler:2017Metrology}, but the use of the six-sided sample switcher allows for a much closer interleaving of lines.

\subsection{Calibration features from transition metals}  \label{sec:cal_transition_metals}

\begin{figure*}
    \centering
    \includegraphics[width=\linewidth,keepaspectratio]{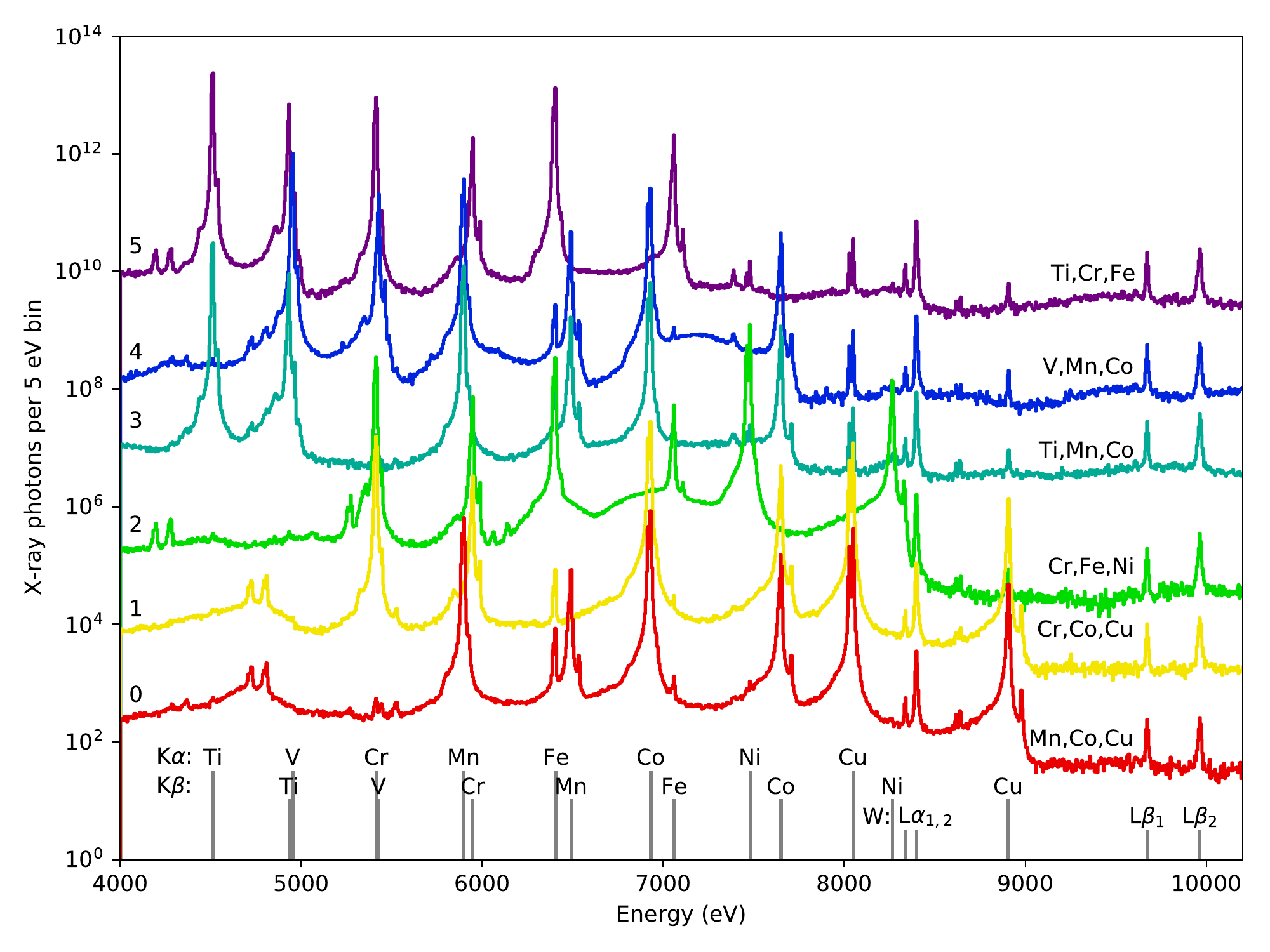}
    \caption{Emission spectra of the six calibration targets, the samples that contain 3d transition metals. Spectra are numbered at the left and named at the right, according to the three transition metals that comprise each target. Vertical lines at the bottom of the figure indicate the energies of the peaks used for calibration. Spectrum 0 is placed at the correct vertical scale; all others are scaled up by a factor of $30^i$ for spectrum $i$, to reduce overlap between the curves. Targets 0--2 are from the higher-energy (April 2018) measurement and 3--5 from the lower-energy (May 2018) one. Additional features (gold-escape peaks and extra fluorescence lines) are discussed in section~\ref{sec:background}.}
    \label{fig:calibrators_full_spectra}
\end{figure*}

The principal lines used for the absolute calibration of energies in this work are the K lines of eight transition metals, with six metals used in each of the two measurement campaigns. The line energies of the six calibrators chosen for the April 2018 campaign of terbium and holmium spanned 5.4\,keV to 8.9\,keV; the energy range of the calibration lines for the May 2018 measurement of praseodymium and neodymium was 4.5\,keV to 7.6\,keV\@. Table~\ref{tab:samples} indicates which elements were combined to form the individual targets. Figure~\ref{fig:calibrators_full_spectra} shows the full energy spectrum of each calibration target, the end result of the calibration procedure.

\begin{figure}
    \centering
    \includegraphics[width=\linewidth, keepaspectratio]{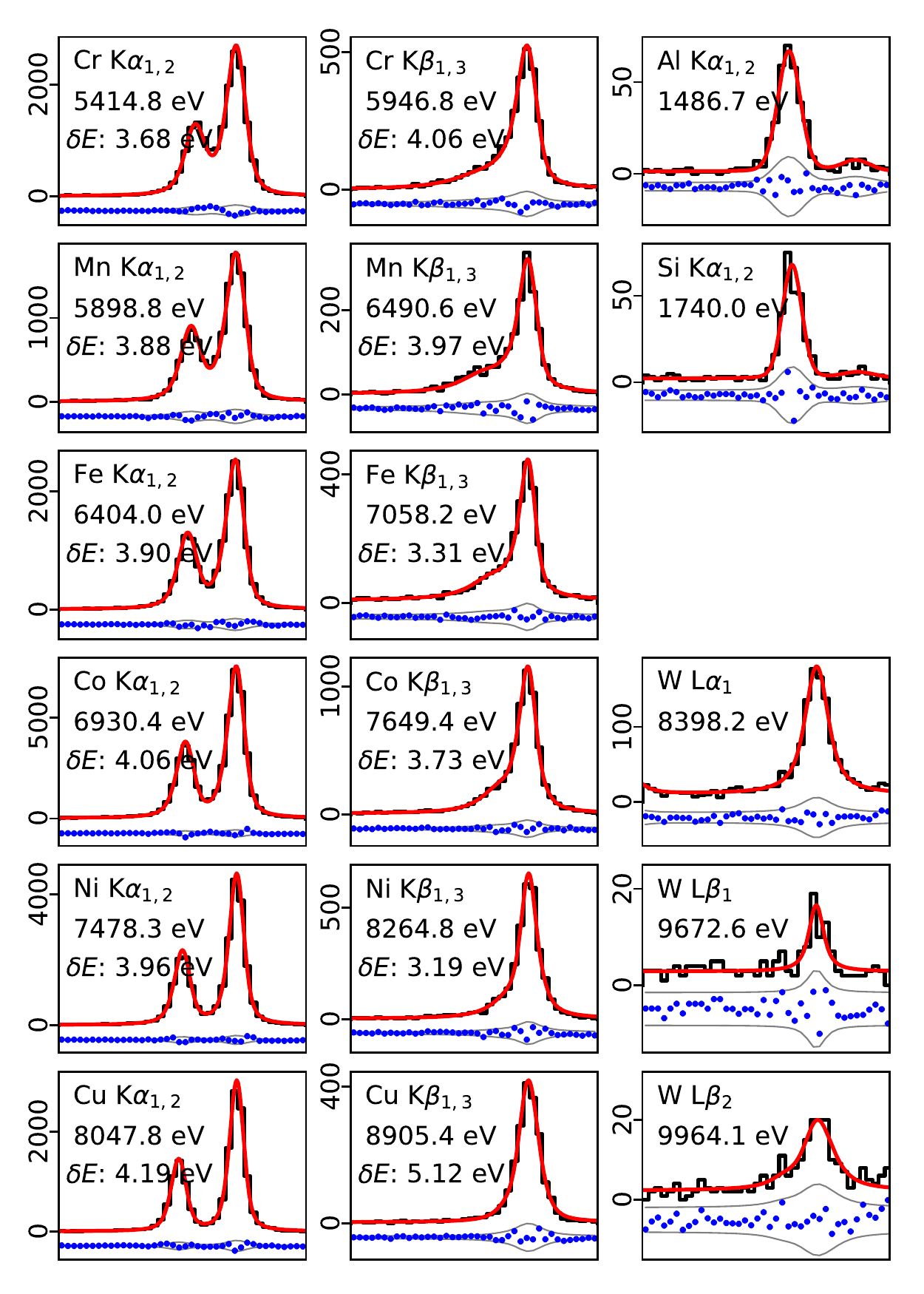}
    \caption{Example fits of the calibration lines for one TES on one day during the April 2018 (higher-energy) measurement of Tb and Ho. Measured values are shown as a histogram (black); the model fit is a smooth curve (red). The tungsten L lines are assumed to be a single Voigt function with width estimated from the all-channel combined spectrum. The x-axis is arbitrary (pulse height) units spanning a range of 1\,\% of their central value, in 50 bins. The text gives the line name, the peak energy, and (where reliable) the energy resolution $\delta E$---the estimated FWHM of the Gaussian energy-response function. The intrinsic shape and width of Al, Si, and W lines are not known well enough to yield a reliable estimate of $\delta E$. The dots beneath each spectrum indicate the residuals (data minus fit), surrounded by gray bands that represent the $\pm 2\sigma$ range given by Poisson counting statistics.}
    \label{fig:example_cal_fits}
\end{figure}

Models for the K line profiles of transition metals are taken from high-resolution wavelength-dispersive measurements~\cite{Holzer:1997ts, Chantler:2006va, Chantler:2013wp}.  Some employed diffracting crystals with unit cell spacing traceable to the SI meter, via lattice-spacing comparators to a crystal measured absolutely by x-ray/optical interferometry. Others used diffracting crystals that are themselves calibrated absolutely by K line emissions. All profiles are characterized as the sum of Voigt functions. All claim absolute calibration accuracy of 0.04\,eV or better, and our repeated measurements agree at this level. Calibration anchor points are thus not the limiting systematic uncertainty in the present work.  Aside from their excellent precision and accuracy, the K line results are usable for our calibration procedure because they include complete models of each line profile. 

A separate calibration curve is estimated for each TES on each day, for a total of 274 calibrations. Each calibration has two steps. First, a model of the K line profiles is fit to the uncalibrated pulse-height spectra, over a 1\,\% energy range near each K-line peak.  The six free parameters of each fit include a line intensity, two parameters of a locally affine transformation from pulse heights to energy, two parameters of a linear background spectrum, and the sensor's Gaussian energy resolution. Figure~\ref{fig:example_cal_fits} shows the fits for a single sensor.  From the results of all fits, a first, approximate calibration curve can be made. It is refined by performing a new fit to each peak in approximate-energy space, where the models better match the measurement. All fits use a Poisson maximum-likelihood fit, which has smaller bias than a fit that minimizes the simpler Neyman's or Pearson's $\chi^2$ statistic~\cite{Fowler:2014JLTP}. The typical statistical uncertainty in the pulse height of the K-line peaks is equivalent to 0.10\,eV for a single TES on a single day.

A model of the energy response function as a Bortels function \cite{Bortels:1987}, a Gaussian convolved with a one-sided exponential, was also attempted for fitting the combined (all-sensor) data. The results were consistent with the exponential having no effect (either zero amplitude or zero exponential scale length). The tail-free response represents an important improvement from the response function of previous TES sensors. We attribute this advance to the use of gold as the x-ray absorbing material~\cite{ONeil:2019} (section~\ref{sec:spectrometer}) instead of evaporated bismuth~\cite{Yan2017,Yan2018} and henceforth take the energy response to be exactly Gaussian.

The fits provide an estimate of the Gaussian energy resolution for each sensor on each date. We find that the median resolution is approximately $\delta E=4$\,eV (Gaussian FWHM), growing slightly with increased energy. Section~\ref{sec:resolution_combine} addresses the question of how to combine data from sensors of slightly different resolution.

In some sensors, the K$\alpha$ lines of aluminum or silicon, or both, are apparent at 1487\,eV and 1740\,eV\@. Both materials are present in the vicinity of the TES sensors and can be made to fluoresce at a low intensity by the x~rays that the science and calibration targets emit. When the pulse-trigger threshold is set low enough, these lines are visible. They are fit and the results included as anchor points in the calibration curves. The aluminum K$\alpha$ line shape is based on several sources~\cite{Lee:2015hh,Wollman:2000kd,Schweppe:1994fc}, and silicon K$\alpha$ is assumed to be the same shape, rescaled to the higher energy of Si emission. Although these points have minimal effect on the lanthanide line energies that are the focus of this work, they do help to validate the procedure we use to generate calibration curves from the anchor points (section~\ref{sec:cal_curves}).

The L lines of tungsten provide additional calibration lines in the energy range above 8\,keV, because the x-ray source that illuminates our targets is a tungsten foil bombarded with energetic electrons. A small but measurable number of photons emitted by the tungsten reflect off the target metals into the spectrometer. Tungsten L$\alpha_1$, L$\beta_1$, and L$\beta_2$ lines (8398\,eV, 9673\,eV, and 9964\,eV~\cite{Deslattes:2003}) are readily fit in each single-sensor, single-day spectrum of pulse heights. The intrinsic line widths (typically 7\,eV FWHM) are estimated from our data, but they are also consistent with published results, where they exist~\cite{Gokhale:1983, Diamant:2001, Mauron:2003}. The tungsten lines substantially constrain the calibration above 8\,keV and are the only constraints above 9\,keV\@.

\subsection{Calibration transfer curves} \label{sec:cal_curves}

\begin{figure}
    \centering
    \includegraphics[width=\linewidth, keepaspectratio]{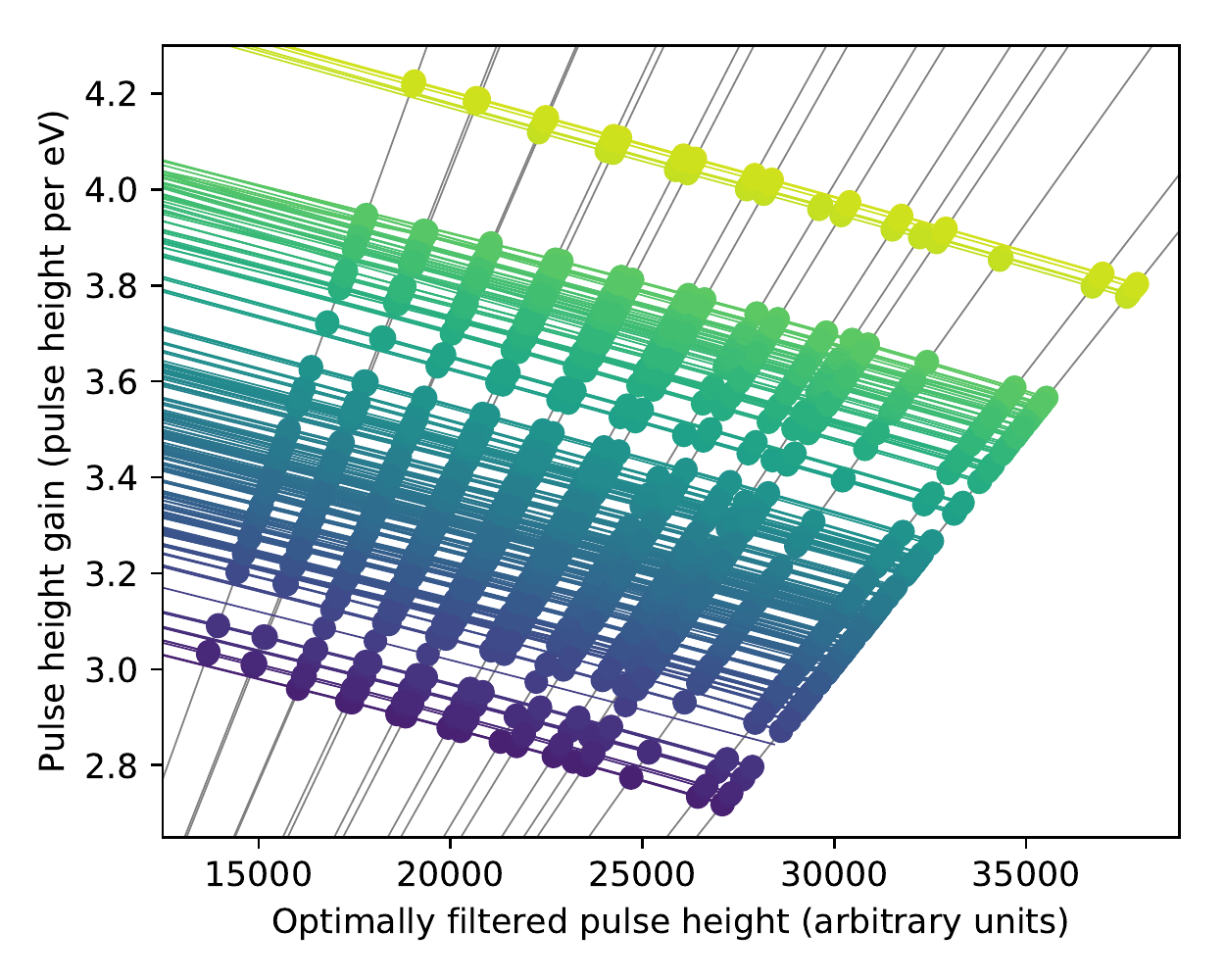}
    \caption{Calibration curves for 274 sensors. Each represents a single sensor as configured for a single day of measurement. Anchor points (dots) are based on fits to K lines of transition metals or L lines of the tungsten primary x-ray source. The negatively sloped curves are the energy calibration curves: splines of gain (pulse height-to-energy ratio) as a function of pulse height. The gray diagonal lines of positive slope represent constant energies, from Ti K$\alpha$ (4511\,eV) at left to W L$\beta_2$ (9964\,eV) at right. As table~\ref{tab:samples} indicates, not all transition metals are present for all measurements.}
    \label{fig:cal_curves}
\end{figure}

We construct the calibration curves with transition-metal K lines and the tungsten L lines serving as the ``anchor points.'' These anchor points have an intrinsic uncertainty on the measured pulse heights, so it is important that we require the calibration curve to pass near but not exactly through them.  Specifically, the calibration curve is $E=P/g(P)$ where the gain $g$ is a cubic \emph{smoothing spline}. The smoothing spline is an approximating, piecewise-cubic function that minimizes a linear combination of the goodness-of-fit statistic $\chi^2$ and the integrated squared curvature (the integral of $|g''(P)|^2$ over the calibrated range of $P$). This balanced optimization is intended to avoid overfitting of noise in the calibration data \cite{Fowler:2017Metrology}. Figure~\ref{fig:cal_curves} shows all calibration curves up to 10\,keV\@.

The choice that the gain $g$ should be a spline function of $P$ instead of $\log(g)$ (as in our previous work) or $1/g$ was determined by comparisons of calibration curves between 5\,keV and 7\,keV and between 6\,keV and 8\,keV\@. In all sensors over either energy range, the function $g(P)$ has less curvature than the alternatives that we have considered, so it is assumed to be the most amenable to the smooth interpolation and extrapolation required.

For comparison of energy linearity between the TES array used in this work to that of other spectrometers, we quantify the local nonlinearity by the \emph{power-law index}. It is defined as $\alpha\equiv\mathrm{d}\log P/\mathrm{d}\log E$, which is the index of the power-law curve tangent to the calibration curve at any energy. The median values of $\alpha$ (over all sensors and all measurement days) at energies 4500\,eV, 6000\,eV, and 7500\,eV are 0.92, 0.89, and 0.87, varying by less than $\pm0.01$ among sensors. In the previous TES-based spectrometer~\cite{Fowler:2017Metrology}, the corresponding median values of the nonlinearity $\alpha$ were 0.82, 0.75, and 0.67.
The much improved linearity of the new detectors, along with the increased number of calibration lines, improves the uncertainty in the absolute calibration scale by a factor of approximately two.

\subsection{Calibration curve assessment and systematic uncertainty} \label{sec:cal_assessment}


We assess the accuracy of the calibration curves through several cross-validation (CV) tests, in which new calibration curves are constructed with either one or two anchor points deliberately omitted. The known energy of the omitted point or points can be compared with the energy that the new curve yields, which tests the fidelity of the curve at each anchor point. When two points are very close (as for example, Cr K$\alpha$ and V K$\beta$ at 5415\,eV and 5427\,eV), both may be omitted to help us learn about calibration uncertainties far from the nearest anchor point.

CV tests are not ideal tools for the estimation of systematic errors, simply because a test curve constructed without a certain point (or pair of points) is structurally different from the curve used for calibration. Furthermore, the systematic error on any measurement depends on multiple factors that can be hard to disentangle; both the energy difference from the nearest anchor point and the energy itself affect the systematic. At higher energies, the uncertainty increases along with the sensor nonlinearity. Above 9\,keV, the calibration relies entirely on the observation of L$\beta$ lines of tungsten, whose very low intensity introduces additional uncertainty to the calibration curve; we estimate an uncertainty of 0.4\,eV at 9.7\,keV\@. Finally, CV tests with limited anchor points cannot completely rule out the possibility that the detailed shape of $R(I, T)$ at the superconducting transition of a TES imposes unseen small-scale features on the calibration curve.

Drop-2 CV tests up to 7.5\,keV show that the dropped pairs of points, which are all approximately 450\,eV from the nearest remaining anchor points, move by no more than 0.17\,eV and often by much less. The drop-1 CV tests at 6400\,eV and above generally agree: they indicate that the shift is 0.1\,eV at 7000\,eV and twice this at 8000\,eV, even when the nearest remaining anchor point is only approximately 200\,eV away. From these observations, we model the calibration systematic uncertainty as the quadrature sum of the following three terms:
\begin{enumerate}
    \item An interpolation uncertainty term, growing with the energy difference between a measurement and the nearest anchor points. The term is scaled to 0.15\,eV uncertainty for a 450\,eV energy difference.
    \item A term equal to zero for any energy less than 6500\,eV but growing linearly with energy above that value, to reach 0.4\,eV uncertainty at 9673\,eV (the W L$\beta_1$ line). The combination of this term and the next is called the \emph{calibration curve absolute uncertainty} in what follows.
    \item A constant 0.07\,eV, to account for both the approximating nature of the calibration interpolation and the possibility of undetected features. The accuracy of drop-2 CV tests strongly suggests this as an upper limit to the size of such features.
\end{enumerate}

\subsection{Combination of spectra from sensors of various energy resolution} \label{sec:resolution_combine}

Each sensor has a unique energy resolution. This fact complicates estimation of the intrinsic line profiles that are the goal of this study, because we combine the spectra of all sensors to achieve high signal-to-background before fitting for line profiles. To preserve a Gaussian instrumental broadening for the combined spectrum, one can include only spectra that have nearly equal resolution.

The individual TES resolutions are determined from fits to the K$\alpha$ and K$\beta$ lines of transition metals (section~\ref{sec:cal_transition_metals}), which yield a typical uncertainty of $\pm0.1$\,eV\@. We assume that these resolutions depend only on the energy being measured and apply equally to calibration K lines and to the L lines under study. From these results, we can standardize a single \emph{nominal energy resolution} as a function of energy:
\begin{equation} \label{eq:nominal_res}
\delta E = 2.8\,\mathrm{eV} + 1.8\,\mathrm{eV}(E/10\,\mathrm{keV}),
\end{equation}
where $\delta E$ is the FWHM of the Gaussian energy-response function for an x~ray of energy $E$. This specific functional form is 0.2\,eV worse than the median resolution of all detectors and amounts to approximately 4.0\,eV in the energy range of interest.

The several K-line fits provide resolutions that are each compared with the nominal resolution and the differences averaged for a sensor.  Any TES having a resolution within $\pm 0.2$\,eV of the nominal value is included directly in the combined, final spectrum. The 15\,\% of TESs with resolution more than 0.2\,eV \emph{worse} than the nominal value, on average, are omitted. The approximately one-quarter of TESs with resolution at least 0.2\,eV \emph{better} than the nominal value are degraded intentionally by the addition of Gaussian random perturbations to their measured energies, such that the resulting resolution follows equation~\ref{eq:nominal_res}. We use this technique to combine spectra, even though the spectra from lanthanide metals are nearly unaffected by this procedure. It ensures that a single width characterizes the energy-response function over small energy ranges, and the combined energy response is most nearly Gaussian.

\subsection{Calibration consistency between samples} \label{sec:cal_consistency}


\begin{figure}
    \centering
    \includegraphics[width=\linewidth, keepaspectratio]{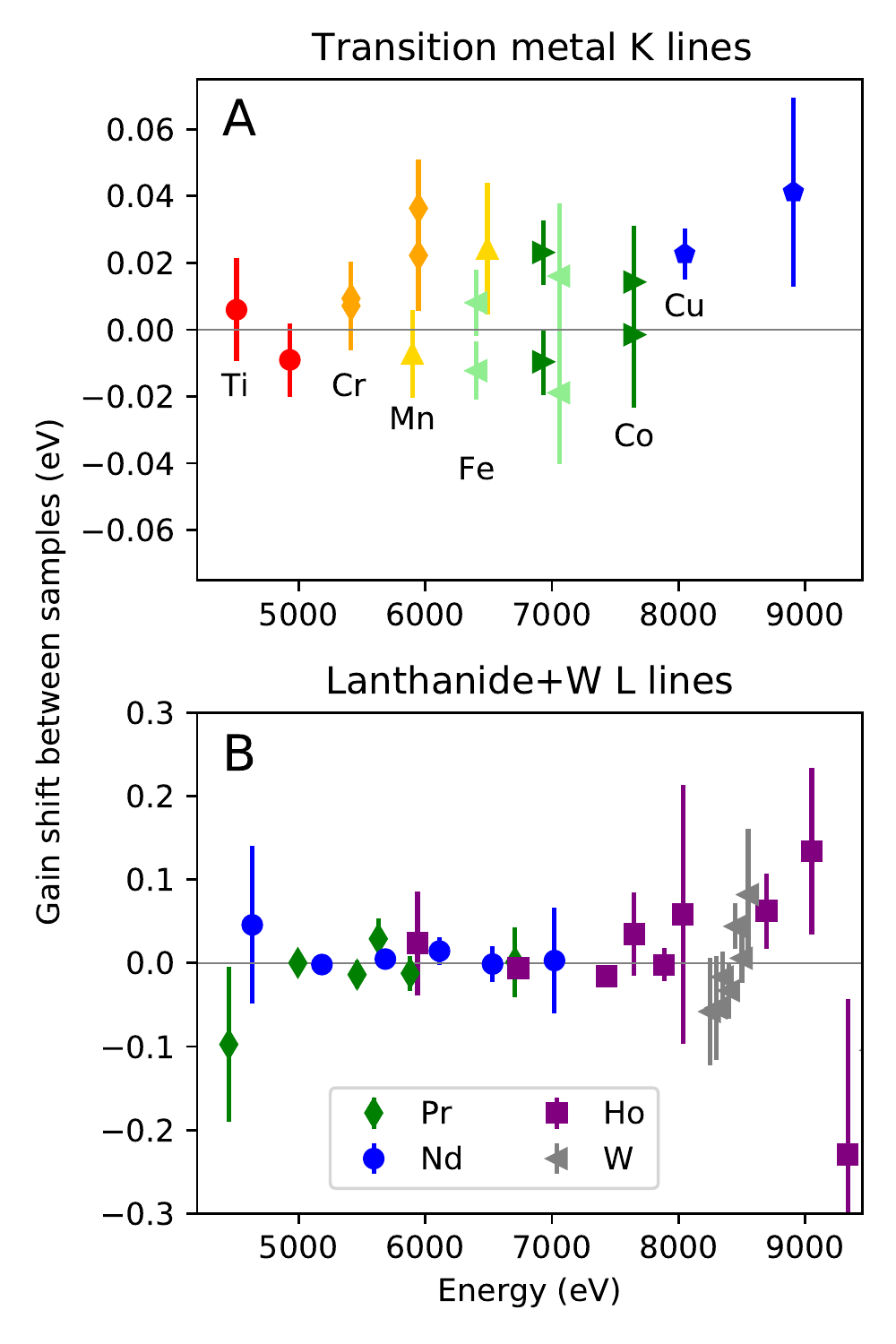}
    \caption{Calibration stability. Each point represents an emission line that appears on spectra from two targets. The vertical value is the difference between the best-fit energy of the (arbitrarily defined) ``first'' appearance on the sample switcher and the second. (A) Transition metals that appear on two targets. Chromium, iron, and cobalt were repeated in both the April and May 2018 measurements, so their K$\alpha$ and K$\beta$ lines appear twice in the figure. (B) Bright lines of the lanthanide metals used in two targets and the tungsten L$\alpha_1$ line. The latter appears on all sample sides, so the points depict each side's energy minus the mean, with an arbitrary horizontal scatter to reduce collisions.}
    \label{fig:cal_consistency}
\end{figure}

Many elements were placed on two sides of the sample holder, as table~\ref{tab:samples} indicates. Elements measured twice provide strong consistency checks, because we use the same calibration curve for all targets. Specifically, lines emitted by repeated elements can verify that the system gain was the same among all targets after the gain-balancing procedure (section~\ref{sec:6SSS}). The repeated elements include most of the transition metal calibrators; three of the lanthanide elements (all except for terbium); and tungsten, which is found in every spectrum.

The repeated transition metals include Cr, Fe, Co, and Cu in the April 2018 measurements, and Ti, Cr, Mn, Fe, and Co in the May 2018 measurements (figure~\ref{fig:cal_consistency}A). They produce eight and ten consistency checks, respectively, because both the K$\alpha$ and K$\beta$ lines of each element can be compared. In all cases but two, the repeated K lines agree to within 0.025\,eV\@. The exceptions are the April measurements of the Cr and Cu K$\beta$ lines, which differ by 0.038\,eV and 0.049\,eV\@. The uncertainties on our estimates of the difference are typically 0.010\,eV\@. 

The repetition of the lanthanide elements Pr, Nd, and Ho on two samples apiece also offers consistency checks, even though the profiles and energies of their L lines are not known with the precision of the transition-metal K lines. As with the K lines, we find the consistency of the L lines to be excellent. We compare 31 lines of the three elements (figure~\ref{fig:cal_consistency}B). Half of these lines are intense enough to constrain the consistency to 0.1\,eV or better. In all cases, the energy difference between repeated, intense lines is less than 0.01\,eV below 7\,keV; less than 0.02\,eV below 8\,keV; and approximately 0.10\,eV to 0.20\,eV between 9\,keV and 10\,keV\@. The appearance of tungsten L lines in all samples at 8398\,eV, 9673\,eV, and 9964\,eV provides additional evidence in the high energy range. The 8398\,eV L$\alpha$ peak is consistent to 0.1\,eV in all samples; the two L$\beta$ peaks are consistent to 0.2\,eV in all samples (figure~\ref{fig:cal_consistency}B does not show the L$\beta$ peaks because of their large uncertainties).

We combine these results on the calibration consistency between samples to assign a systematic uncertainty that depends only on the energy $E$ of the emission line. It is the larger of 0.01\,eV or $0.09\,\mathrm{eV}[(E-7.4\,\mathrm{keV})/1\,\mathrm{keV}]$.

In summary, numerous improvements in the measurement combine to simplify the instrumental broadening and make the calibration technique more effective than in a previous study that also employed cryogenic microcalorimeters. A combination of gold x-ray absorbers and better analysis techniques to identify membrane hits eliminates the non-Gaussian features in the instrumental response. The switchable sample holder allows for calibration and science lines arbitrarily close in energy, which reduces the interpolation uncertainty in the calibration procedure by a factor of two or more.

\section{Lanthanide Fluorescence Line Results} \label{sec:results}

\begin{figure*}
    \centering
    \includegraphics[width=\linewidth,keepaspectratio]{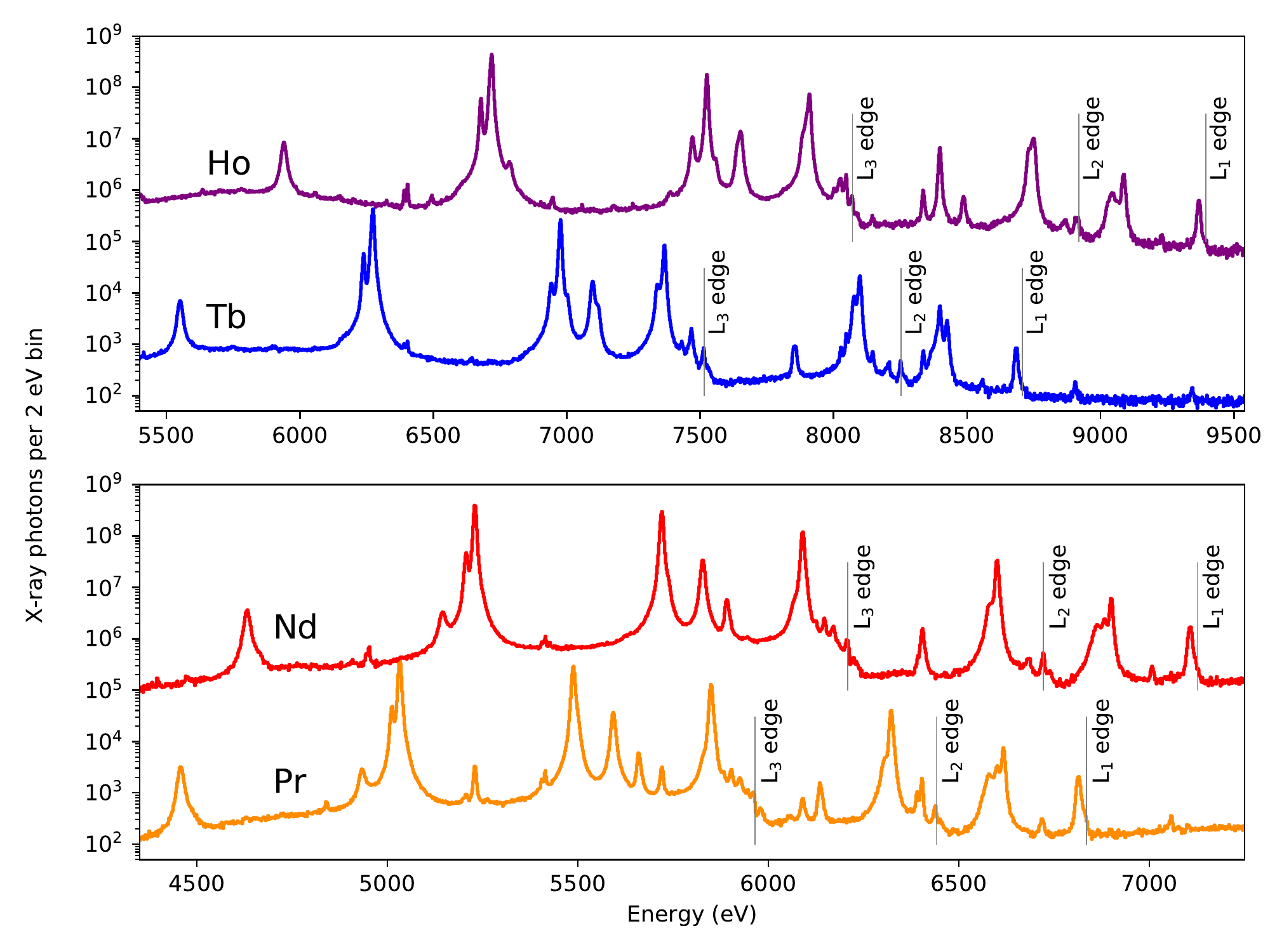}
    \caption{Emission spectra of the four lanthanide samples. The full range of L lines is depicted, with vertical marks at each L edge (from~\cite{Deslattes:2003}). The Ho and Nd spectra are scaled up by a factor of 1000 for clarity. Individual transitions are identified in the tables that follow.}
    \label{fig:lanthanide_full_spectra}
\end{figure*}

Characterization of the photoinduced x-ray emission of the lanthanide metals Pr, Nd, Tb, and Ho is the primary subject of this work, including absolute energies and the full line shape. Figure~\ref{fig:lanthanide_full_spectra} shows the four emission spectra. An online supplement to this publication provides the spectra numerically.

We determine the line profiles from the data, first by characterizing any interfering background features (section~\ref{sec:background}), then by approximating the L line shapes as a sum of multiple Voigt functions in excess of the background features (section~\ref{sec:shape}). We estimate the peak energies of the lines from the line profile models (section~\ref{sec:peak_energy}). Sections~\ref{sec:stat_uncert} and \ref{sec:systematics} discuss statistical and systematic uncertainties, respectively.

\subsection{Estimation of background features} \label{sec:background}

Numerous background features are apparent in the calibration and the lanthanide samples. These include gold \emph{escape peaks} and emissions from trace elements. These features are assessed quantitatively, as a few coincide with the energies of certain of the lanthanide L lines. This model of background features can later be effectively subtracted from the spectra when the lanthanide line profiles are analyzed.

The escape peaks are a pair of dim echoes of any line at approximately 2\,keV below the primary line and approximately 0.2\,\% times its intensity. They result when the capture of an x-ray photon in the gold absorbing material is followed by the re-emission of a gold M~line that happens to escape the absorber. The result is a pulse with a missing thermal energy most often equal to the energy of either the M$\alpha$ or M$\beta$ line of gold: 2123\,eV or 2205\,eV, respectively~\cite{Zschornack:2007wu}. Examples are apparent in figure~\ref{fig:calibrators_full_spectra} as the line pairs near 4200\,eV that echo the Fe K$\alpha$ line in samples containing iron and the pairs near 4700\,eV in samples containing cobalt.

We model the escape peak effect as a copy of the primary spectrum, first convolved with a Gaussian of width 10\,eV (approximately 1.5 times the width of gold M lines at this resolution~\cite{Zschornack:2007wu,Campbell:2001}). This smoothed spectrum is then rescaled and shifted to lower energy, repeating the process to capture both M lines of gold. The energy shifts that best fit the several cleanly detected escape peaks are 2125\,eV and 2206\,eV, with a decrease in the shift of 0.3\,eV per 1\,keV increase in photon energy. The small variation in shift with energy, as well as the difference between the shifts and the literature values of the gold M lines, together suggest problems with the absolute calibration of up to 1\,eV in the energy range of 2\,keV to 3.5\,keV\@: most escape peaks appear in a range where there are no anchor points, so excellent calibration was not possible. The best-fit intensity scaling is approximately $2\times 10^{-3}$ for the M$\alpha$ peak and $1.5\times 10^{-3}$ for the M$\beta$.

\begin{table}
    \centering
    \begin{tabular}{llrr}
    & & \multicolumn{2}{c}{Fraction} \\
    Sample & Impurity & Measured & Assay \\ \hline
         $_{59}$Pr & $_{58}$Ce & $6\times10^{-4}$ & $3.7\times10^{-4}$ \\
                   & $_{60}$Nd & $70\times10^{-4}$ & $36.2\times10^{-4}$ \\ \hline
         $_{65}$Tb & $_{73}$Ta & $7\times10^{-4}$ & ``Present'' \\ \hline
         $_{67}$Ho & $_{62}$Sm & $2\times10^{-4}$ & --- \\
                   & $_{64}$Gd & $2\times10^{-4}$ & --- \\
                   & $_{66}$Dy & $5\times10^{-4}$ & $<1\times10^{-4}$ \\
                   & $_{68}$Er & $5\times10^{-4}$ & $6\times10^{-4}$ \\
                   & $_{69}$Tm & $2\times10^{-4}$ & $<1\times10^{-4}$ \\
                   & $_{73}$Ta & $2\times10^{-4}$ & ``Present'' \\
    \end{tabular}
    \caption{Lanthanide impurities detected in our samples by their L$\alpha$ and L$\beta_1$ emission. The measured fractions were estimated as the relative amplitudes of the L$\alpha$ peaks for the impurity and the primary element in each sample and have uncertainty of $\pm10^{-4}$ or $\pm10\,\%$ of the value, whichever is larger. The last column gives the relevant values from the materials assays provided by the vendor. For the impurity estimation, we assume equal absolute  fluorescence yield and equal self-absorption effects for all elements. The Nd, Tb, and Ho samples were stated to be 99.9\,\% pure and the Pr 99.5\,\%, values approximately consistent with our data.}
    \label{tab:impurities}
\end{table}

We also detect fluorescence lines of trace elements. This category includes parasitic emission from stainless steel vacuum parts and the copper sample chamber, plus impurities in the lanthanide samples. The largest impurity is found in the Pr sample; according to the materials assay, it contains 0.4\,\% Nd, though we estimate 0.7\,\% Nd. Other lanthanide metals are detectable as impurities in three of the samples at the level of $10^{-3}$ or lower (table~\ref{tab:impurities}).  This represents a very sensitive XRF detection of 500 ppm impurities of metals differing in $Z$ by only $\pm 1$, a feat that would not be possible with solid-state energy-dispersive detectors of modest resolution.

Other trace element lines detected include the K$\alpha$ and K$\beta$ lines of Fe and Cu (in all samples) and of Cr (in the Pr and Nd spectra only). The most intense L lines of the science samples contain at least 200 times as many photons as the most prominent background lines from transition metals.

We estimate the contribution to each spectrum by background features for all escape peaks and trace-element emission lines. These estimates are used in the next stage to prevent background features from biasing the estimation of line shape of the lanthanide L lines.

\subsection{Line shape estimation} \label{sec:shape}

Atomic fluorescence lines in general and the L lines of lanthanide metals in particular can have very complex shapes~\cite{Mauron:2003}. While the diagram lines are identified with specific final and initial states of a single-electron transition, the underlying atomic physics can yield an asymmetric emission feature that actually consists of contributions from many thousands of slightly different transitions, depending on the exact configuration of all ``spectator'' electrons before and after the transition~\cite{Deslattes:2003,Deutsch:2004,Lowe:2010,Chantler:2013theory,Guerra2015,Pham:2016,Zeeshan:2019}. The initial excitation event changes atomic potentials, which in turn often excites a second electron to a higher shell (``shake-up'' processes) or to the continuum (``shake-off'')~\cite{Chantler:2013theory,Pham:2016,Ito:2016,Ito:2018}. The double excitation causes different fluorescence energies. Relaxation processes involving two or more electrons can add other satellite lines and radiative Auger-effect lines. In the current measurement, lanthanide atoms are excited by tungsten L lines below 10\,keV\@. Because this energy is less than twice the L$_3$ edge energies of our samples, double-L-shell ionization and the ``hyper-satellite'' lines that result are not expected~\cite{Deutsch:2004,Zeeshan:2019}.

A perfect absolute calibration of the energy scale is not by itself sufficient to establish a metrological transfer standard for the L lines; the line shapes are also necessary. Ideally, these shapes should be conveyed as analytic, easily used summaries of the potentially complex spectra. The task of line-profile estimation is to represent a  fluorescence line profile in a way that balances simplicity and parsimony of expression against completeness and accuracy.

An analytic form often used to describe fluorescence line shapes is the sum of Lorentzian (also called Cauchy) distributions. In the simplest possible case of a single-electron transition between two states with a well-defined lifetime and no excited spectator electrons, a single Lorentzian is an excellent approximation to the theoretical line shape. Even in the vastly more complex case of real-world fluorescence lines, sum-of-Lorentzian expressions have several virtues: they are easily written and computed; the three free parameters per component are readily understood as a line center, a width, and an intensity or amplitude; and their long tails are consistent with the expected and observed tails of emission lines. Furthermore, when the effect of instrumental broadening is (or can be approximated as) convolution with a Gaussian, then the Lorentzian distribution becomes a Voigt function~\cite[Eq.~7.19]{NIST:DLMF}. The Voigt function (or the Faddeeva function, of which it is the real part) is widely available in numerical computing libraries.

Lorentzian or Voigt profiles have been used for decades by various groups working to understand certain lines theoretically or to establish lines as metrological standards~\cite[among many examples]{Krause:1975,Klauber:1993a,Klauber:1993b,Schweppe:1994fc,Deutsch:1995,Holzer:1997ts,Diamant:2001,Mauron:2003,Anagnostopoulos:2003,Chantler:2006va,Chantler:2013wp,Illig:2013,Smale:2013,Mendenhall2017}. When instrumental broadening is small compared to any features in a line, the fit is made to a sum of Lorentzians; otherwise, a sum of Voigts is used. The long history and widespread adoption of sum-of-Voigts representations in the community argues for its use in the current work on the lanthanide L lines. We use Voigts, with one departure from convention: additional Gaussian broadening as a free parameter of the model. We do not attempt to identify each Voigt component with a specific cause grounded in atomic physics; this is an empirical description of the spectral profile.

One can easily adjust the Gaussian width used in a sum-of-Voigts model, adapting the model to a different instrument with different resolution. Setting this width to a value \emph{smaller} than the width appropriate to the instrument that acquired the spectrum, even to zero, is effectively a deconvolution.  It comes with all the practical problems that deconvolution always does: sensitivity both to noise in the data and to inaccurate estimation of the original instrumental response function~\cite[\S13.1.2]{NumRec3:2007}. 

\begin{figure}
    \centering
    
    \includegraphics[width=\linewidth,keepaspectratio]{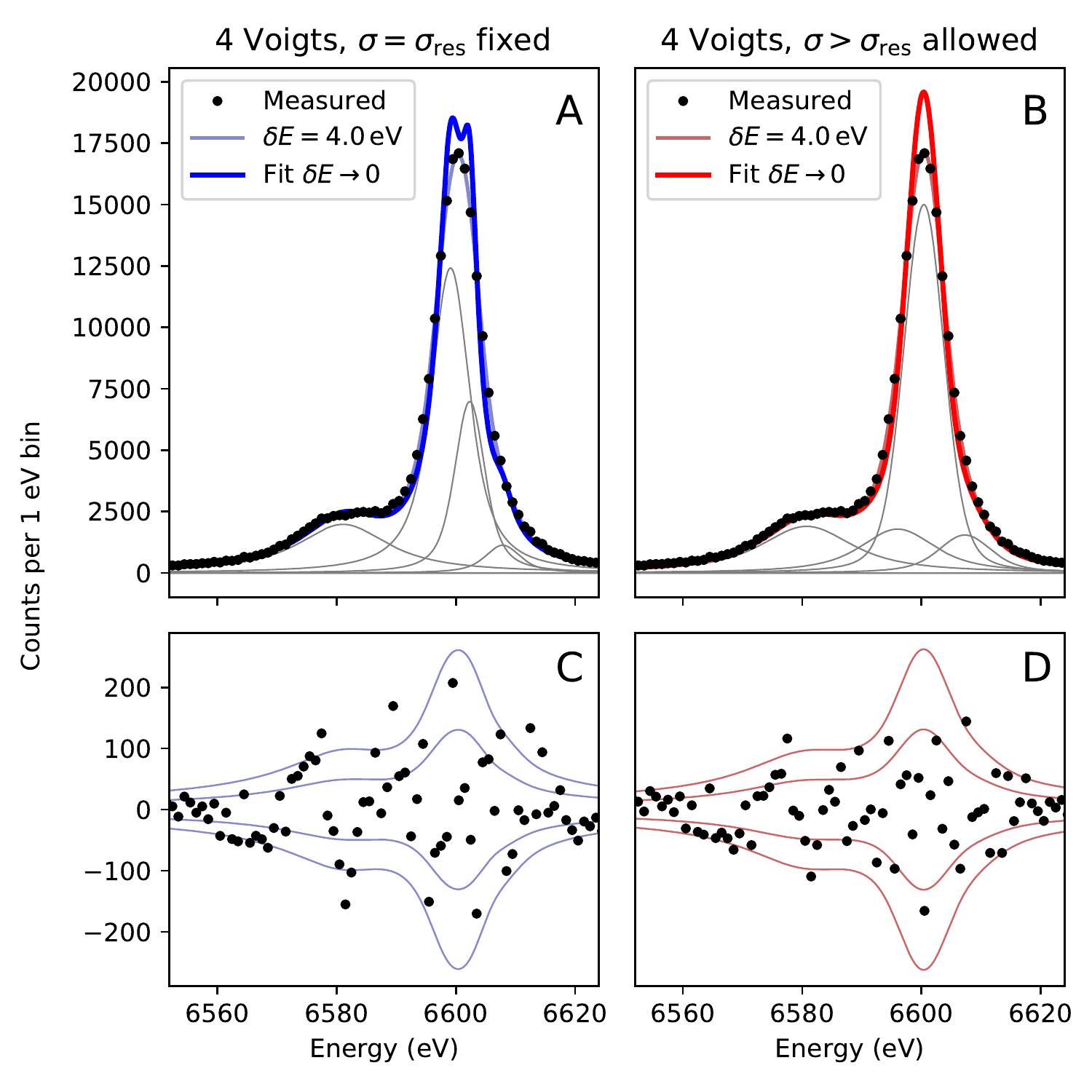}
    \caption{The Nd L$\gamma_1$ (L$_2$N$_{4,5}$) line fits: (A) with no zero-resolution model Gaussian width ($\sigma_0=0$), and (B) with model Gaussian width as a free parameter of the fit ($\sigma_0=2.39$\,eV). In both cases, the model consists of nine Voigt functions and a linear background, with fitting over the range 6331\,eV to 6732\,eV\@. A narrower range is shown for clarity. Four of the nine Voigt components are centered in this range and are shown in gray. The background is very small (fewer than 100 counts per bin) in each fit. Panels A and B show the measured spectrum (identical in both panels) as points; the best-fit model as a thin line; and the same model with background removed and the resolution set to $\delta E=0$, or $\sigma=\sigma_0$ (equation~\ref{eq:sigma}). (C) The residuals (measured minus fit) with curves indicating the $\pm 1\sqrt{N}$ and $\pm 2\sqrt{N}$ Poisson standard deviations for the $\sigma_0=0$ case. (D) Residuals for the $\sigma_0=2.39$\,eV case. The double-peaked model in the $\delta E=0$ limit (A) is an example of the unexpected spectral features that a nonzero $\sigma_0$ can prevent. The $\chi^2$ value of the fit is 536 for 372 degrees of freedom when $\sigma_0=0$ and 464 for 371 degrees of freedom when $\sigma_0>0$ is allowed.}
    \label{fig:example_excess_gauss}
\end{figure}

Consider the specific case of Nd L$\gamma_1$. Figure~\ref{fig:example_excess_gauss}A shows a model of four Voigt functions fit to the data, where $\delta E$=4.00\,eV, as well as the $\delta E=0$ limit of the same model. While we cannot say with certainty that the sharp, asymmetric peak in the latter curve is unphysical, we judge it more likely to be an artifact of deconvolution.

Only with great caution should one use a sum-of-Voigts model with $\delta E$ smaller than that of the instrument used to produce the data to which the model was fit. Because we intend to create a transfer standard of the broadest possible applicability, though, we want the models to be as well-behaved as possible for the case of $\delta E<4$\,eV\@.
Therefore, we have explored a range of strategies that aim to minimize artifacts in the $\delta E=0$ case without requiring significant compromises in fit quality.  These strategies included the use of various regularization penalties, asymmetric line components, and certain purely linear models. Strategies not chosen and our reasoning are described more fully in \ref{sec:failed_strategies}.

The most satisfactory approach we found was to modify the sum-of-Voigts model slightly, by allowing the fits to employ a finite \emph{zero-resolution model Gaussian width}. That is, we do not require the total Gaussian width in each Voigt component to obey $\sigma = \delta E/\sqrt{8 \ln 2}$, but instead allow  $\sigma$ to exceed this resolution-only value.  Unlike regularization, which deliberately nudges a fit function away from the maximum likelihood values, this approach introduces an extra  degree of freedom to \emph{improve} the agreement between model and data. At the same time, we find empirically that the zero-resolution limit of the resulting model contains unphysical features far less often. Figure~\ref{fig:example_excess_gauss}B shows the same data fit with the additional freedom, both at the actual as-measured resolution and in the limit of zero instrumental broadening.

We believe that besides improving the fits, this additional Gaussian width captures an essential fact about the underlying atomic physics: that a full theoretical model of any given fluorescence line is actually a sum over hundreds or thousands~\cite{Lowe:2010,Chantler:2013theory,Pham:2016,Menesguen:2019,Zeeshan:2019} of distinct transitions between various initial and final states, each a separate Lorentzian. A model consisting of a few Voigt functions must be an approximation, as it cannot possibly capture the full complexity of the relaxation of a many-electron system like an excited lanthanide atom. Conversely, realistic empirical data cannot constrain a model with as many degenerate parameters as hundreds of Voigt functions would present, not without copious theoretical guidance about the intrinsic line shape. The heuristic of additional Gaussian width allows each Voigt component in our model to represent many actual transitions over a narrow range of energies. The reason to use a single value per fitting ROI for this additional width is practical: it suffices for our current purposes, and more free parameters would make the nonlinear fitting process more fragile.

Specifically, the fit in each spectral region uses the signal model 
\begin{equation}\label{eq:signal_model}
S(E) \equiv \sum_{i=1}^N\,I_i V(E; c_i, \Gamma_i, \sigma),
\end{equation}
where $N$ is the number of Voigt components in the model. $V(E; c, \Gamma, \sigma)$ represents a normalized Voigt function centered at energy $E=c$ with Lorentzian and Gaussian width $\Gamma$ and $\sigma$. $I_i$ is the integrated intensity of component $i$, normalized such that $\sum_i\,I_i$ is the number of photons expected over all energies in an ideal sensor. The Gaussian width $\sigma$ (which is common to all $N$ Voigt components) is the quadrature sum of the resolution term and the zero-resolution model width:
\begin{equation}\label{eq:sigma}
    \sigma^2 = \sigma_\mathrm{res}^2 + \sigma_0^2\hspace{5mm}\mathrm{where}\ \sigma_\mathrm{res}\equiv \delta E/\sqrt{8 \ln 2}.
\end{equation}
The value $\sigma_0\ge 0$ is allowed to vary in the fit, as are the $N$ amplitudes, center energies, and Lorentzian widths of each component. The full model is the product of the signal and the quantum efficiency plus backgrounds. There are $3N+3$ free parameters in each fit: three per Voigt, the common $\sigma_0$, and two to describe the continuum background.  \ref{sec:fit_details} details our convention for the Voigt distribution and our handling of background features and quantum efficiency. This model of the measured spectra departs from previous practice in only one way: $\sigma_0=0$ and $\sigma=\sigma_\mathrm{res}$ has always been implicit. As with the fits between measured histograms and K-line models for calibration data, these fits for shape estimation maximize the Poisson likelihood of the data given the model.

The models given by equations~\ref{eq:signal_model} and \ref{eq:sigma} constitute a transferable standard, which should be applicable to any instrument whose energy-response function is known. Of course, there will be higher (and hard-to-characterize) uncertainty for instruments where $\sigma_\mathrm{res}$ is smaller than that given here. Such cases are equivalent to a partial deconvolution of the energy response (or a full deconvolution, if $\sigma_\mathrm{res}=0$ and $\sigma=\sigma_0$).

\begin{figure*}
    \centering
    \includegraphics[width=\linewidth]{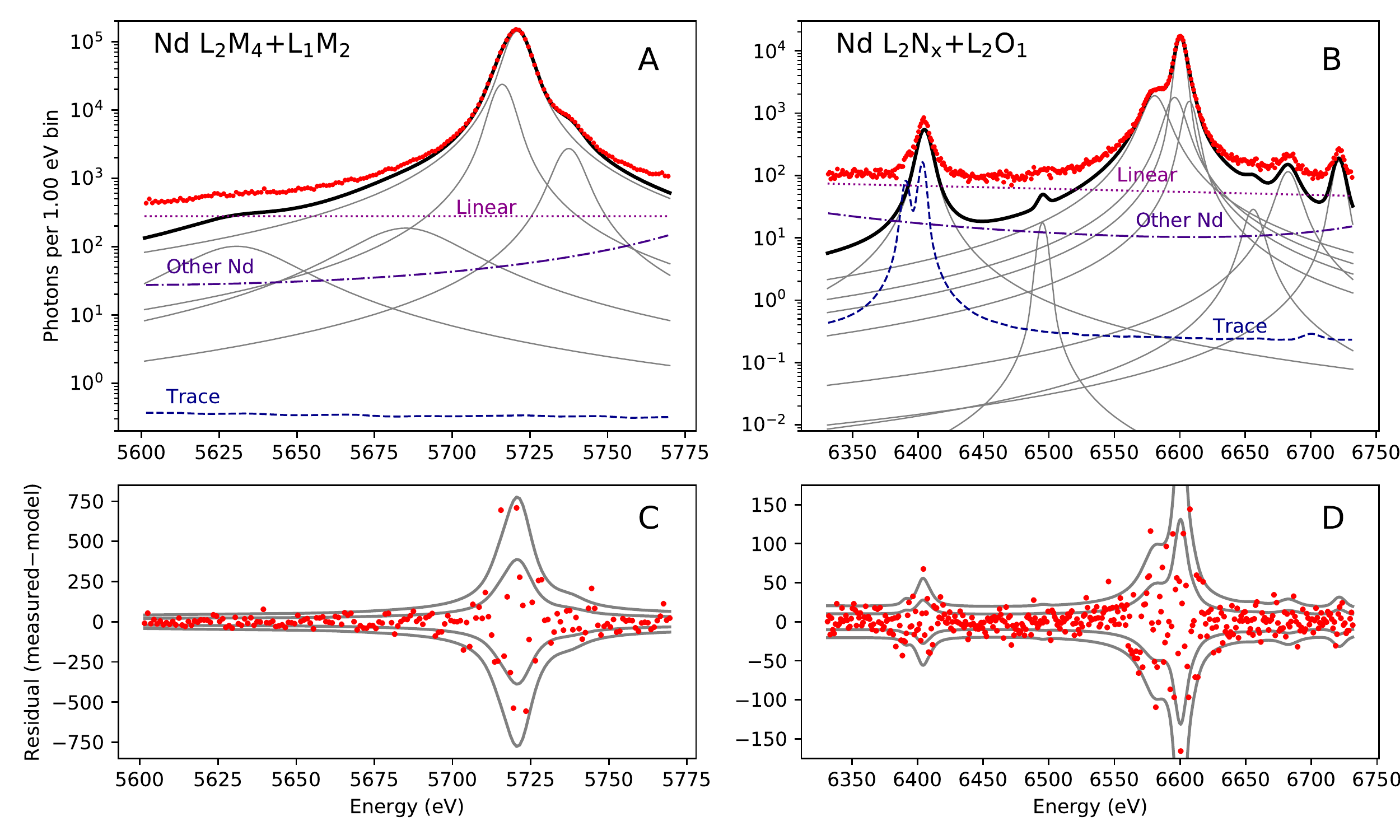}
    \caption{
    Examples of the line profile fitting procedure in two regions of interest (ROIs) for the Nd spectrum: (A) the L$\beta_{1,4}$ unresolved pair and (B) the L$\gamma_1$ and other peaks near the L$_2$ edge. Raw measurements are indicated by dots. The best-fit model (not shown) consists of a sum of several Voigt-like components (thin gray lines) plus three background terms: emission by trace elements (``Trace'', \emph{dashed}); the extrapolated tails of other Nd emission lines (``Other Nd'', \emph{dot-dashed}); and the linear background model (``Linear'', \emph{dotted}). Of the three background terms, only the linear term is varied in the fit. Each Voigt-like component is the product of a Voigt function and the quantum efficiency model $Q(E)$. The fits have three free parameters per Voigt function, plus the model Gaussian width $\sigma_0$ and the two parameters of the linear background. The solid, thick line indicates the sum of the Voigt-like components, excluding all background terms.
    (C) The difference between the measured spectrum and the best-fit model for L$\beta_{1,4}$, the fit in (A). The gray lines indicate the $\pm 1\sigma$ and $\pm 2\sigma$ Poisson standard deviations. (D) Measured minus model for the fit in (B).
}
    \label{fig:example_shape_fits}
\end{figure*}

Several regions of interest (ROIs) have been identified in each spectrum. Each ROI contains one or more fluorescence lines. Figure~\ref{fig:example_shape_fits} shows two examples for the Nd spectrum, including the background components and several Voigt functions. The difference between measurement and model is broadly consistent with Poisson counting statistics. Analysis of the line shapes to extract an estimate of the peak energy is the topic of the next section. Further details of the ROIs and complete numerical results for all fits (the background levels found, the parameters of the Voigt components fit in each region, and the quality of each fit) are given in \ref{sec:fit_details}.

\begin{figure*}
    \centering
    \includegraphics[width=\linewidth]{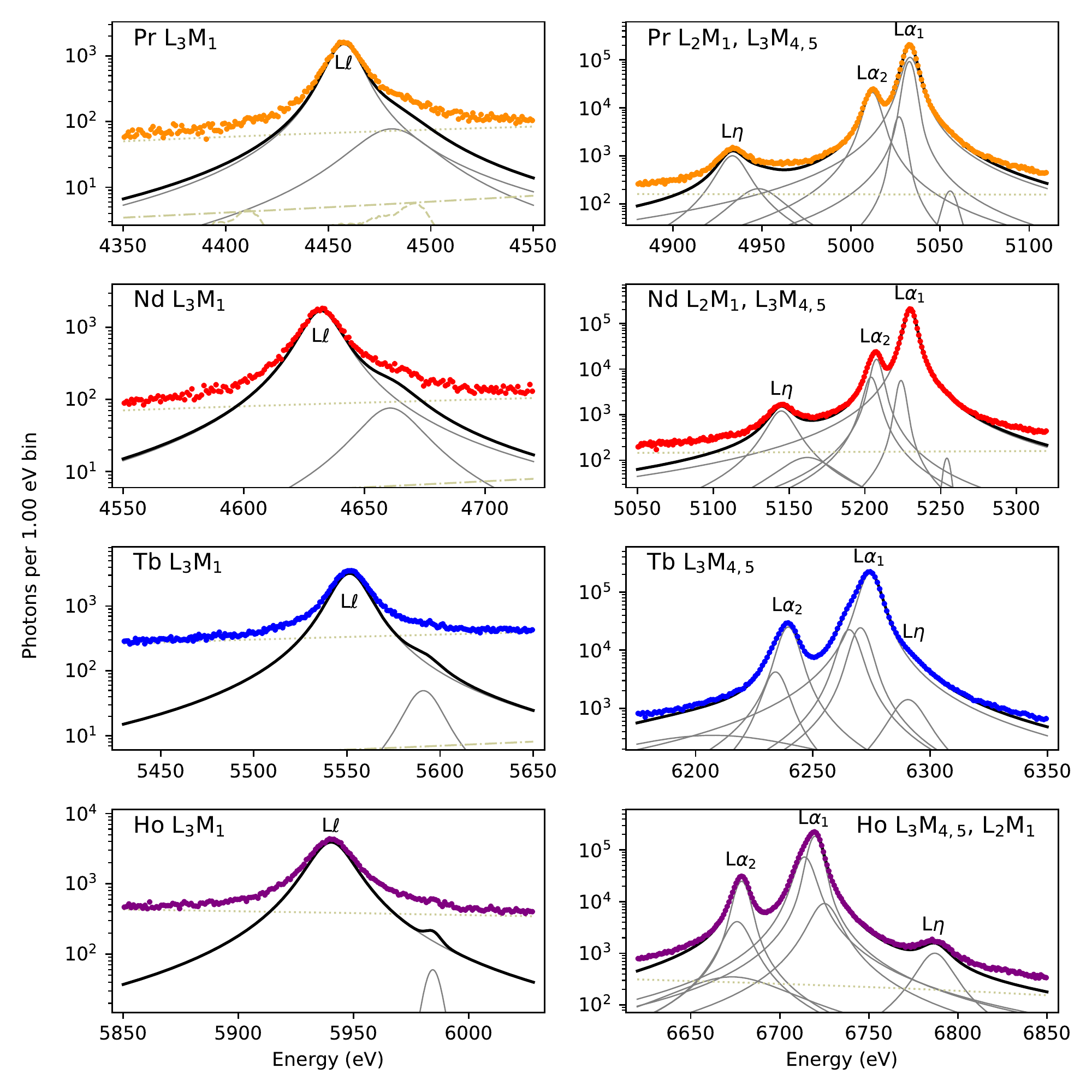}
    \caption{Two ROIs per element with the L$_3$M lines. \emph{Left:} The L$_3$M$_1$ (L$\ell$) line. \emph{Right:} The L$_3$M$_4$-L$_3$M$_5$ doublet (L$\alpha_{2,1}$). For Pr, Nd, and Ho, L$_2$M$_1$ (L$\eta$) is also visible; the L$\eta$ of Tb overlaps the L$\alpha_1$ peak. The curves follow the same convention as in the upper panels of figure~\ref{fig:example_shape_fits} and in the next four figures: the raw spectrum, counts per 1.0\,eV bin, are dots; the Voigt-like components (Voigts times quantum efficiency) are thin gray lines; the sum of all Voigt-like components is the thicker black line; the three background terms are shown separately as the linear background fit (\emph{dotted}), trace element emission (\emph{dashed}), and the tails of any emission lines centered beyond the ROI (\emph{dash-dot}). Background components are not all intense enough to appear in every panel. Residuals are not shown, but goodness-of-fit is assessed in \ref{sec:fit_details}.}
    \label{fig:allL3M}
\end{figure*}

\begin{figure*}
    \centering
    \includegraphics[width=\linewidth]{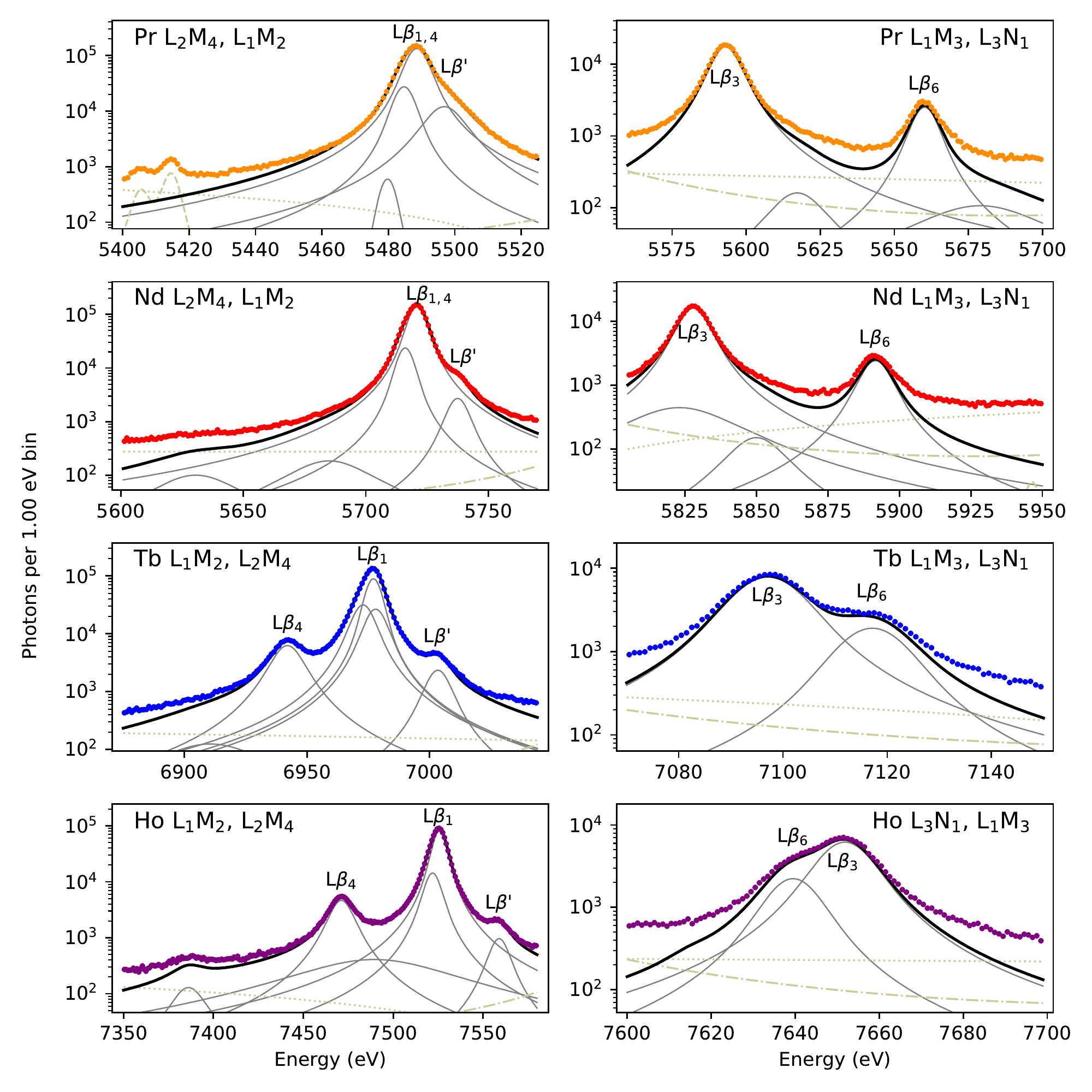}
    \caption{\emph{Left: }The ROIs with the L$_2$M$_4$ (L$\beta_{1}$) and L$_1$M$_2$ (L$\beta_4$) diagram lines and the non-diagram L$\beta'$; these are unresolved in the Pr and Nd spectra. \emph{Right:} ROIs with the L$_1$M$_3$ (L$\beta_3$) and L$_3$N$_1$ (L$\beta_6$) lines, which are not fully resolved in the Tb and Ho spectra. All components shown are described in figure~\ref{fig:allL3M}.}
    \label{fig:allLbeta}
\end{figure*}

\begin{figure*}
    \centering
    \includegraphics[width=\linewidth]{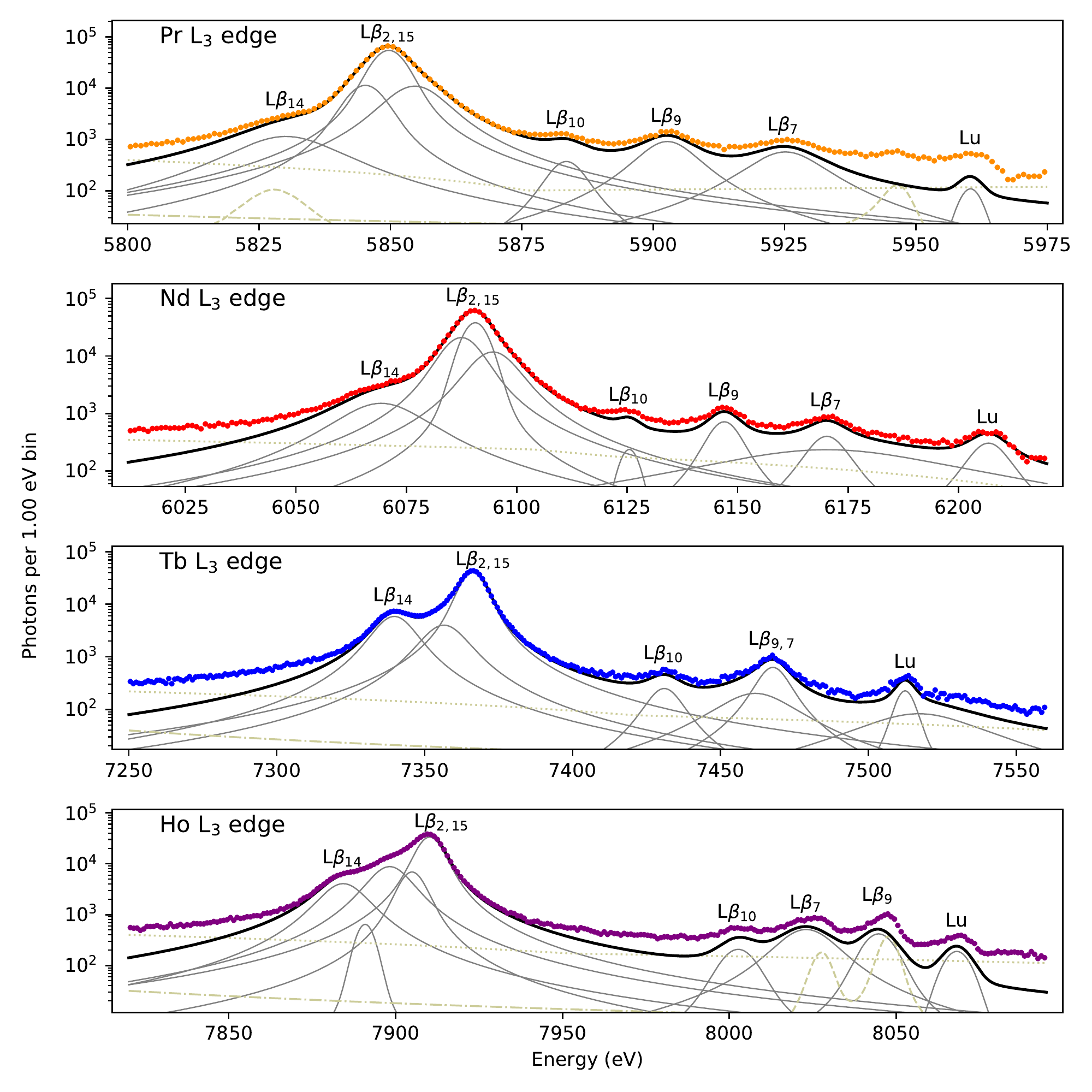}
    \caption{The ROIs near each L$_3$ edge. The most prominent peak in each is the L$_3$N$_{4,5}$ (L$\beta_{2,15}$) line. Other lines just barely detected include the L$_1$M$_4$ (L$\beta_{10}$) and L$_1$M$_5$ (L$\beta_9$), L$_3$O$_1$ (L$\beta_7$), and L$_3$N$_6$ (Lu) lines. The ROIs labelled E and F, though fit separately, are plotted together here as one panel per element for clarity. All components shown are described in figure~\ref{fig:allL3M}.}
    \label{fig:allL3edges}
\end{figure*}

\begin{figure*}
    \centering
    \includegraphics[width=\linewidth]{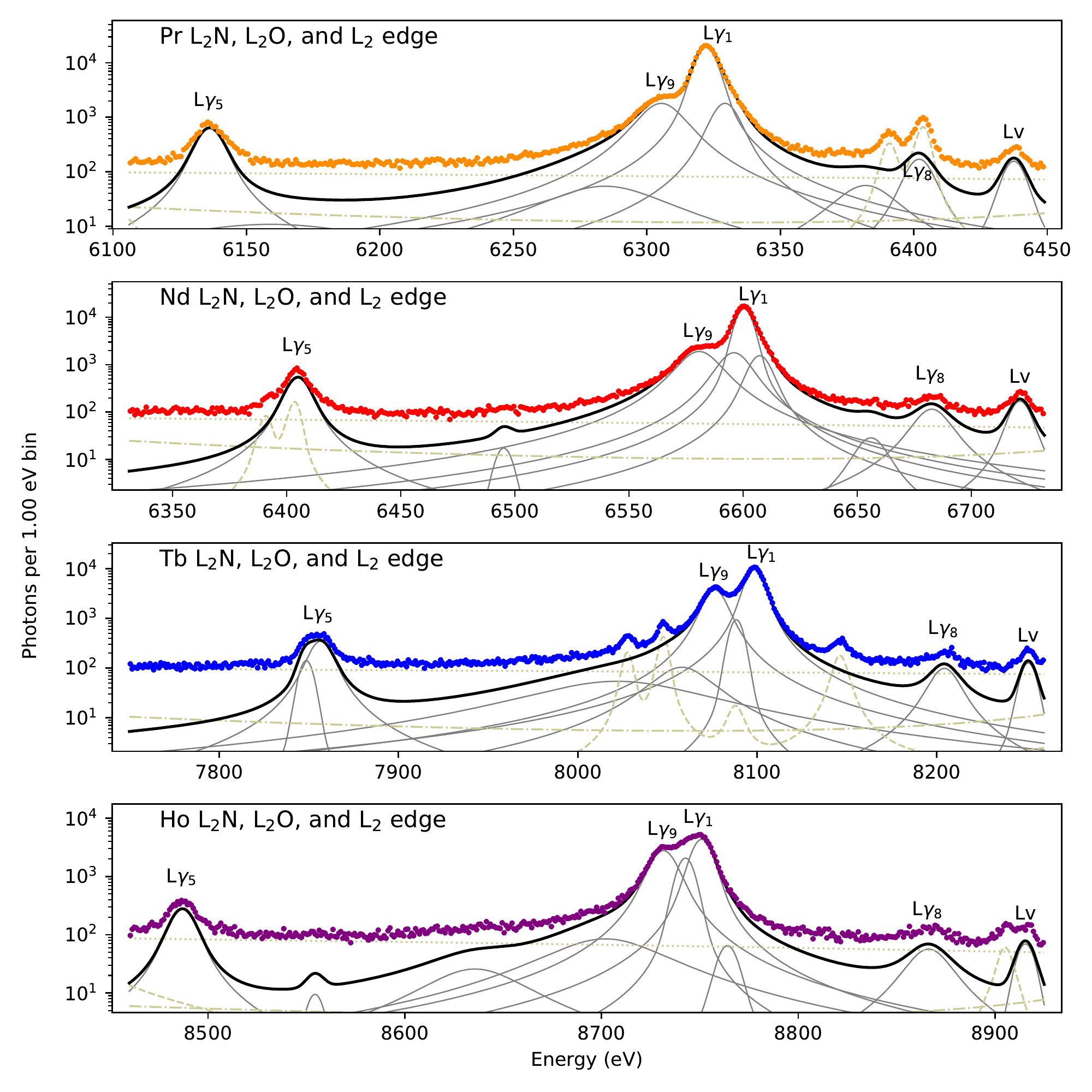}
    \caption{The ROIs near each L$_2$ edge. The most prominent peak in each is the L$_2$N$_4$ (L$\gamma_1$) line. Also visible are the L$_2$N$_1$ (L$\gamma_5$), L$_2$O$_1$ (L$\gamma_8$), and L$_2$N$_6$ (Lv) lines, as well as Fe K$\alpha$ emission in the Pr and Nd spectra and Cu K$\alpha$ emission in the Tb spectrum. All components shown are described in figure~\ref{fig:allL3M}.}
    \label{fig:allL2edges}
\end{figure*}

\begin{figure*}
    \centering
    \includegraphics[width=\linewidth]{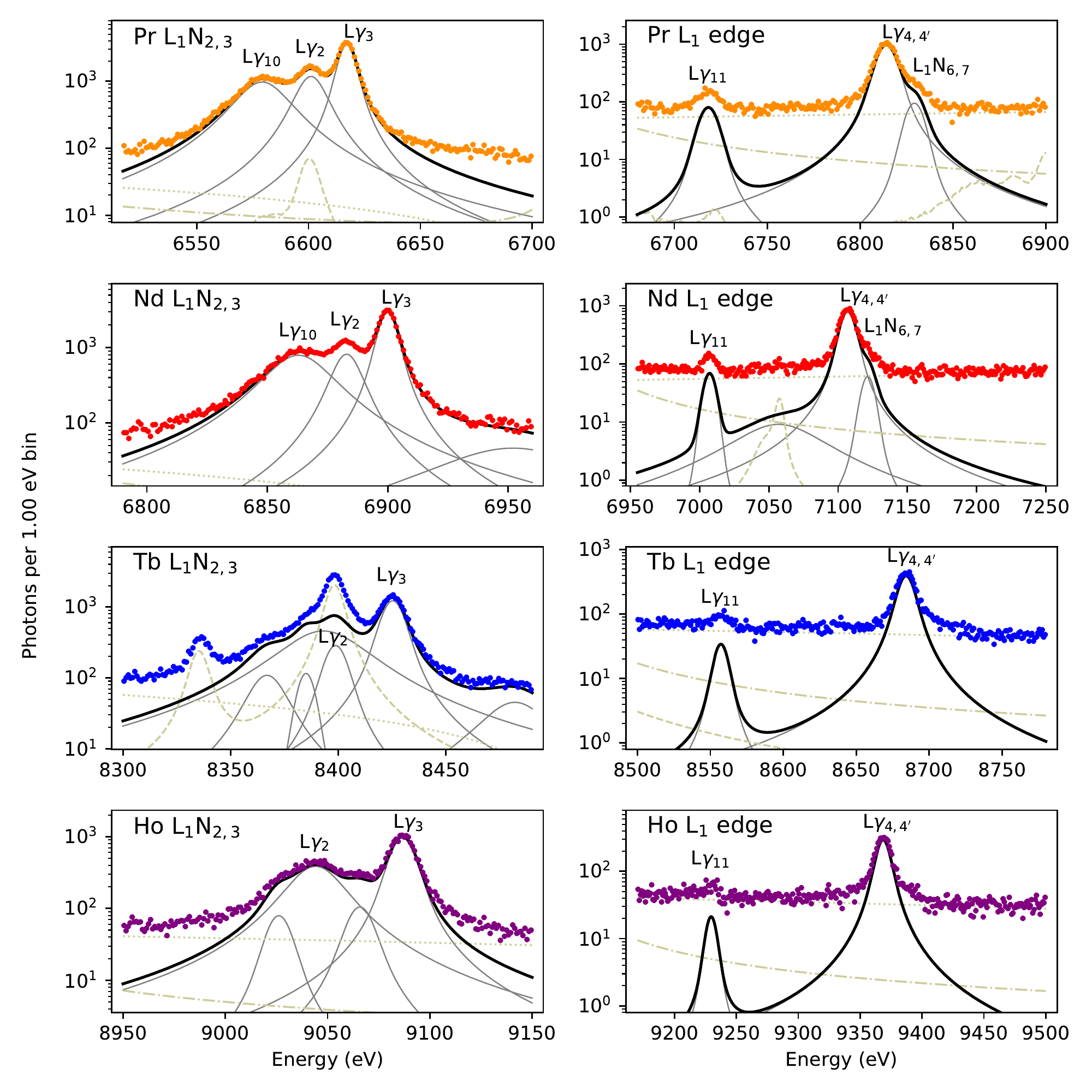}
    \caption{\emph{Left: }The ROIs with the L$_1$N$_2$ (L$\gamma_2$) and L$_1$N$_3$ (L$\gamma_3$) ``doublet,'' which actually appears as three peaks in the Pr, Nd, and possibly the Tb spectra, owing to the L$\gamma_{10}$ non-diagram line. \emph{Right:} ROIs near the L$_1$ edge, including the L$_1$N$_{4,5}$ (L$\gamma_{11}$) and an unresolved combination of L$_1$O$_{2,3}$ (L$\gamma_{4,4'}$) and L$_1$N$_{6,7}$. All components shown are described in figure~\ref{fig:allL3M}.}
    \label{fig:allL1edges}
\end{figure*}

All ROIs are shown in five figures, including the Voigt-like components, both separately and combined; all background components; and the measured spectra. Figure~\ref{fig:allL3M} shows the L$_3$M lines, including the most intense feature in each element's spectrum, the L$\alpha_{2,1}$ doublet. Figure~\ref{fig:allLbeta} shows several lines of the L$\beta$ series. Figure~\ref{fig:allL3edges} shows the L$\beta_2$ line and all features just below the L$_3$ edge. Figures~\ref{fig:allL2edges} and \ref{fig:allL1edges} show the regions near the L$_2$ and L$_1$ edges, respectively.

\subsection{Estimation of line energies} \label{sec:peak_energy}

\begin{table*}
\centering
\include{Tables/peak_data_Pr}
\caption{All fluorescence peaks identified in the praseodymium spectrum. Diagram lines are named by both IUPAC and Siegbahn conventions; non-diagram lines have only a Siegbahn name. The sparkline illustration \emph{Spark} depicts a small region of the spectrum as measured by the TES spectrometer; it has a linear vertical scale, is centered on the peak, and covers an energy range of $\pm0.25\,\%$. \emph{Reference} gives the value and the 1$\sigma$ uncertainty of Deslattes et al.~\cite{Deslattes:2003} when available. Other sources of reference data are marked: ``CS'' from Cauchois \& S\'en\'emaud~\cite{Cauchois:1978,Jonnard:2011}, with uncertainties for diagram lines as estimated by Zschornack~\cite{Zschornack:2007wu}; ``Zsc'' found in~\cite{Zschornack:2007wu} only. \emph{Peak Estimate} indicates the peak measured in this work, estimated by the procedure explained in section~\ref{sec:peak_energy} and \ref{sec:peak_corrections}. Values marked with $\star$ correspond to lines that do not have a local maximum in the spectrum but are inferred from structure in the best-fit model; those marked in italics have no or nearly no local maximum and are assigned additional systematics. Columns \emph{Stat} is the statistical uncertainty on the peak energy (section~\ref{sec:stat_uncert}); \emph{Sys} is the systematic uncertainty on the energy scale (section~\ref{sec:systematics}). The \emph{Peak} column is the additional systematic uncertainty arising from the peak-estimation procedure (section~\ref{sec:peak_energy}) and applies only when the peak estimate is used in place of the full profile.}
\label{tab:peaksPr}
\end{table*}

\begin{table*}
\centering
\include{Tables/peak_data_Nd}
\caption{All fluorescence peaks identified in the neodymium spectrum. Columns are the same as in table~\ref{tab:peaksPr}.}
\label{tab:peaksNd}
\end{table*}

\begin{table*}
\centering
\include{Tables/peak_data_Tb}
\caption{All fluorescence peaks identified in the terbium spectrum. Columns are the same as in table~\ref{tab:peaksPr}. Reference data marked ``Th'' are the values labelled as ``theory energy'' in Deslattes et al., used where no experimental value is given there or elsewhere.}
\label{tab:peaksTb}
\end{table*}

\begin{table*}
\centering
\include{Tables/peak_data_Ho}
\caption{All fluorescence peaks identified in the holmium spectrum. Columns are the same as in table~\ref{tab:peaksPr}.}
\label{tab:peaksHo}
\end{table*}

The local maximum of a line profile fixes a line's peak energy, a definition that presents serious problems under typical measurement conditions. The asymmetric profiles of the more intense L lines mean that the energy of this local maximum is an unknown and potentially strong function of a spectrometer's energy response. Even when that energy response is purely Gaussian, it is impossible to work from a measurement at finite resolution and estimate the peak energy in the preferred limit $\delta E\rightarrow0$ with complete confidence. This limitation is not unique to microcalorimeters: the possible existence of features in the true spectrum at energy scales much smaller than $\delta E$ simply cannot be ruled out.

Alternative definitions of a line's energy are possible, besides the local maximum of a noise-free, perfect-resolution spectrum. When a profile is fit as a sum of several components and one is identified with the diagram line, the center of that most prominent component can be used to summarize energy. Line energy has also been defined as the center at 2/3 of maximum~\cite{Kawai:1994}, a more robust estimator than a local maximum. For solid-state detectors with modest resolution, the centroid is the most useful number, though modeling of backgrounds strongly influences the estimation of a centroid.

In spite of the various definitions of a line energy and the real-world problems associated with the estimation of each, many existing collections of reference data---and specifically NIST SRD-128---summarize even the most complex lines by their peak energy. We cannot compare our measurements to earlier results, then, until we attempt to estimate the energy at which the spectral profiles would have peaked in the case of an idealized, noise-free, zero-resolution measurement.  We stress again that the most robust and precise results of the current study are the complete line profiles described in the previous section; the peak energies computed here are intended only for comparison to published peak energies that are not accompanied by models of line profiles.

We define the unknown \emph{peak shift} as the shift in a peak's energy due to resolution effects. Specifically, it is the difference between the actual peak as measured (with $\delta E$ given by equation~\ref{eq:nominal_res}, approximately 4\,eV) and the idealized peak. We can get a sense of the approximate scale of the shift by comparison of peak energies in the measured spectrum to peaks found at similar but poorer resolutions---we choose a resolution of 8\,eV for this comparison. Peaks that shift the most between resolutions of 4\,eV and 8\,eV  are likely to be affected by a peak shift between 4\,eV and 0\,eV resolutions of similar magnitude.

For the majority of lines we have studied, the asymmetry is fortunately small enough that we believe this peak shift to be smaller than 0.1\,eV\@. This level is smaller than the other systematic uncertainties on the calibrated energy scale. We can never definitively rule out a conspiracy of asymmetries at small scales that cause arbitrarily large offset, but most lines appear to be quite symmetric and suggest a small shift. In a few cases, however, asymmetry of the profile suggests that the peak shift is as large as 1\,eV\@.  Again, this is not unique to energy-dispersive microcalorimeters; even a wavelength-dispersive spectrometer with $\delta E=1.0$\,eV would produce shifts up to 0.1\,eV\@.

One obvious way to estimate the peak energy in the limit of $\delta E\rightarrow0$ is to find the maximum of the best-fit sum-of-Voigts model in this very limit. A simulation of the fitting procedure for those few lines for which high-resolution measurements are available suggests that this limit---in effect, a deconvolution of the resolution from a fitted model---tends to understate the size of the shift. That is, the fitting procedure effectively makes the conservative assumption that all unresolved structure in a line profile is as symmetric as possible. The zero-resolution limit of a model will therefore give a biased estimate of the peak energy.

In search of better estimation methods, we have explored high-resolution data for both the K lines of transition metals (section~\ref{sec:cal_transition_metals}) and for ten L-line measurements made by Deslattes' team. Given these examples, we find the most success in estimating the peak shift by a different method: we assume that the peak energy is a linear function of resolution between the measured $\delta E\approx 4$\,eV and the fixed value of $\delta E=6$\,eV\@. The peak at these two resolutions can be reliably estimated, as the convolution of data or models to make the energy resolution \emph{worse} (larger) tends to tame noise and is safe. The linear function can be extrapolated to the desired $\delta E=0$ to estimate the peak shift. The use of 6\,eV resolution as the reference point and of a linear function are choices that minimize the mean square error over the several K and L lines available for testing the correction (see \ref{sec:peak_corrections}). In these test cases, the root-mean-square error in peak energies is 0.18\,eV after the correction procedure. For the lanthanide L lines, we therefore assign a systematic uncertainty inherent to the peak-finding correction of 0.18\,eV, although the error distribution is assuredly not Gaussian. This peak-finding uncertainty is the dominant systematic for most peaks at energies of 7.5\,keV and below.

The meaning of peak energy is clear enough when a peak is isolated far from other spectral features. When two features overlap yet retain distinct local maxima, such as the L$\alpha_{2,1}$ doublets (figure~\ref{fig:allL3M}), we still define the peak as the local maximum of the composite profile. Such an empirical definition yields peak energies that must depend somewhat on the relative intensities of the nearby lines. This ratio depends in turn on the excitation mechanism and spectrum, at least when the overlapping features result from transitions that fill different L subshells (such as the L$_2$M$_4$ and L$_1$M$_2$ transitions of terbium, figure~\ref{fig:allLbeta}). Nevertheless, we persist in the local-maximum definition, thereby avoiding any need to definitively allocate the fitted Voigt components in any ROI between the overlapping lines. 

When two lines overlap enough that only one local maximum is seen (such as the L$_1$M$_3$ and L$_3$N$_1$ transitions of terbium and holmium, figure~\ref{fig:allLbeta}), we use the local maximum definition for the more intense line. For the feature without a peak of its own, however, we state the peak of only that Voigt component (or components) in the energy range nearest the non-peaking subsidiary feature. Such cases are indicated in the results and are assigned an arbitrary additional 1\,eV systematic uncertainty on the peak energy. Intermediate cases of lines with small peaks on a large tail are assigned smaller systematics of a fraction of 1\,eV\@.

Tables~\ref{tab:peaksPr}, \ref{tab:peaksNd}, \ref{tab:peaksTb}, and \ref{tab:peaksHo} give the corrected peak energies along with the uncertainties outlined in this and the next two sections, plus the SRD-128 reference data for comparison. Peaks found in this way are compared with existing data sets in section~\ref{sec:results_discussion}. More detailed justification of the correction procedure and specifics of the larger estimated peak shifts appear in \ref{sec:peak_corrections}.

\subsection{Statistical uncertainties on line energies} \label{sec:stat_uncert}

The statistical, or Type A~\cite{GUM:2008}, uncertainty on each line energy is a function of the number of photons in the line, the width of the line, and the signal-to-background ratio at that energy. It is a statement about the uncertainty that follows from the finite amount of data and that we would expect to improve if repeated, substantially identical measurements were made.  For purposes of estimating statistical uncertainty, we assume that the line profiles fit in section~\ref{sec:shape} are correct. 
The line profile fitter estimates the statistical uncertainty of the line energy directly for the simplest emission lines, where a single Voigt function fits the data. It is the standard error on the Voigt function's center parameter. Most lines, though, are best fit by two or more Voigt components; standard errors of the fit become much less useful as a guide to the uncertainty. However we might define ``the energy'' of a compound profile, its uncertainty will depend on multiple standard errors and correlations of parameters whose distribution is unlikely to be multivariate Gaussian. In short, it is not possible to work directly from the results of a complete line profile fit to an estimate of the energy uncertainty of a compound line.

Instead, we make this estimate from a different fit, a fit in which line profiles are fixed and are allowed only to shift in energy. We break each ROI into one or more named ``lines'' and fit the data with a simpler model: the Voigt widths from the full profile fit are retained, and only the center energies and intensities are allowed to vary. When there are multiple Voigt components in a line, the centers of these components are constrained to shift equally, and the relative intensities of the components are also fixed. As a result, each line's \emph{shape} is preserved. This fit therefore has two parameters for the continuum background plus two per line  (energy and intensity). Now the reported standard error of each shift parameter is an estimate of the statistical uncertainty on the line energies.

Standard errors estimated by a nonlinear fitting package are not always reliable. The log-likelihood function in the neighborhood of the best fit model might be approximated poorly by a quadratic, with unpredictable consequences for the reported errors. We have performed a form of audit for the reported uncertainties by simulating data and performing the shift-only fit on the simulated spectra. Specifically, we simulated five of the Nd ROIs, containing fifteen distinct lines, 1000 times apiece. The standard deviation of the line energies in the shift fits agreed with the reported standard error on the fits to the actual measurement, to within 10\,\% in most cases and to 20\,\% in all cases. This level of agreement indicates that the potential perils of standard errors from a nonlinear fit did not cause problems in this instance.

As a second check on the uncertainties, we performed similar shift-only fits of the full line profile to the subset of the data obtained from a single TES on a single day. Such subsets contain approximately 0.7\,\% of all data. These fits are completely consistent with zero shift between the subsets, and the standard deviations of the shifts agree with the estimated statistical uncertainties, scaled up appropriately given the reduced number of photons in the subsets.

Finally, we find that the statistical uncertainties are at the expected level, given the number of photons observed and the line widths. A reasonable model for the energy uncertainty on the long-tailed fluorescence lines is the maximum likelihood estimator of the center of a Lorentzian distribution. Such an estimator for $N$ samples from a Lorentzian with half-width at half maximum $\Gamma$, has a standard deviation of 1.40$\,\Gamma/\sqrt{N}$. We increase this expectation by a factor of $\sqrt{1+B/N}$ in the presence $B$ background photons. The fitter's estimated statistical uncertainty is typically equal to this rule-of-thumb expectation, times a factor between 1 and 2. The statistical uncertainties range from 0.004\,eV for the L$\alpha_1$ lines, with between two and three million photons detected, up to 0.8\,eV for a few peaks with fewer than 2000 photons detected and a very low signal-to-background ratio.

\subsection{Systematic uncertainties on line energies} \label{sec:systematics}

We have studied many systematic uncertainties on the energy scale for our results. The dominant contributions have been described above. A complete list includes the following:

\begin{enumerate}[label=\arabic*]
    \item The energy resolution is estimated by equation~\ref{eq:nominal_res} (approximately 4\,eV FWHM), based on the K-line spectra of transition metals (section~\ref{sec:cal_transition_metals}). Our estimated uncertainty on the energy resolution of $\pm 5\,\%$ produces an uncertainty the energy scale of only $\pm$2\,meV\@.
    \item The spectrometer's quantum efficiency is a decreasing function of photon energy for energies exceeding 6\,keV (figure~\ref{fig:qe}). The energy of emission line features changes by only $\pm$4\,meV if the slope of the QE model is rescaled by a factor of 0.7 or 1.4, the widest realistic range of model uncertainty.
    \item Our model of the continuum background in each fitting region is linear. If we fit with a quadratic model instead, emission line features change by approximately $\pm$4\,meV or less.
    \item We assume the low-energy tail in the TES energy response function to be zero (section~\ref{sec:cal_transition_metals}). The contribution of such a tail to the energy response is  not larger than 2\,\%, an amount that would shift energy estimates by typically $\pm$10\,meV\@.
    \item \label{systematic:gains} The small possible inconsistency in gains from one side to another of sample switcher yields uncertainties less than 0.1\,eV at most energies, or up to 0.2\,eV at the highest energies (section~\ref{sec:cal_consistency}).
    \item \label{systematic:interpolation} Interpolation of calibration curves produces an energy-dependent uncertainty (section~\ref{sec:cal_assessment}).
    \item \label{systematic:calcurves} Calibration curves have an absolute uncertainty, because some of the calibration anchor points themselves have finite energy uncertainty (section~\ref{sec:cal_assessment}).
    \item \label{systematic:tails} The tails of neighboring lines have uncertain intensities.
    \item \label{systematic:features} The background features such as K lines of Cr, Fe, and Cu have uncertain intensities (section~\ref{sec:background}).
\end{enumerate}

\begin{figure}
    \centering
    \includegraphics[width=\linewidth]{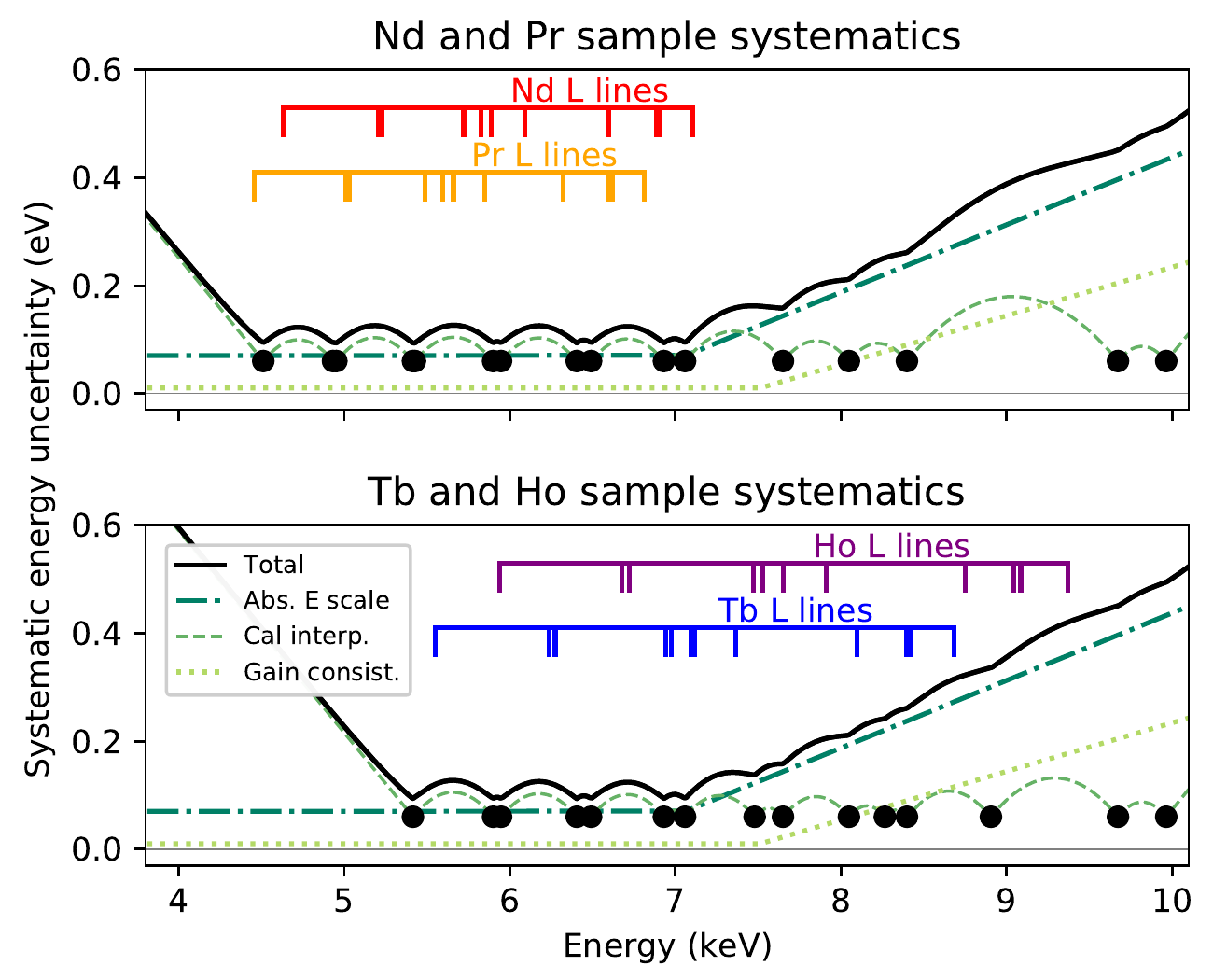}
    \caption{Systematic energy uncertainties. The vertical hash marks indicate the energy of the most prominent L lines of each element. The circles indicate the energies of the calibration curve anchor points. The solid curve in each panel indicates the overall systematic energy uncertainty as a function of energy. The other curves in each panel separately depict the three dominant constituents: the calibration absolute energy-scale uncertainty, the interpolation of calibration curves, and the gain consistency (items 7, 6, and 5 in the text). The overall systematic uncertainty is the sum (in quadrature) of the three indicated components, along with the several constant, sub-dominant terms described in section~\ref{sec:systematics}. }
    \label{fig:systematics}
\end{figure}

For most lines, items \ref{systematic:gains}--\ref{systematic:calcurves} are the dominant sources of systematic uncertainty. Figure~\ref{fig:systematics} depicts terms \ref{systematic:gains}, \ref{systematic:interpolation}, and \ref{systematic:calcurves} separately, and the total uncertainty for all of items 1--7, as a function of energy.

The size of the last two items varies widely between lines. Item~\ref{systematic:tails}, uncertain intensity of lines just outside a given ROI, is never dominant and causes systematic uncertainty up to 0.03\,eV in the worst cases. Item~\ref{systematic:features}, the unknown intensity of trace element emission, is the dominant systematic for five low-intensity lines: Pr L$_2$O$_1$ (2.0\,eV uncertainty because of the Fe K$\alpha$ line); Nd L$_2$N$_1$ (0.3\,eV because of Fe K$\alpha$); Tb L$_1$N$_2$ (3.0\,eV because of W L$\alpha_1$); and Ho L$_1$M$_4$, L$_3$O$_1$, and L$_1$M$_5$ (0.2, 1.0, and 3.0\,eV because of Cu K$\alpha$).

A tenth source of systematic uncertainty is the one discussed in section~\ref{sec:peak_energy}: that the peak energy of a fluorescence line cannot be estimated perfectly from measurements with non-zero noise or energy resolution. This additional uncertainty on the absolute energy does \emph{not} apply to the model profiles, but only to the peak energy extracted from these profiles. Because it applies only to peak energies, we tabulate it as a systematic uncertainty distinct from all the others in tables~\ref{tab:peaksPr}-\ref{tab:peaksHo}. In other words, readers who use line \emph{profiles} from these tables should assume energy uncertainty equal to the quadrature sum of only the statistical and systematic columns. Readers interested in the \emph{peak energy} values should assume uncertainty equal to the quadrature sum of all three uncertainties.
 
\subsection{Discussion of results} \label{sec:results_discussion}

We have measured the line profiles of 83 distinct fluorescence emission diagram lines, as well as 14 emission features identified by Cauchois and S\'en\'emaud as non-diagram lines~\cite{Cauchois:1978}. Most of these 97 features appear as peaks in our data with its resolution of $\delta E$=4\,eV FWHM, though 13 are detected only as asymmetric components of more intense nearby lines. Statistical uncertainties range from 0.004\,eV for the most intense L$\alpha$ peaks to more than 0.5\,eV for those barely seen above the background. The energy-scale systematic uncertainty is less than 0.2\,eV over much of the energy range. Additional systematics of 0.18\,eV or more apply specifically to our estimation of peak energies.

The line profiles have been fit by a sum of Voigt functions, where the Gaussian component includes both instrumental broadening and an additional zero-resolution width $\sigma_0$, chosen for the best fit to the data. Most lines are well fit by one or two Voigt components, but some of the higher intensity lines with visible asymmetry such as the L$\alpha_1$ or L$\beta_1$ lines require up to five Voigt components. Complete fit results appear in \ref{sec:fit_details}.

The 83 measured diagram lines include 62 that have experimental results in the Deslattes et al. reference table/SRD-128~\cite{Deslattes:2003}. An additional 15 appear in Cauchois and S\'en\'emaud; all 15 are ultimately based on measurements reported in Bearden~\cite{Bearden:1967tg}. Four others contain only a theoretical estimate in SRD-128, and two appear in neither table. If we consider the 77 lines that are in either collection with measured values, 64 cite no sources more recent than 1980, and 50 of these rely entirely on the Bearden compilation~\cite{Bearden:1967tg}.
Only 13 lines cite \emph{any} measurements fewer than 40 years old: the L$_1$M$_4$ and L$_1$M$_5$ lines of Ho were measured in 2002 by a von Hamos spectrometer~\cite{Raboud:2002}, and six lines of Nd and five of Ho were measured in the 1990s by the NIST VDCS double-crystal spectrometer~\cite{Deslattes:2003}. 

Thirteen of the 14 non-diagram lines we detect are also found in Cauchois and S\'en\'emaud; only Pr L$\beta'$ is absent. However, none of the line energies listed therein cites a reference nor states an uncertainty. Comparable results for diagram lines suggest that uncertainties of at least 1\,eV to 2\,eV are likely.

Some of our line identifications are uncertain. For instance, the features we have identified as the non-diagram L$\beta'$ (figure~\ref{fig:allLbeta}) is within approximately 10\,eV of the SRD-128 theoretical values for the forbidden L$_2$M$_5$ diagram line. It is possible that this feature is the L$_2$M$_5$ line, or a combination of it and the L$\beta'$ satellite. The L$\beta_9$ and L$\beta_7$ lines of terbium near 7470\,eV (figure~\ref{fig:allL3edges}) are expected to coincide, and we have labelled them as a single, unresolved L$\beta_{9,7}$ line, even though two Voigt components are required for the fit.  Also, the assignment of the non-diagram L$\gamma_{10}$ to the feature just below the L$\gamma_{2,3}$ doublet for Pr and Nd (figure~\ref{fig:allL1edges}) matches the energy given by Cauchois and S\'en\'emaud, but L$\gamma_{10}$ cannot be identified for Ho or Tb. The L$\gamma_{10}$ and L$\gamma_2$ features of Ho appear to be unresolved in our data, so we label them simply as L$\gamma_2$. The presence of W L$\alpha_1$ emission in the relevant portion of the Tb spectrum leaves the L$\gamma_{10,2}$ question ambiguous for that element.

The median uncertainty on our estimates of peak energy (statistical and systematic combined) is 0.30\,eV on the 97 measured lines; it is 0.24\,eV on the 62 that also appear in SRD-128. For the latter subset, the median uncertainty in SRD-128 is 0.95\,eV, and there is considerable evidence that many of these uncertainties are underestimated. The combined uncertainty is smaller than that of the SRD-128 reference data in the case of 43 of the 62 common lines.

\begin{table}
    \centering
    \begin{tabular}{lrrl}
    Data sets & $\chi^2$ & d.o.f. & PTE \\ \hline
Present - Mooney  & 11.231 & 11 & 0.424 \\
Present - Fowler 2017 & 18.088 & 15 & 0.258 \\
Fowler 2017 - Mooney & 8.596 & 13 & 0.803 \\
Present - Mauron      & 82.238 & 12 & $2\times 10^{-12}$ \\
Fowler 2017 - Mauron  & 69.428 & 11 & $2\times 10^{-10}$ \\
Mooney - Mauron    & 240.008 &  9 & $1\times 10^{-46}$ \\
Present - All SRD-128  & 284.483 & 62 & $3\times 10^{-30}$ \\
    \end{tabular}
    \caption{Agreement between data sets on the peak energies of lanthanide L lines measured in both. \emph{Data sets} gives the two data sets being compared, which are two of: the present work, the Mooney subset of Deslattes 2003~\cite{Deslattes:2003}, Fowler 2017~\cite{Fowler:2017Metrology}, Mauron 2003~\cite{Mauron:2003}, and the full (experimental) data in Deslattes. $\chi^2$ is the (uncertainty-weighted) sum of squared energy differences, and \emph{d.o.f.} is the number of common entries and thus the number of degrees of freedom. \emph{PTE} (also known as $p$-value) gives the probability to exceed the measured $\chi^2$ value, assuming Gaussian-distributed errors with the stated standard deviations. Both TES-based data sets and the subset of Deslattes reference are all mutually consistent. The complete Deslattes data set, with numerous values established before 1970, is not consistent with the new work. The Mauron data are not consistent but use a distinct photoexcitation mechanism and may not be directly comparable to the others.
    }
    \label{tab:peak_agreement}
\end{table}

\begin{figure*}
    \centering
    \includegraphics[width=\linewidth]{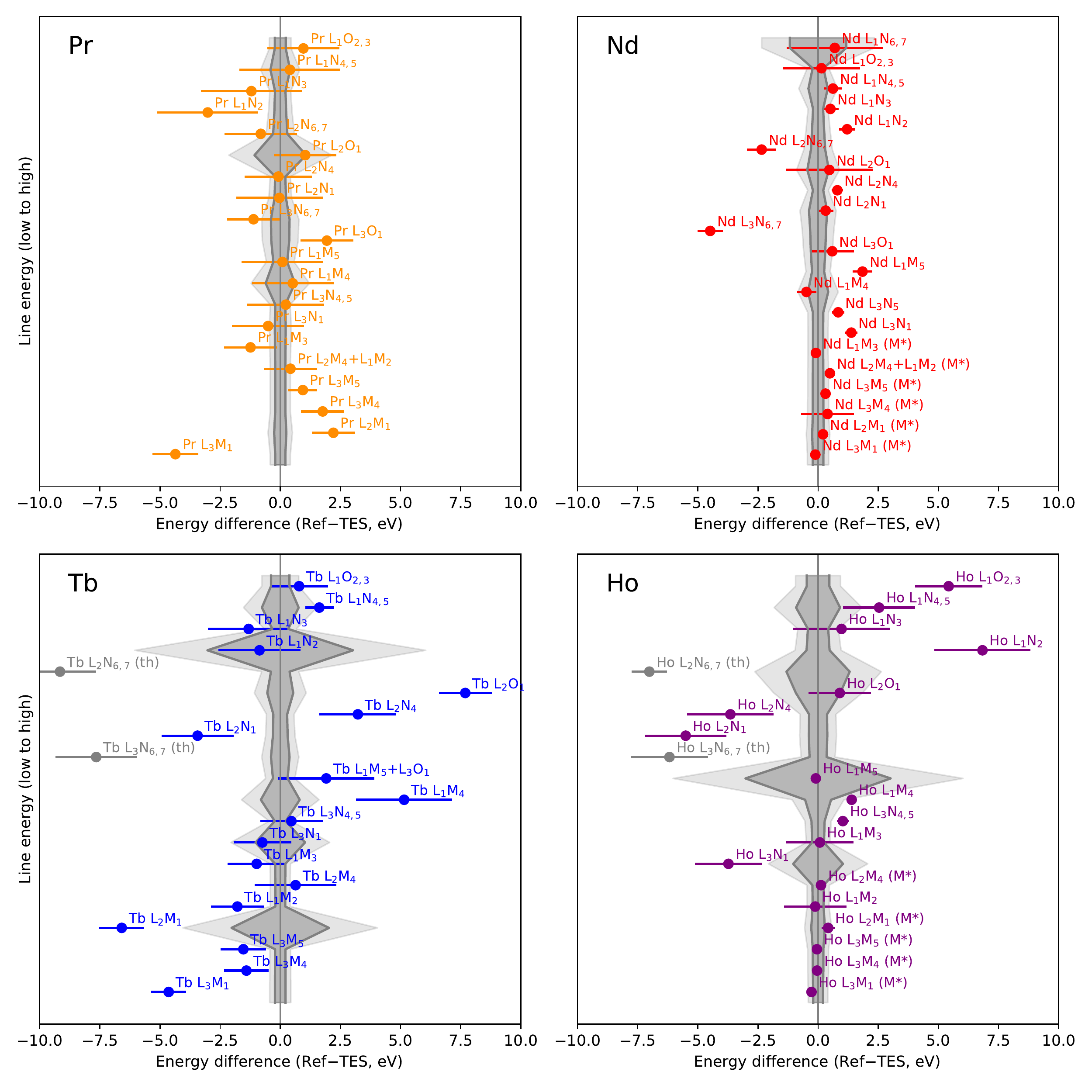}
    \caption{Differences in peak energy estimates between Deslattes et al.~\cite{Deslattes:2003} and the present work. The $1\sigma$ error bars are those given in the reference data. The darker and lighter gray bands around 0 represent the $1\sigma$ and $2\sigma$ error bars of the present work (statistical and systematic combined). Each panel represents a different element, and lines are sorted by energy (lowest energy is at the bottom). The four lines with only a theoretical value are noted with \emph{(th)} and shown in light gray on the figure. Eleven lines from the best subset of Deslattes (the Mooney subset) are marked with \emph{(M$\star$)} and are also featured in figure~\ref{fig:peak_compare_mooney}.}
    \label{fig:peak_compare_deslattes}
\end{figure*}

\begin{figure}
    \centering
    \includegraphics[width=\linewidth]{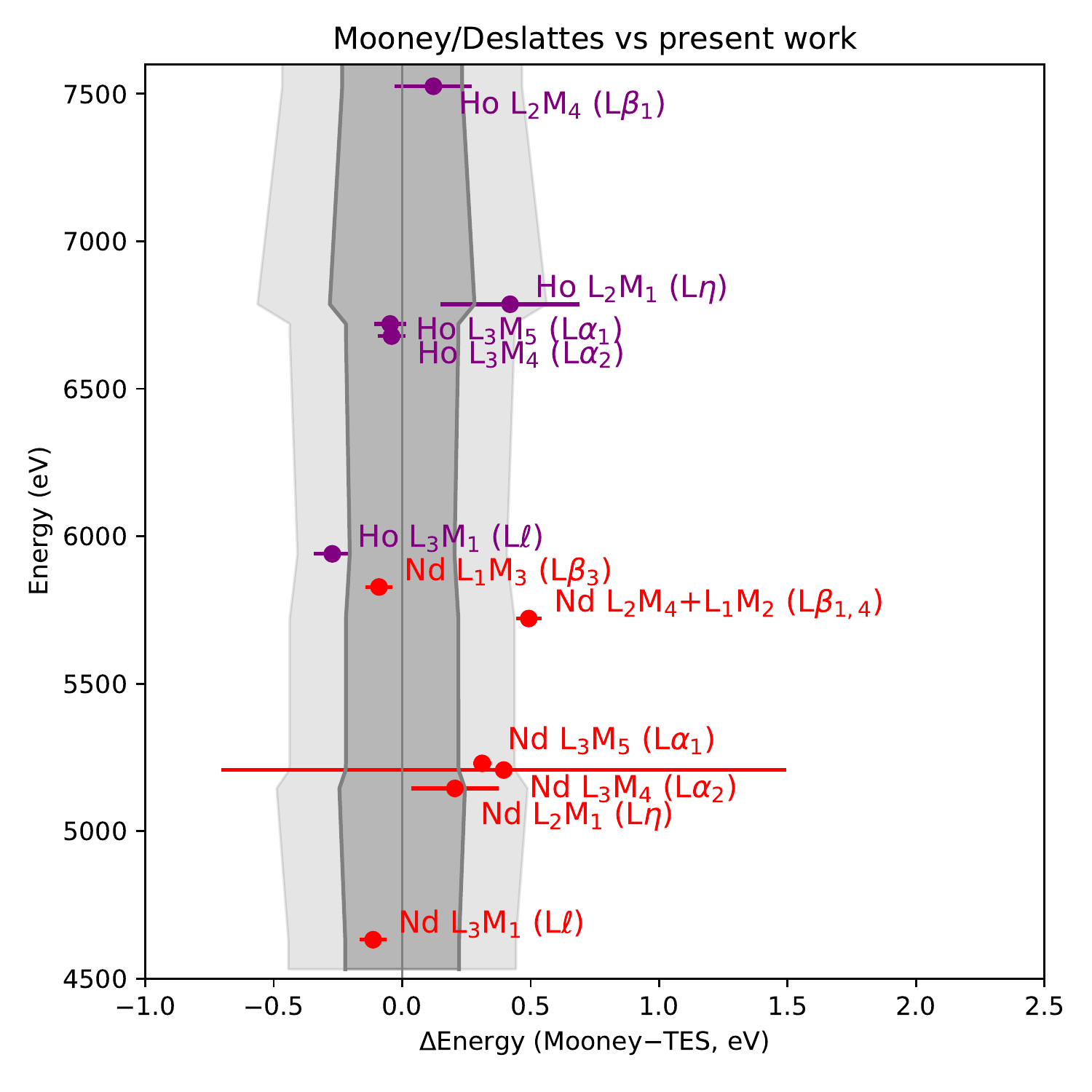}
    \caption{Differences in peak energy estimates between the Mooney subset of data from Deslattes et al.~\cite{Deslattes:2003} and the present work. The $1\sigma$ error bars are those given in the reference data. The darker and lighter gray bands around 0 represent the $1\sigma$ and $2\sigma$ error bars of the present work (statistical and systematic combined). Line energy is shown on the vertical axis.}
    \label{fig:peak_compare_mooney}
\end{figure}

\begin{figure}
    \centering
    \includegraphics[width=\linewidth]{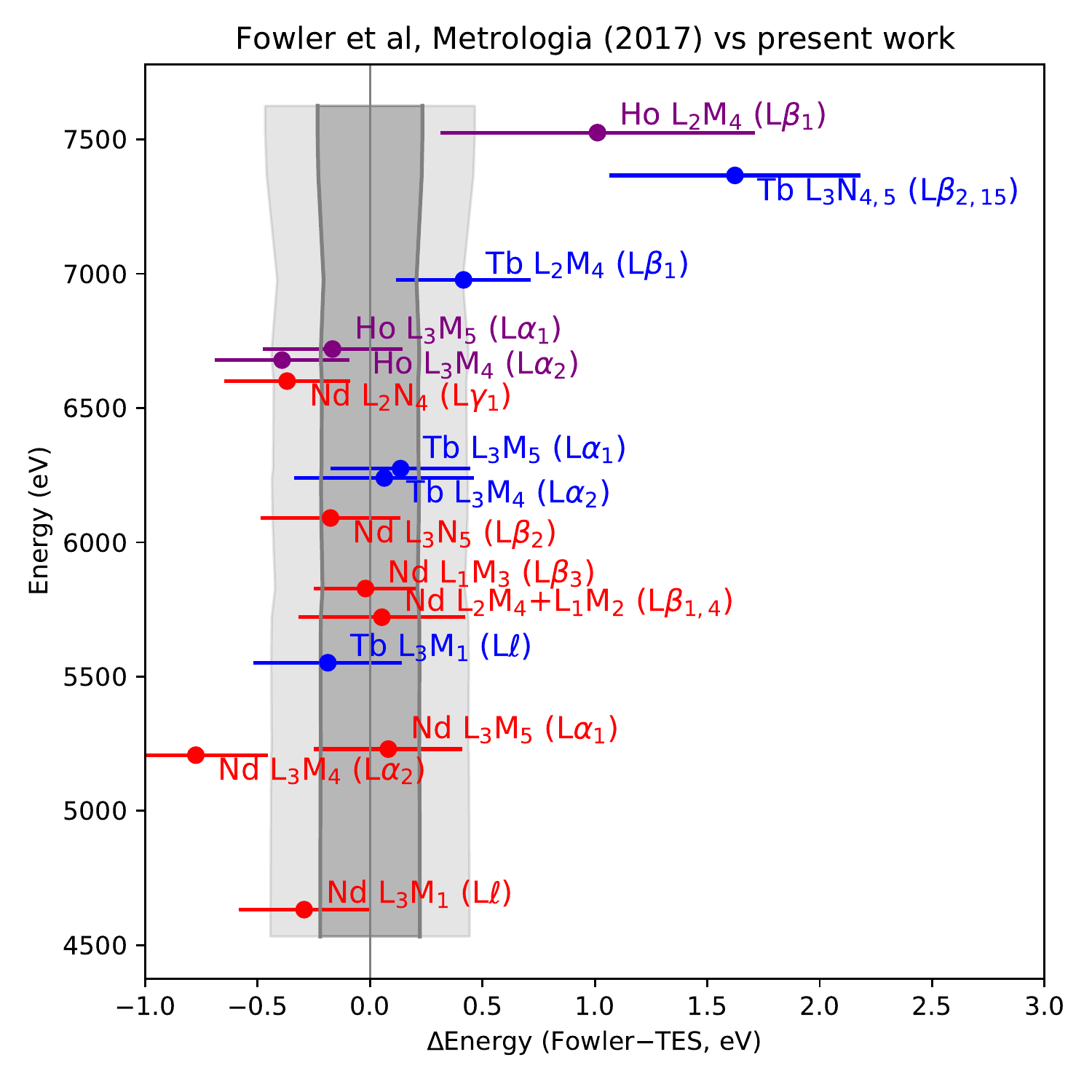}
    \caption{Differences in peak energy estimates between our previous work~\cite{Fowler:2017Metrology} and the present work. Error bars and gray bands are as in figure~\ref{fig:peak_compare_mooney}.}
    \label{fig:peak_compare_fowler}
\end{figure}

We compare our new measurements for consistency with the peak energy values found in two other sources: the VDCS lines that make up the most reliable and modern subset of the data in Deslattes et al.~\cite{Deslattes:2003}, and our earlier measurement~\cite{Fowler:2017Metrology}, also based on TES calorimeters. We have 11 and 15 lines in common with these sources, respectively. For purposes of comparison, we add in quadrature all relevant stated uncertainties: statistical and (when stated) systematics. In the case of the present work, the energy scale and the peak-finding systematics are both included. Assuming that all uncertainties are Gaussian-distributed, the chi-squared test is appropriate for assessing differences in peak energies. In table~\ref{tab:peak_agreement}, we give the value of the $\chi^2$ statistic and the number of peak energies compared, which is the number of degrees of freedom. The resulting probability-to-exceed (PTE) should be uniformly distributed in [0,1] when two data sets are consistent and have correctly evaluated uncertainties~\cite{Cochran:1952}. Values larger than approximately 0.05 suggest consistency, and much smaller values indicate inconsistency.

We find that the two TES-based measurements are statistically consistent with one another and with Deslattes' modern data. The full SRD-128 database, on the other hand, is not at all consistent with our results. If the uncertainties in SRD-128 are inflated by a factor of 2.5 for the lines other than the 11 most modern, then agreement is restored (PTE=0.19). This suggests that the older data could have uncertainties that are a factor of 2 to 3 too small. Line-by-line comparisons are shown in figure~\ref{fig:peak_compare_deslattes} for the full SRD-128 database, figure~\ref{fig:peak_compare_mooney} for the modern subset of Deslattes et al., and figure~\ref{fig:peak_compare_fowler} for the earlier TES measurement.

\begin{figure}
    \centering
    \includegraphics[width=\linewidth]{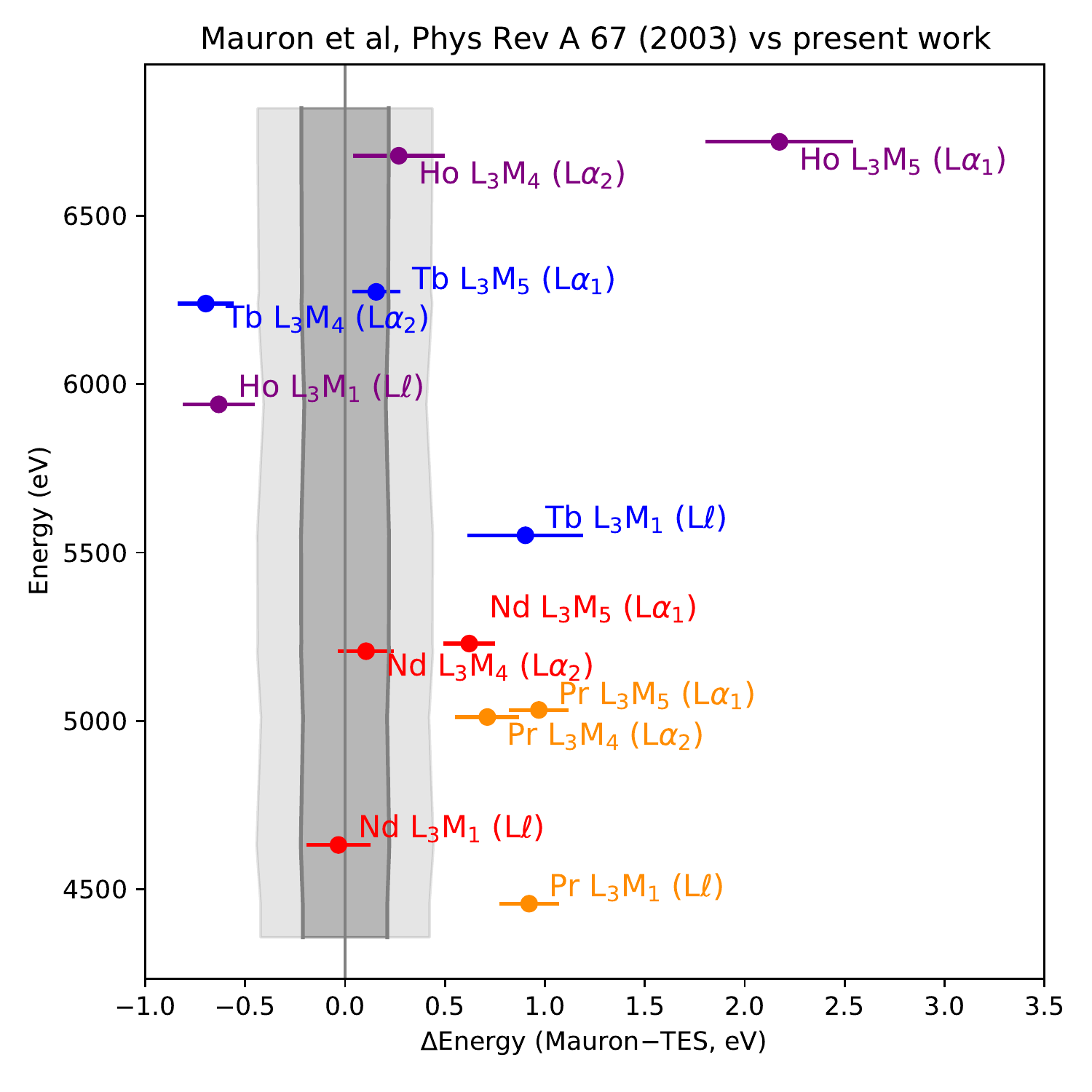}
    \caption{Differences in peak energy estimates between the results of Mauron~\cite{Mauron:2003} and the present work. Error bars and gray bands are as in figure~\ref{fig:peak_compare_mooney}.}
    \label{fig:peak_compare_mauron}
\end{figure}

We also compare our measurements to the values found in Mauron~\cite{Mauron:2003}, which just predates the compilation of Deslattes et al. This work primarily concerns the L$_3$ line widths of several lanthanide elements but also estimates the peak energy of their L$_3$M$_1$, L$_3$M$_4$, and L$_3$M$_5$ lines. Figure~\ref{fig:peak_compare_mauron} shows a line-by-line comparison, and table~\ref{tab:peak_agreement} shows that they are not statistically consistent with either TES measurement or SRD-128. While the other measurements used very broad-band excitation from a tube source, Mauron measured the various L$_3$M transitions at a synchrotron, with excitation energy specifically tuned below the L$_2$ edge. This choice simplifies the emission line profiles by eliminating much of the unresolved satellite structure, but it also yields profiles and peak energies not directly comparable to those produced to photo-excitation by a broad-band source. A broad-band excitation would be expected to broaden and shift L lines to higher energy; Mauron shows both effects to be of order 1\,eV~\cite{Mauron:2003}. We observe the opposite effect: a peak at lower energy. This discrepancy remains unresolved. The difficulty of explaining such discrepancies highlights the need for both transparent uncertainty analyses and the publication of full line profiles to identify subtle differences in peak-shape fits.

\subsection{Microcalorimeter results as new reference data} \label{sec:new_reference}

We have argued that a new set of x-ray emission data should be considered to supersede an existing reference set if it improves on its predecessor in self-consistency, accuracy, and precision. Our new measurements meet these criteria. 

The data were taken in two experiments over a continuous ten-day period. We used an array of 50 optimally functioning TES microcalorimeters with science and calibration data repeatedly interleaved, allowing us both to estimate the size of and to minimize the gain differences between fluorescence targets. Self-consistency was further ensured by simultaneous measurement of the entire spectral band spanning a factor of two in energy. The existing reference data from SRD-128 and Cauchois and S\'en\'emaud, on the other hand, come from a diverse array of sources. The reference data were measured over a period of at least 50 years, with a wide variety of instrumentation, calibration techniques, and analysis choices. How the original experimenters handled the estimation of uncertainties and of peak energies in asymmetric lines is varied, and often not known. The self-consistency of the current data set clearly improves upon existing reference collections.

The new data also have excellent accuracy, shown by the limited number of cases where this accuracy can be tested by comparison to superior, earlier measurements. The 11 lines of Nd and Ho that make up the Mooney subset of SRD-128 are the best high-precision, high-accuracy results ever published on the L emission of these elements under broad-band excitation. They are also statistically consistent with our measurements (table~\ref{tab:peak_agreement}).

Finally, the overall uncertainty of our results is better than that found in the reference data for most lines, in many cases much better. The combined statistical and systematic uncertainty in our peak energy estimates is smaller in 43 of the 62 lines listed in SRD-128, and it improves on SRD-128 plus all other reference data for 77 of the 97 lines overall. The median stated uncertainty for the 62 lines found in SRD-128 is 0.94\,eV, four times the median of 0.24\,eV in this work. If the reference uncertainties of older measurements are increased by a factor of 2.5, as the previous section suggests, then their median becomes 2.34\,eV, or ten times the median of our results.

Our measurements are also traceable to the SI, because they rely on anchor points in the calibration curve that are themselves SI-traceable. Although the connections are indirect, they apply uniformly to all the lines in this work. The reference data include a mix of indirectly traceable and non-traceable results, many of the latter corrected long after publication based on refined understandings of the traditional ``X unit'' of wavelength.

In summary, the fluorescence line profiles we have measured improve on the existing reference data. They constitute an accurate, precise, SI-traceable, and self-consistent collection of complete line profiles. While a small subset of SRD-128 is measured to better precision than the microcalorimeter array is capable of, a much larger portion is half a century old and can now benefit from recent advances in instrumentation and methods. We believe that the TES technology has already made important improvements on the existing reference data, and it has enormous capacity to broaden its reach in the near future.

\subsection{Prospects for future measurements} \label{sec:prospects}

The current measurements improve on our previous effort with TES microcalorimeters in several critical ways. The sensors are more linear, they have better energy resolution, they have a much more nearly Gaussian energy-response function, and we have used an automated sample switcher to achieve excellent gain matching of multiple fluorescence targets. Nevertheless, we plan further improvements for future measurements. First, we can increase the opacity of the x-ray absorber by use of a thicker layer of gold or addition of electroplated bismuth atop the existing 1\,$\mu$m of gold. Either step would greatly reduce the number of photons that are currently stopped in the silicon membrane from the current value of 1.5\,\%, which could eliminate the need to identify and remove membrane-stopping events. Development of a very high-speed sample switcher is also being pursued. Switching between multiple samples on timescales fast compared with the millisecond-scale response times of the TES detectors would guarantee the exact equality of gains between the switched samples. Finally, the TES is capable of energy resolution more than a factor of two better than that achieved in the current work. Gaussian resolutions of 1.58\,eV have been demonstrated in the 6\,keV energy range~\cite{Miniussi:2018}.

We also anticipate measurements of L lines from additional elements, both in the lanthanide series and at atomic numbers up to approximately 83. We expect that K lines of elements such as phosphorous, sulfur, chlorine, and potassium can be measured by the high-accuracy, high-resolution VDCS\@. Traceable and modern measurements of K lines in the energy range 2\,keV to 3.5\,keV to serve as primary reference standards would enable extension of TES technique into this energy range. This, in turn, would allow measurements of L lines of the $40\lesssim Z\lesssim 60$ elements, as well as M lines for $Z\gtrsim73$.

TES spectrometers have recently been shown to be sensitive to chemical effects in the K lines of titanium~\cite{Miaja-Avila:2020}. These effects include shifts in the Ti K$\alpha$ line of less than 1\,eV and large variations in the valence-to-core satellite transitions K$\beta''$ and K$\beta_{2,5}$, both of which depend on the oxidation state and ligand chemistry of the titanium sample. This demonstration opens the prospect of many future measurements, such as the comparison of K, L, or M lines of an element in its various chemical states. Because many XRF applications use molecular samples, a chemistry-specific database of line energies and profiles could be a valuable tool. A TES spectrometer with 2\,eV resolution could generate such data.

Finally, the relative fluorescence intensity is implicit in measurements like the ones presented here. Correction of self-absorption effects in the emitting material is required, unless measurements are made on thin films. We plan to pursue both approaches.

Future arrays with at least one thousand TESs are planned. Large arrays would be more sensitive to lines of low intensity. Other energy ranges are possible than the 4\,keV to 10\,keV studied here. TESs have already been used at synchrotrons at energies as low as 270\,eV \cite{Ullom:2014er,Fowler:2015MPF} and in the 1\,keV to 2\,keV range~\cite{Lee:2019}, where 0.75\,eV resolution has been demonstrated~\cite{Morgan:2019}. With the addition of macroscopic pieces of tin to each TES~\cite{Zink:2006,Bennett:2012kf}, excellent energy resolution for photons up to 200\,keV has also been achieved~\cite{Winkler:2015,Mates:2017}. The superconducting sensors used in this work can therefore be applied to study a large portion of all x-ray fluorescence lines, and to estimate their line profiles, absolute energies, and relative intensities.

\section{Conclusions} \label{sec:conclusions}

We have used a spectrometer consisting of 50 TES microcalorimeters to record the fluorescence emission of four lanthanide elements in the energy range 4\,keV to 10\,keV\@. The resulting spectra allow us to derive expressions to measure the line profiles of 97 L lines of Pr, Nd, Tb, and Ho with absolute calibration known to between 0.1\,eV at the lower energies and 0.4\,eV at the highest. We summarize these profiles by a sum-of-Voigts representation, with all relevant parameters given in \ref{sec:fit_details}.

The microcalorimeter spectrometer is capable of great sensitivity over a very broad energy range. The 97 L lines measured in this work outnumber all the L lines appearing in the SRD-128 reference survey that were measured in the 25 years preceding its publication. The total number of L-line measurements in that survey across the entire periodic table is 1263, only 13 times the number of lines in tables~\ref{tab:peaksPr} to \ref{tab:peaksHo}. Most of these lines are accessible to future measurements with microcalorimeters. So are lines not previously measured, satellite features, and M lines that are absent from SRD-128.

These data represent a self-consistent survey of the two dozen most intense L lines of four lanthanide-series elements, both diagram and non-diagram lines. The survey has thoroughly characterized uncertainties and an energy calibration linked back to SI standards via the K lines of the 3d transition metals. Previous reference compilations such as SRD-128 collect measurements of diverse origin, taken across many decades often with minimal quantitative assessment of uncertainties and limited reference to the SI scale. We recommend the results of this study as the contemporary x-ray wavelength and energy standards for the L-line profiles and energies of the elements Pr, Nd, Tb, and Ho. 



\ack

Authors LH and CS work in the laboratory that was once used by Deslattes and Mooney. We thank these predecessors for their lab notebooks and unpublished data on the L lines of Nd, Sm, and Ho, taken with the spectrometer described in~\cite{Deslattes:1967,Mooney:1992}. We have learned much from participants in the several FP Initiative workshops of recent years, which were organized by the European X-ray Spectrometry Association. We appreciate helpful discussions with Burkhard Beckhoff, Chris Chantler, Brianna Ganly, Mauro Guerra, Philipp H\"onicke, Joanna Hoszowska, Haibo Huang, Michael Kolbe, Marie-Christine L\'epy, Jos\'e Paulo Santos, Timo Wolff, and Charalampos Zarkadas.

We thank Yuri Ralchenko and Dan Becker for reviewing this manuscript and James Hays-Wehle for laboratory measurements on the first ``sidecar absorber'' TES designs. We gratefully acknowledge financial support from the NIST Innovations in Measurement Science Program. CS performed this work under the financial assistance awards 70NANB15H051 and 70NANB19H157 from U.S. Department of Commerce, National Institute of Standards and Technology.


\printbibliography

\newpage
\appendix

\section{Complete results of fits to Lanthanide spectra} \label{sec:fit_details}

\begin{table*}
    \centering
    \include{Tables/roi_table}
    \caption{All fitting regions of interest. Fits are performed on spectra binned into histograms with one bin per eV\@. \emph{E range} gives the range of energies in each ROI (in eV). $N_\mathrm{V}$ is the number of Voigt components fit. \emph{BG low} and \emph{BG high} are the background values, in photons per bin, at the lowest and highest energies in the range; the background model is linear between these endpoints. The value of $\chi^2$ is defined as $\chi^2\equiv-2\log\cal{L}$ with $\cal{L}$ the maximum Poisson likelihood, which asymptotically follows the $\chi^2$ distribution with the given number of degrees of freedom~\cite{Baker:1984} (\emph{dof}). Assuming this limit is reached yields the given \emph{p-value}. }
    \label{tab:ROIs}
\end{table*}

The lanthanide fluorescence spectra are fit in nine separate regions of interest (ROI) for each element, shown in figures~\ref{fig:allL3M} through \ref{fig:allL1edges}. There are eight panels per element instead of nine, because figure~\ref{fig:allL3edges} merges ROIs E and F into one panel per element. Each fit model consists of all a priori known backgrounds, plus a linear background, plus the quantum efficiency times the sum of two or more Voigt components. Table~\ref{tab:ROIs} gives details of the 36 ROIs, including the energy range, the number of Voigt components used, the zero-resolution Gaussian width $\sigma_0$, the background, and the quality of fit.

The $p$-value measure shows that the model fits very well those ROIs with relatively weak or symmetric emission lines. ROIs with very intense lines, where complex line shapes can be seen with high significance, tend not to fit perfectly. Our understanding is that these regions would be best modeled with additional Voigt components, but the fragility of nonlinear function minimization in the presence of partial degeneracies between parameters prevented us from increasing the number of components.

The model of the signal (equation~\ref{eq:signal_model}) relies on the normalized Voigt function centered at energy $E$, with Gaussian and Lorentzian widths $\sigma$ and $\Gamma$. The Voigt function is the convolution of the normalized Gaussian and Lorentzian distributions $G$ and $L$:
\begin{equation} \label{eq:voigt}
V(x; E,\Gamma,\sigma) \equiv \int_{-\infty}^\infty\,\mathrm{d}y\, G(y; \sigma)\,L(x-y; E, \Gamma),
\end{equation}
where we use the width conventions
\begin{equation} \label{eq:gauss}
G(x; \sigma) \equiv \frac{1}{\sqrt{2\pi}\sigma} \exp\left[-\frac{x^2}{2\sigma^2}\right]
\end{equation}
and
\begin{equation} \label{eq:lorentz}
L(x; E, \Gamma) \equiv \frac{\Gamma/\pi}{(x-E)^2+\Gamma^2}.
\end{equation}
The signal model $S(E)$ is the sum of multiple Voigts, with various amplitudes, widths, and center energies. For fitting our data, we have used the nominal resolution (equation~\ref{eq:nominal_res}) to compute the total $\sigma$ in equation~\ref{eq:sigma}, but the signal formula can be also used to predict the spectral profile for any background-free measurement with a Gaussian instrumental broadening if one replaces $\delta E$ with the appropriate value. The use of any $\delta E \lesssim 4$\,eV represents a deconvolution of our results, of course, and is likely to yield less reliable results.

The overall data model $f(E)$ fit in each spectral region of interest is the quantum efficiency times the signal, plus two background terms:
\begin{equation}
f(E) = Q(E)S(E) + b_c(E) + b_f(E).
\end{equation}
Quantum efficiency $Q(E)$ is the product of the sensor quantum efficiency (figure~\ref{fig:qe}) and a correction factor of 0.98 to 0.99 to account for the energy-dependent effect of the cut used to eliminate membrane hits (section~\ref{sec:pulse_analysis}). 
The $b_c(E)$ is a 2-parameter model of the continuum background, linear in the energy $E$ and constrained such that $b_c\ge 0$ throughout the fit range, and $b_f(E)$ is the sum of all background features. (The $b_f$ model has no free parameters; see section~\ref{sec:background}.) Values of $b_c(E)$ at either end of the ROI appear in table~\ref{tab:ROIs}.

The statistic minimized in the fits between model and data was $\chi^2_\mathrm{ML}\equiv -2\log\,\mathcal{L}$, where $\mathcal{L}$ is the Poisson likelihood. This choice avoids the biases associated with the use of weighted least squares when data are Poisson-distributed. We used the LMFit~\cite{LMFit} package for function minimization, with $\chi^2_\mathrm{ML}$ replacing the usual target function. In all fits, we imposed the limits 0.6\,eV $\le \Gamma_i \le$\,50\,eV\@. All $E_i$ were constrained to lie within the energy ROI. The background was required to be non-negative at each end of the ROI\@.

\begin{table*}
    \centering
\include{Tables/shape_data_Pr}
    \caption{Results from maximum likelihood fits of a multi-Voigt model in each ROI of praseodymium. The ROI labels A--I refer to energy ranges found in table~\ref{tab:ROIs}. Each entry corresponds to a single Voigt component of the model. The zero-resolution Gaussian width (in eV) is $\sigma_0$, the same for all components in a single ROI\@. Equations~\ref{eq:nominal_res} and \ref{eq:sigma} define $\sigma_\mathrm{res}$ and $\sigma$. The center energy (in eV) is $E_i$. The Lorentzian half-width at half maximum (in eV) is $\Gamma_i$. The intensity $I_i$ is the integrated number of photons in each Voigt component in an idealized instrument with 100\,\% efficiency; the spectrometer quantum efficiency (figure~\ref{fig:qe}) reduces the detected number of photons relative to this value by a factor of 4 to 6, depending on energy. An online supplement to this publication provides the values from this table and the following three tables in digital form.}
    \label{tab:shape_fits_Pr}
\end{table*}

\begin{table*}
    \centering
\include{Tables/shape_data_Nd}
    \caption{Maximum likelihood fit results for a multi-Voigt model in each ROI of neodymium. Columns are as in table~\ref{tab:shape_fits_Pr}.}
    \label{tab:shape_fits_Nd}
\end{table*}

\begin{table*}
    \centering
\include{Tables/shape_data_Tb}
    \caption{Maximum likelihood fit results for a multi-Voigt model in each ROI of terbium. Columns are as in table~\ref{tab:shape_fits_Pr}.}
    \label{tab:shape_fits_Tb}
\end{table*}

\begin{table*}
    \centering
\include{Tables/shape_data_Ho}
    \caption{Maximum likelihood fit results for a multi-Voigt model in each ROI of holmium. Columns are as in table~\ref{tab:shape_fits_Pr}.}
    \label{tab:shape_fits_Ho}
\end{table*}

The signal model $S(E)$ in each ROI for the Pr emission spectrum can be reconstructed from the values in table~\ref{tab:shape_fits_Pr}. Tables~\ref{tab:shape_fits_Nd}, \ref{tab:shape_fits_Tb}, and \ref{tab:shape_fits_Ho} give the corresponding results for Nd, Tb, and Ho spectra, respectively. The center energy $E_i$, Lorentzian half-width at half-maximum $\Gamma_i$, and integrated intensity $I_i$ are the parameters of each Voigt component (equation~\ref{eq:signal_model}). The value given for $\sigma_0$ in the table is combined with the instrumental resolution (by equation~\ref{eq:sigma}) to find the combined Gaussian width $\sigma$. The resolution used can be that of our spectrometer (equation~\ref{eq:nominal_res}), or that of any other. Fit uncertainties are not included in the tables, because many strong correlations between parameters are present, particularly when an asymmetric peak is modeled as two or more Voigts. Full estimates of parameter covariance are included with the online supplementary material.

\section{Corrections to the estimates of peak energy}\label{sec:peak_corrections}

\begin{table*}
\centering
\include{Tables/peak_data_supplement}
\caption{Corrections required to estimate the peak energy. Lines with a correction magnitude of less than 0.1\,eV are omitted. The columns \emph{6\,eV} and \emph{raw} indicate the energy at which the spectral model peaks, with resolutions of $\delta E=6$\,eV and the as-measured resolution (typically $\delta E\approx 4$\,eV). \emph{Shift} is the correction to the raw peak required to estimate the zero-resolution peak, on the assumption that shift changes linearly between $\delta E=6$\,eV and $\delta E=0$. \emph{Peak Corrected} is the final estimate of the peak when $\delta E=0$.}
\label{tab:peaks_supplement}
\end{table*}

For the purposes of comparison to existing reference data, we have attempted to estimate the true peak energy of each line profile. By \emph{peak energy}, we mean the local maximum in the spectrum for each line that would be seen in an ideal measurement with perfect energy resolution and no noise. Given that we measure the profiles with 4\,eV resolution, we evaluated three methods of estimation:
\begin{itemize}
\item Assume the sum-of-Voigts model is correct even after deconvolution. Set $\delta E=0$ and $\sigma=\sigma_0$, and find peaks from the model.
\item Assume that instrumental broadening does not shift peaks, and make no correction. Use the sum-of-Voigts model with the appropriate resolution from equation~\ref{eq:nominal_res}. Find peaks from this as-measured model.
\item Assume peaks shift linearly as a function of $\delta E$.
\end{itemize}
We find that deconvolution of the model systematically underestimates the peak shifts (section~\ref{sec:peak_energy}), so only the second and third methods are compared in this appendix. The third proves superior.

We chose between the methods based on a set of high-resolution data on the profiles of 28 K and L lines. Specifically, these lines are:
\begin{itemize}
    \item K$\alpha_1$, K$\alpha_2$, and K$\beta_{1,3}$ lines of Cr, Mn, Fe, Co, Ni, and Cu~\cite{Holzer:1997ts}. From this set, we omit Cr K$\alpha_2$, because it is only marginally resolved at $\delta E$=6\,eV\@.
        \item Ti K$\alpha_1$ \cite{Chantler:2006va}.
\item L$\alpha_1$, L$\alpha_2$, and L$\beta_1$  lines of Nd, Sm, and Ho, plus Ho L$\beta_3$.
\end{itemize}
The 10 lanthanide L line profiles are unpublished data taken by Mooney, Deslattes, and others with the VDCS spectrometer~\cite{Deslattes:1967,Mooney:1992}. We smooth these data by fitting them as a linear combination of a Lorentzian ($\Gamma$=5\,eV) and its first ten derivatives with respect to energy. Such fits capture line asymmetries, eliminate much of the measurement noise, and have reasonable extrapolations beyond the measured energy regions.

\begin{figure}
    \centering
    \includegraphics[width=\linewidth]{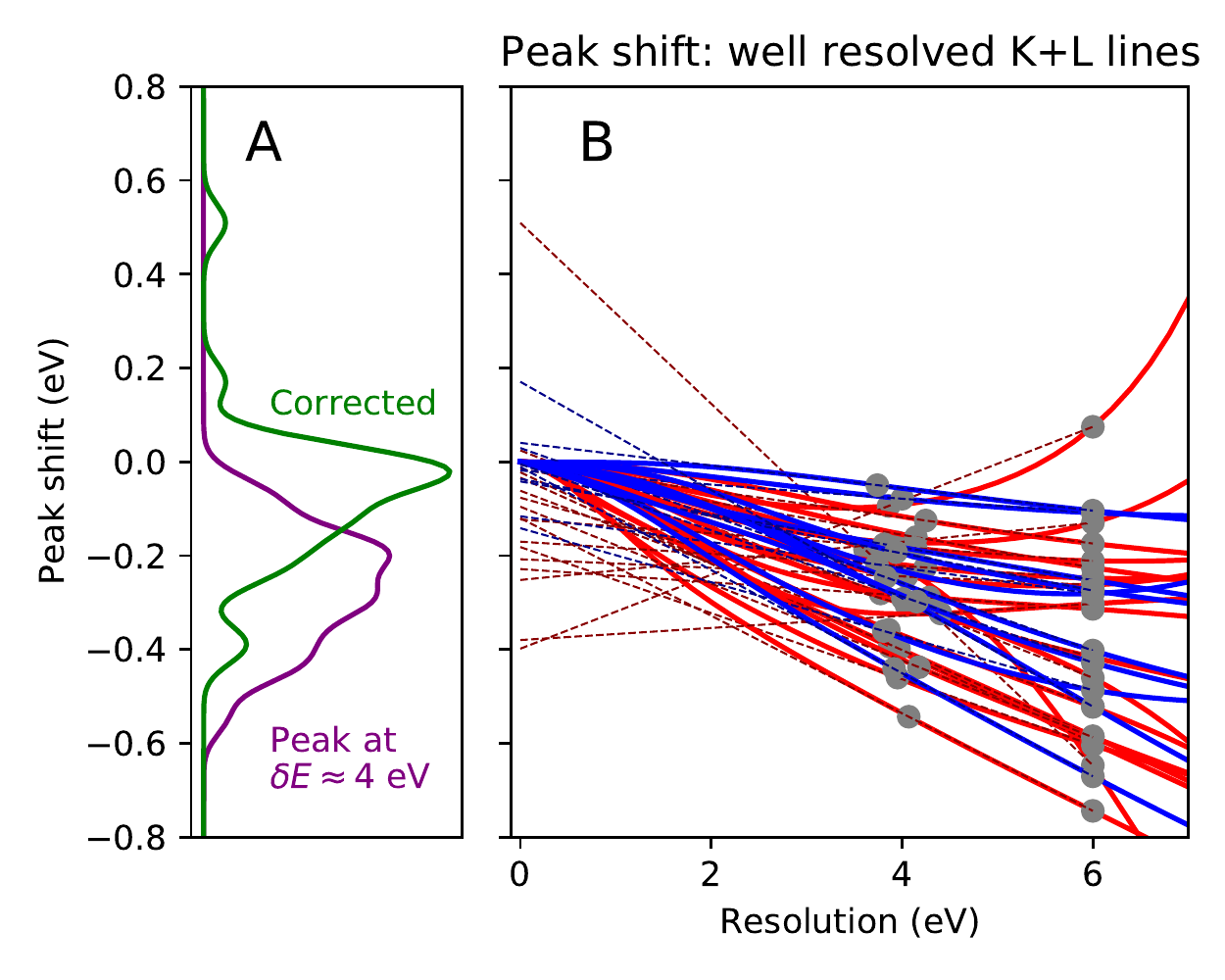}
    \caption{(B) Peak shift, relative to peak energy at resolution $\delta E=0$, for the 18 K lines (red) and 10 L lines (blue) listed in \ref{sec:peak_corrections}. The dashed lines show the linear extrapolation of the peak at 6\,eV and at nominal resolution (dots near 4\,eV) down to the desired $\delta E=0$. The y-intercept of the dashed lines shows the peak bias that our method of correction would give for each profile.
    (A) Distributions of the 28 peak shifts, corrected and at the nominal resolution of $\delta E\approx 4$\,eV\@. Each point is convolved with a Gaussian kernel of $\sigma=0.04$\,eV to make a smoother curve. These distributions establish that the corrected peak has less bias.
    \label{fig:peak_shift_KL_examples}
    }
\end{figure}

Given high-resolution line profile models for these 28 fluorescence lines, we can compute the peak energy shift as a function of resolution, from the original instrumental resolution of $\delta E\lesssim 0.4$\,eV up to any higher value (figure~\ref{fig:peak_shift_KL_examples}). If these lines had been measured with the TES spectrometer, no data would be available to us at resolutions lower than the nominal value (equation~\ref{eq:nominal_res}, approximately 4\,eV); the peak shift could be estimated using only information at nominal and higher resolutions, just as we have to do with the lanthanide lines of interest in this work. 

We estimate the peak by assuming that peak shift is a linear function of resolution between $\delta E=0$ and $\delta E=6$\,eV\@. We can sample this function at the nominal resolution (approximately 4\,eV) and 6\,eV by our sum-of-Voigts model. Linear extrapolation to $\delta E=0$ yields the estimated peak energy. The dashed lines in figure~\ref{fig:peak_shift_KL_examples} show this extrapolation. Although the extrapolation to $\delta E=0$ is not perfect, it shows both smaller bias and (apart from a few outliers) smaller variation than the (uncorrected) peak at 4\,eV resolution. The median bias (error) in the extrapolated peak estimates is $-0.052$\,eV, while it is five times as large ($-0.263$\,eV) for the collection of peaks at the TES nominal resolution. A robust estimator of variation, the interquartile range, is 0.14\,eV for the extrapolated estimates, versus 0.18\,eV for the shifts at nominal resolution. Therefore, we choose the method of linear extrapolation to $\delta E=0$ to estimate the ideal line peak energies for the lanthanide line profiles in this study. To capture both variation and any small remaining bias, we assign 0.18\,eV as the systematic uncertainty on the peak correction method. This value is the root-mean-square of the 28 biases.

The peak shift is a slightly stronger function of energy among the available K lines than the L lines, and the uncorrected biases are therefore larger. If this observation were true for the L lines generally---which after all have shorter lifetimes and greater intrinsic width than K lines at a similar energy---then our estimated 0.18\,eV systematic uncertainty may be conservative.

With the assumption that peaks depend linearly on $\delta E$, we can compute the corrections required to estimate the peak energy for the lanthanide-metal L lines. Most corrections are very small. Those that exceed 0.1\,eV in magnitude are given in table~\ref{tab:peaks_supplement}. All peak estimates in tables~\ref{tab:peaksPr}-\ref{tab:peaksHo} are the corrected values.

\section{Fit strategies other than Voigts with additional Gaussian width}\label{sec:failed_strategies}

We undertook this work to create transferable, absolutely calibrated emission line standards of the broadest possible applicability. Therefore, we explored a wide range of strategies that aimed to minimize unphysical artifacts in the $\delta E\rightarrow0$ limit without requiring significant compromises in fit quality. The approach we chose (section~\ref{sec:shape}) is to fit each line as a sum of one or more Voigt functions with a common width $\sigma$ that exceeds the width given by the instrumental broadening alone. Here we describe the range of ideas tried but ultimately \emph{not} chosen, ideas that include the use of regularization, asymmetric line components, direct deconvolution, and certain linear models.

Regularization strategies employ a penalty term added to the usual quality-of-fit statistic (in these fits, the statistic being minimized was $\chi^2_\mathrm{ML}\equiv -2\log\,\mathcal{L}$ with $\mathcal{L}$ the Poisson likelihood). It was hoped that if the proper quantity could be identified and penalized, then the unphysical features in the $\delta E\rightarrow0$ limit of a fit would be suppressed at insignificant cost to the fit quality. In fact, we found no such quantity. We tried several penalties computed globally on the zero-resolution fit function $f(E)$: its integrated squared curvature; its spectral entropy; the inverse of its expected probability density; and the difference in energy of $f$ between the instrumental and the zero-resolution versions. We also tried multiple penalties that acted effectively as Bayesian priors on the widths of the individual Lorentzian components, favoring wider components. In most cases, it was possible to find at least one region of interest, when fit both with and without penalties, where the preferred function (the one with an apparently more physical shape) was counterintuitively assigned a \emph{higher} penalty. Worse, the addition of a penalty term large enough to tame the unphysical behavior inevitably led to poor fits.

We also tried replacing the symmetric fit function with asymmetric components. These included ad hoc models in which a standard Lorentzian was modified by either a smooth or sudden change in its width parameter near or at its center, as well as the physically motivated Doniach-\v{S}unji\'{c} (DS) model~\cite{Doniach:1970}. In each case, the asymmetric model was convolved with a Gaussian to account for the spectrometer's finite resolution. Although the DS model was proposed specifically to explain the asymmetry in x-ray K$\alpha$ lines, none of the models that employed asymmetric components led to good fits of the L-line data with fewer free parameters than the corresponding Voigt fits, nor to a reduction in structure as $\delta E\rightarrow0$.

We also attempted deconvolution of the spectra, either the raw histograms of x-ray counts, or smoothed representations of them. The Richardson-Lucy deconvolution~\cite{Richardson:1972,Lucy:1974} of the spectra was highly influenced by noise, as is true of deconvolution generally. No clear relationship was identified between the peak energies found after deconvolution and the zero-resolution peaks in those cases where the latter were known ahead of time. As with the zero-resolution limit of the sum-of-Voigts representation, the deconvolution tended to understate the effects of asymmetry.

Purely linear models were also considered. In one, each line was fit as a sum of a single Voigt function and its first several derivatives. In another, a line was fit as sum of many Voigt functions with common width and centered on a regular grid. This grid could be made much more dense with the use of an L0 regularization penalty. Both linear fits produced a great deal of unwanted structure in the zero-resolution limit.  Furthermore, tests of the grid-of-Voigts method applied to the known shapes of several K$\alpha$ lines demonstrated both bias in the peak location and a sensitivity to details like the center and width of the initial Voigt or the grid of them.

\end{document}

%% file: Tables/peak_data_Pr.tex
%
\begin{tabular}{lllllrlrr}
\multicolumn{2}{c}{Line name} & & & & Peak & \multicolumn{3}{c}{Uncertainty} \\
IUPAC & Siegbahn & Spark & &  Reference &  Estimate & Stat & Sys & Peak  \\
\hline
Pr L$_3$M$_1$             & L$\ell$          & \begin{sparkline}{6}
    \spark 0.00 0.335 0.02 0.357 0.04 0.380 0.06 0.404 0.08 0.431 0.10 0.458 0.12 0.488 0.14 0.519 0.16 0.551 0.18 0.584 0.20 0.619 0.22 0.654 0.24 0.690 0.26 0.726 0.28 0.762 0.30 0.797 0.32 0.831 0.34 0.863 0.36 0.892 0.38 0.919 0.40 0.943 0.42 0.963 0.44 0.979 0.46 0.991 0.48 0.998 0.50 1.000 0.52 0.998 0.54 0.990 0.56 0.979 0.58 0.962 0.60 0.942 0.62 0.919 0.64 0.892 0.66 0.863 0.68 0.831 0.70 0.798 0.72 0.764 0.74 0.729 0.76 0.694 0.78 0.659 0.80 0.625 0.82 0.592 0.84 0.559 0.86 0.528 0.88 0.499 0.90 0.471 0.92 0.445 0.94 0.420 0.96 0.397 0.98 0.375 1.00 0.355 /
\end{sparkline}
 &     &  4453.2\phantom{0} $\pm$ 0.9\phantom{0} & 4457.59 & $\pm$0.03  & $\pm$0.11  & $\pm$0.18  \\
Pr L$_2$M$_1$             & L$\eta$          & \begin{sparkline}{6}
    \spark 0.00 0.347 0.02 0.363 0.04 0.379 0.06 0.397 0.08 0.417 0.10 0.438 0.12 0.461 0.14 0.486 0.16 0.512 0.18 0.540 0.20 0.570 0.22 0.601 0.24 0.635 0.26 0.669 0.28 0.705 0.30 0.741 0.32 0.778 0.34 0.814 0.36 0.849 0.38 0.882 0.40 0.912 0.42 0.939 0.44 0.961 0.46 0.979 0.48 0.992 0.50 0.999 0.52 1.000 0.54 0.996 0.56 0.986 0.58 0.972 0.60 0.953 0.62 0.931 0.64 0.907 0.66 0.880 0.68 0.852 0.70 0.824 0.72 0.796 0.74 0.768 0.76 0.742 0.78 0.717 0.80 0.693 0.82 0.671 0.84 0.651 0.86 0.632 0.88 0.615 0.90 0.599 0.92 0.585 0.94 0.572 0.96 0.560 0.98 0.550 1.00 0.540 /
\end{sparkline}
 &     &  4935.6\phantom{0} $\pm$ 0.9\phantom{0} & 4933.39 & $\pm$0.15  & $\pm$0.09  & $\pm$0.18  \\
Pr L$_3$M$_4$             & L$\alpha_2$      & \begin{sparkline}{6}
    \spark 0.00 0.095 0.02 0.101 0.04 0.108 0.06 0.115 0.08 0.124 0.10 0.134 0.12 0.145 0.14 0.159 0.16 0.175 0.18 0.193 0.20 0.215 0.22 0.241 0.24 0.272 0.26 0.308 0.28 0.350 0.30 0.398 0.32 0.451 0.34 0.509 0.36 0.570 0.38 0.632 0.40 0.691 0.42 0.745 0.44 0.790 0.46 0.822 0.48 0.840 0.50 0.843 0.52 0.831 0.54 0.806 0.56 0.769 0.58 0.725 0.60 0.677 0.62 0.628 0.64 0.580 0.66 0.538 0.68 0.501 0.70 0.471 0.72 0.448 0.74 0.433 0.76 0.424 0.78 0.423 0.80 0.428 0.82 0.440 0.84 0.460 0.86 0.487 0.88 0.524 0.90 0.570 0.92 0.627 0.94 0.697 0.96 0.781 0.98 0.882 1.00 1.000 /
\end{sparkline}
 &     &  5013.6\phantom{0} $\pm$ 0.9\phantom{0} & 5011.88 & $\pm$0.02  & $\pm$0.11  & $\pm$0.18  \\
Pr L$_3$M$_5$             & L$\alpha_1$      & \begin{sparkline}{6}
    \spark 0.00 0.062 0.02 0.065 0.04 0.068 0.06 0.073 0.08 0.080 0.10 0.088 0.12 0.098 0.14 0.110 0.16 0.124 0.18 0.140 0.20 0.160 0.22 0.183 0.24 0.211 0.26 0.244 0.28 0.284 0.30 0.333 0.32 0.392 0.34 0.460 0.36 0.539 0.38 0.625 0.40 0.715 0.42 0.803 0.44 0.882 0.46 0.945 0.48 0.986 0.50 1.000 0.52 0.986 0.54 0.946 0.56 0.883 0.58 0.805 0.60 0.717 0.62 0.626 0.64 0.538 0.66 0.458 0.68 0.386 0.70 0.324 0.72 0.273 0.74 0.230 0.76 0.195 0.78 0.166 0.80 0.143 0.82 0.124 0.84 0.109 0.86 0.096 0.88 0.085 0.90 0.076 0.92 0.069 0.94 0.062 0.96 0.057 0.98 0.052 1.00 0.047 /
\end{sparkline}
 &     &  5033.8\phantom{0} $\pm$ 0.6\phantom{0} & 5032.85 & $\pm$0.004 & $\pm$0.11  & $\pm$0.18  \\
Pr L$_2$M$_4$+L$_1$M$_2$  & L$\beta_{1,4}$   & \begin{sparkline}{6}
    \spark 0.00 0.052 0.02 0.057 0.04 0.063 0.06 0.069 0.08 0.077 0.10 0.086 0.12 0.097 0.14 0.110 0.16 0.126 0.18 0.145 0.20 0.169 0.22 0.197 0.24 0.231 0.26 0.272 0.28 0.320 0.30 0.375 0.32 0.437 0.34 0.505 0.36 0.578 0.38 0.655 0.40 0.733 0.42 0.810 0.44 0.882 0.46 0.941 0.48 0.982 0.50 1.000 0.52 0.991 0.54 0.955 0.56 0.895 0.58 0.818 0.60 0.731 0.62 0.641 0.64 0.556 0.66 0.479 0.68 0.413 0.70 0.358 0.72 0.314 0.74 0.278 0.76 0.250 0.78 0.227 0.80 0.208 0.82 0.191 0.84 0.177 0.86 0.164 0.88 0.152 0.90 0.140 0.92 0.130 0.94 0.119 0.96 0.110 0.98 0.101 1.00 0.092 /
\end{sparkline}
 &     &  5488.9\phantom{0} $\pm$ 1.1\phantom{0} & 5488.48 & $\pm$0.008 & $\pm$0.11  & $\pm$0.18  \\
Pr ---                    & L$\beta$'        & \begin{sparkline}{6}
    \spark 0.00 0.418 0.02 0.484 0.04 0.556 0.06 0.632 0.08 0.711 0.10 0.789 0.12 0.863 0.14 0.927 0.16 0.974 0.18 1.000 0.20 0.999 0.22 0.971 0.24 0.918 0.26 0.845 0.28 0.759 0.30 0.669 0.32 0.582 0.34 0.502 0.36 0.432 0.38 0.374 0.40 0.326 0.42 0.288 0.44 0.258 0.46 0.233 0.48 0.213 0.50 0.196 0.52 0.181 0.54 0.168 0.56 0.155 0.58 0.144 0.60 0.133 0.62 0.122 0.64 0.112 0.66 0.103 0.68 0.095 0.70 0.087 0.72 0.080 0.74 0.073 0.76 0.067 0.78 0.062 0.80 0.057 0.82 0.053 0.84 0.049 0.86 0.046 0.88 0.042 0.90 0.040 0.92 0.037 0.94 0.035 0.96 0.033 0.98 0.031 1.00 0.029 /
\end{sparkline}
 &     &                  & $\star$\ \emph{5496.8\phantom{0} }& $\pm$0.10  & $\pm$0.11  & $\pm$1.0\phantom{0}  \\
Pr L$_1$M$_3$             & L$\beta_3$       & \begin{sparkline}{6}
    \spark 0.00 0.126 0.02 0.135 0.04 0.145 0.06 0.156 0.08 0.169 0.10 0.184 0.12 0.200 0.14 0.219 0.16 0.241 0.18 0.266 0.20 0.295 0.22 0.327 0.24 0.365 0.26 0.407 0.28 0.455 0.30 0.508 0.32 0.566 0.34 0.628 0.36 0.693 0.38 0.758 0.40 0.821 0.42 0.880 0.44 0.929 0.46 0.968 0.48 0.992 0.50 1.000 0.52 0.992 0.54 0.968 0.56 0.929 0.58 0.880 0.60 0.821 0.62 0.758 0.64 0.693 0.66 0.628 0.68 0.566 0.70 0.508 0.72 0.454 0.74 0.407 0.76 0.364 0.78 0.327 0.80 0.294 0.82 0.266 0.84 0.241 0.86 0.219 0.88 0.200 0.90 0.183 0.92 0.169 0.94 0.156 0.96 0.145 0.98 0.135 1.00 0.126 /
\end{sparkline}
 &     &  5591.8\phantom{0} $\pm$ 1.1\phantom{0} & 5593.04 & $\pm$0.02  & $\pm$0.12  & $\pm$0.18  \\
Pr L$_3$N$_1$             & L$\beta_6$       & \begin{sparkline}{6}
    \spark 0.00 0.183 0.02 0.188 0.04 0.194 0.06 0.201 0.08 0.210 0.10 0.219 0.12 0.231 0.14 0.244 0.16 0.260 0.18 0.278 0.20 0.300 0.22 0.326 0.24 0.357 0.26 0.393 0.28 0.435 0.30 0.483 0.32 0.538 0.34 0.598 0.36 0.664 0.38 0.732 0.40 0.800 0.42 0.863 0.44 0.919 0.46 0.963 0.48 0.990 0.50 1.000 0.52 0.991 0.54 0.963 0.56 0.920 0.58 0.864 0.60 0.800 0.62 0.732 0.64 0.663 0.66 0.597 0.68 0.536 0.70 0.480 0.72 0.431 0.74 0.389 0.76 0.352 0.78 0.321 0.80 0.294 0.82 0.271 0.84 0.252 0.86 0.235 0.88 0.221 0.90 0.208 0.92 0.197 0.94 0.188 0.96 0.180 0.98 0.173 1.00 0.166 /
\end{sparkline}
 &     &  5659.7\phantom{0} $\pm$ 1.5\phantom{0} & 5660.20 & $\pm$0.05  & $\pm$0.13  & $\pm$0.18  \\
Pr ---                    & L$\beta_{14}$    & \begin{sparkline}{6}
    \spark 0.00 0.029 0.02 0.030 0.04 0.032 0.06 0.033 0.08 0.035 0.10 0.036 0.12 0.038 0.14 0.040 0.16 0.042 0.18 0.044 0.20 0.046 0.22 0.049 0.24 0.051 0.26 0.054 0.28 0.056 0.30 0.059 0.32 0.062 0.34 0.065 0.36 0.068 0.38 0.071 0.40 0.074 0.42 0.077 0.44 0.080 0.46 0.083 0.48 0.086 0.50 0.089 0.52 0.092 0.54 0.095 0.56 0.098 0.58 0.102 0.60 0.105 0.62 0.110 0.64 0.115 0.66 0.121 0.68 0.128 0.70 0.136 0.72 0.147 0.74 0.159 0.76 0.175 0.78 0.195 0.80 0.220 0.82 0.250 0.84 0.289 0.86 0.335 0.88 0.392 0.90 0.460 0.92 0.540 0.94 0.632 0.96 0.739 0.98 0.861 1.00 1.000 /
\end{sparkline}
 & CS  &  5830.6          & $\star$\ \emph{5830.1\phantom{0} }& $\pm$0.35  & $\pm$0.11  & $\pm$1.0\phantom{0}  \\
Pr L$_3$N$_\mathrm{4,5}$  & L$\beta_{2,15}$  & \begin{sparkline}{6}
    \spark 0.00 0.055 0.02 0.058 0.04 0.062 0.06 0.067 0.08 0.073 0.10 0.081 0.12 0.090 0.14 0.102 0.16 0.117 0.18 0.135 0.20 0.158 0.22 0.185 0.24 0.217 0.26 0.255 0.28 0.298 0.30 0.348 0.32 0.406 0.34 0.471 0.36 0.544 0.38 0.624 0.40 0.709 0.42 0.795 0.44 0.874 0.46 0.940 0.48 0.984 0.50 1.000 0.52 0.986 0.54 0.944 0.56 0.878 0.58 0.796 0.60 0.707 0.62 0.618 0.64 0.536 0.66 0.463 0.68 0.401 0.70 0.349 0.72 0.306 0.74 0.270 0.76 0.239 0.78 0.212 0.80 0.188 0.82 0.167 0.84 0.148 0.86 0.132 0.88 0.117 0.90 0.104 0.92 0.093 0.94 0.084 0.96 0.075 0.98 0.068 1.00 0.061 /
\end{sparkline}
 &     &  5849.9\phantom{0} $\pm$ 1.6\phantom{0} & 5849.67 & $\pm$0.012 & $\pm$0.10  & $\pm$0.18  \\
Pr L$_1$M$_4$             & L$\beta_{10}$    & \begin{sparkline}{6}
    \spark 0.00 1.000 0.02 0.935 0.04 0.877 0.06 0.824 0.08 0.777 0.10 0.735 0.12 0.696 0.14 0.661 0.16 0.630 0.18 0.601 0.20 0.576 0.22 0.553 0.24 0.533 0.26 0.515 0.28 0.500 0.30 0.489 0.32 0.480 0.34 0.475 0.36 0.473 0.38 0.475 0.40 0.479 0.42 0.486 0.44 0.493 0.46 0.497 0.48 0.498 0.50 0.493 0.52 0.481 0.54 0.464 0.56 0.442 0.58 0.419 0.60 0.395 0.62 0.373 0.64 0.353 0.66 0.337 0.68 0.323 0.70 0.313 0.72 0.305 0.74 0.299 0.76 0.296 0.78 0.294 0.80 0.294 0.82 0.296 0.84 0.300 0.86 0.306 0.88 0.314 0.90 0.324 0.92 0.337 0.94 0.353 0.96 0.372 0.98 0.393 1.00 0.417 /
\end{sparkline}
 &     &  5884.0\phantom{0} $\pm$ 1.7\phantom{0} & \emph{5883.5\phantom{0} }& $\pm$0.5\phantom{0}  & $\pm$0.10  & $\pm$0.35  \\
Pr L$_1$M$_5$             & L$\beta_{9}$     & \begin{sparkline}{6}
    \spark 0.00 0.612 0.02 0.585 0.04 0.563 0.06 0.546 0.08 0.533 0.10 0.524 0.12 0.519 0.14 0.517 0.16 0.519 0.18 0.523 0.20 0.531 0.22 0.543 0.24 0.558 0.26 0.578 0.28 0.603 0.30 0.632 0.32 0.667 0.34 0.707 0.36 0.751 0.38 0.798 0.40 0.846 0.42 0.893 0.44 0.935 0.46 0.968 0.48 0.991 0.50 1.000 0.52 0.994 0.54 0.974 0.56 0.940 0.58 0.897 0.60 0.847 0.62 0.794 0.64 0.740 0.66 0.688 0.68 0.640 0.70 0.597 0.72 0.559 0.74 0.526 0.76 0.497 0.78 0.473 0.80 0.454 0.82 0.438 0.84 0.426 0.86 0.416 0.88 0.410 0.90 0.406 0.92 0.405 0.94 0.406 0.96 0.410 0.98 0.417 1.00 0.425 /
\end{sparkline}
 &     &  5902.8\phantom{0} $\pm$ 1.7\phantom{0} & 5902.71 & $\pm$0.21  & $\pm$0.09  & $\pm$0.18  \\
Pr L$_3$O$_1$             & L$\beta_7$       & \begin{sparkline}{6}
    \spark 0.00 0.837 0.02 0.794 0.04 0.758 0.06 0.728 0.08 0.705 0.10 0.686 0.12 0.673 0.14 0.665 0.16 0.660 0.18 0.661 0.20 0.665 0.22 0.673 0.24 0.685 0.26 0.701 0.28 0.721 0.30 0.745 0.32 0.772 0.34 0.801 0.36 0.833 0.38 0.866 0.40 0.899 0.42 0.930 0.44 0.957 0.46 0.979 0.48 0.993 0.50 1.000 0.52 0.998 0.54 0.986 0.56 0.965 0.58 0.937 0.60 0.901 0.62 0.861 0.64 0.818 0.66 0.773 0.68 0.728 0.70 0.683 0.72 0.641 0.74 0.601 0.76 0.564 0.78 0.529 0.80 0.497 0.82 0.468 0.84 0.442 0.86 0.418 0.88 0.396 0.90 0.376 0.92 0.358 0.94 0.342 0.96 0.326 0.98 0.313 1.00 0.300 /
\end{sparkline}
 & CS  &  5927.0\phantom{0} $\pm$ 1.1\phantom{0} & 5925.06 & $\pm$0.31  & $\pm$0.10  & $\pm$0.18  \\
Pr L$_3$N$_\mathrm{6,7}$  & Lu               & \begin{sparkline}{6}
    \spark 0.00 0.788 0.02 0.767 0.04 0.748 0.06 0.730 0.08 0.713 0.10 0.697 0.12 0.682 0.14 0.667 0.16 0.654 0.18 0.642 0.20 0.630 0.22 0.620 0.24 0.610 0.26 0.602 0.28 0.595 0.30 0.591 0.32 0.590 0.34 0.594 0.36 0.607 0.38 0.634 0.40 0.677 0.42 0.739 0.44 0.817 0.46 0.898 0.48 0.966 0.50 1.000 0.52 0.990 0.54 0.937 0.56 0.853 0.58 0.757 0.60 0.667 0.62 0.594 0.64 0.539 0.66 0.500 0.68 0.474 0.70 0.456 0.72 0.443 0.74 0.432 0.76 0.424 0.78 0.416 0.80 0.410 0.82 0.404 0.84 0.398 0.86 0.393 0.88 0.388 0.90 0.383 0.92 0.379 0.94 0.374 0.96 0.370 0.98 0.366 1.00 0.362 /
\end{sparkline}
 & CS  &  5959.5\phantom{0} $\pm$ 1.1\phantom{0} & 5960.61 & $\pm$0.33  & $\pm$0.10  & $\pm$0.18  \\
Pr L$_2$N$_1$             & L$\gamma_5$      & \begin{sparkline}{6}
    \spark 0.00 0.117 0.02 0.123 0.04 0.130 0.06 0.138 0.08 0.147 0.10 0.158 0.12 0.171 0.14 0.185 0.16 0.203 0.18 0.224 0.20 0.248 0.22 0.278 0.24 0.313 0.26 0.354 0.28 0.402 0.30 0.456 0.32 0.517 0.34 0.584 0.36 0.655 0.38 0.727 0.40 0.798 0.42 0.864 0.44 0.920 0.46 0.963 0.48 0.991 0.50 1.000 0.52 0.991 0.54 0.963 0.56 0.920 0.58 0.864 0.60 0.798 0.62 0.727 0.64 0.655 0.66 0.585 0.68 0.518 0.70 0.458 0.72 0.404 0.74 0.356 0.76 0.316 0.78 0.281 0.80 0.252 0.82 0.228 0.84 0.207 0.86 0.190 0.88 0.175 0.90 0.163 0.92 0.153 0.94 0.144 0.96 0.136 0.98 0.130 1.00 0.124 /
\end{sparkline}
 &     &  6136.2\phantom{0} $\pm$ 1.8\phantom{0} & 6136.24 & $\pm$0.09  & $\pm$0.12  & $\pm$0.18  \\
Pr ---                    & L$\gamma_9$      & \begin{sparkline}{6}
    \spark 0.00 0.029 0.02 0.031 0.04 0.033 0.06 0.035 0.08 0.037 0.10 0.039 0.12 0.041 0.14 0.044 0.16 0.047 0.18 0.051 0.20 0.054 0.22 0.058 0.24 0.062 0.26 0.067 0.28 0.071 0.30 0.076 0.32 0.081 0.34 0.086 0.36 0.091 0.38 0.096 0.40 0.100 0.42 0.105 0.44 0.108 0.46 0.112 0.48 0.114 0.50 0.116 0.52 0.118 0.54 0.119 0.56 0.119 0.58 0.120 0.60 0.120 0.62 0.121 0.64 0.121 0.66 0.123 0.68 0.127 0.70 0.132 0.72 0.140 0.74 0.151 0.76 0.168 0.78 0.192 0.80 0.224 0.82 0.266 0.84 0.320 0.86 0.387 0.88 0.467 0.90 0.559 0.92 0.658 0.94 0.759 0.96 0.855 0.98 0.938 1.00 1.000 /
\end{sparkline}
 & CS  &  6305.4          & \emph{6305.4\phantom{0} }& $\pm$0.09  & $\pm$0.11  & $\pm$0.5\phantom{0}  \\
Pr L$_2$N$_4$             & L$\gamma_1$      & \begin{sparkline}{6}
    \spark 0.00 0.114 0.02 0.114 0.04 0.115 0.06 0.115 0.08 0.115 0.10 0.116 0.12 0.118 0.14 0.120 0.16 0.125 0.18 0.131 0.20 0.141 0.22 0.156 0.24 0.176 0.26 0.204 0.28 0.242 0.30 0.290 0.32 0.351 0.34 0.425 0.36 0.510 0.38 0.603 0.40 0.700 0.42 0.795 0.44 0.879 0.46 0.945 0.48 0.987 0.50 1.000 0.52 0.984 0.54 0.940 0.56 0.873 0.58 0.792 0.60 0.702 0.62 0.611 0.64 0.525 0.66 0.447 0.68 0.380 0.70 0.324 0.72 0.278 0.74 0.240 0.76 0.209 0.78 0.183 0.80 0.161 0.82 0.142 0.84 0.125 0.86 0.110 0.88 0.097 0.90 0.086 0.92 0.076 0.94 0.067 0.96 0.060 0.98 0.054 1.00 0.048 /
\end{sparkline}
 &     &  6322.1\phantom{0} $\pm$ 1.4\phantom{0} & 6322.18 & $\pm$0.012 & $\pm$0.11  & $\pm$0.18  \\
Pr L$_2$O$_1$             & L$\gamma_8$      & \begin{sparkline}{6}
    \spark 0.00 0.554 0.02 0.547 0.04 0.540 0.06 0.533 0.08 0.526 0.10 0.521 0.12 0.516 0.14 0.513 0.16 0.513 0.18 0.514 0.20 0.519 0.22 0.527 0.24 0.540 0.26 0.558 0.28 0.581 0.30 0.611 0.32 0.646 0.34 0.688 0.36 0.735 0.38 0.785 0.40 0.837 0.42 0.886 0.44 0.931 0.46 0.966 0.48 0.990 0.50 1.000 0.52 0.994 0.54 0.973 0.56 0.938 0.58 0.891 0.60 0.836 0.62 0.776 0.64 0.714 0.66 0.653 0.68 0.595 0.70 0.543 0.72 0.496 0.74 0.456 0.76 0.421 0.78 0.391 0.80 0.366 0.82 0.345 0.84 0.327 0.86 0.312 0.88 0.299 0.90 0.288 0.92 0.278 0.94 0.270 0.96 0.263 0.98 0.257 1.00 0.252 /
\end{sparkline}
 & CS  &  6403.0\phantom{0} $\pm$ 1.3\phantom{0} & 6402.0\phantom{0} & $\pm$0.33  & $\pm$1.0\phantom{0}  & $\pm$0.18  \\
Pr L$_2$N$_\mathrm{6,7}$  & Lv               & \begin{sparkline}{6}
    \spark 0.00 0.278 0.02 0.276 0.04 0.275 0.06 0.274 0.08 0.274 0.10 0.275 0.12 0.277 0.14 0.280 0.16 0.284 0.18 0.291 0.20 0.301 0.22 0.314 0.24 0.331 0.26 0.355 0.28 0.385 0.30 0.424 0.32 0.473 0.34 0.531 0.36 0.599 0.38 0.673 0.40 0.751 0.42 0.827 0.44 0.896 0.46 0.951 0.48 0.987 0.50 1.000 0.52 0.988 0.54 0.953 0.56 0.898 0.58 0.829 0.60 0.750 0.62 0.670 0.64 0.592 0.66 0.521 0.68 0.459 0.70 0.406 0.72 0.363 0.74 0.328 0.76 0.301 0.78 0.280 0.80 0.263 0.82 0.250 0.84 0.240 0.86 0.232 0.88 0.225 0.90 0.220 0.92 0.215 0.94 0.211 0.96 0.208 0.98 0.205 1.00 0.202 /
\end{sparkline}
 & CS  &  6436.7\phantom{0} $\pm$ 1.5\phantom{0} & 6437.51 & $\pm$0.21  & $\pm$0.10  & $\pm$0.18  \\
Pr ---                    & L$\gamma_{10}$   & \begin{sparkline}{6}
    \spark 0.00 0.339 0.02 0.356 0.04 0.374 0.06 0.393 0.08 0.414 0.10 0.436 0.12 0.459 0.14 0.483 0.16 0.509 0.18 0.536 0.20 0.564 0.22 0.592 0.24 0.622 0.26 0.653 0.28 0.684 0.30 0.714 0.32 0.745 0.34 0.775 0.36 0.803 0.38 0.830 0.40 0.854 0.42 0.876 0.44 0.895 0.46 0.910 0.48 0.921 0.50 0.928 0.52 0.932 0.54 0.931 0.56 0.927 0.58 0.920 0.60 0.910 0.62 0.898 0.64 0.884 0.66 0.869 0.68 0.854 0.70 0.839 0.72 0.825 0.74 0.812 0.76 0.801 0.78 0.793 0.80 0.788 0.82 0.787 0.84 0.790 0.86 0.797 0.88 0.809 0.90 0.827 0.92 0.850 0.94 0.879 0.96 0.914 0.98 0.955 1.00 1.000 /
\end{sparkline}
 & CS  &  6577.2          & 6579.26 & $\pm$0.16  & $\pm$0.11  & $\pm$0.18  \\
Pr L$_1$N$_2$             & L$\gamma_2$      & \begin{sparkline}{6}
    \spark 0.00 0.273 0.02 0.268 0.04 0.263 0.06 0.259 0.08 0.255 0.10 0.252 0.12 0.250 0.14 0.248 0.16 0.248 0.18 0.250 0.20 0.253 0.22 0.257 0.24 0.263 0.26 0.271 0.28 0.281 0.30 0.293 0.32 0.306 0.34 0.321 0.36 0.337 0.38 0.353 0.40 0.369 0.42 0.383 0.44 0.395 0.46 0.405 0.48 0.410 0.50 0.412 0.52 0.410 0.54 0.403 0.56 0.394 0.58 0.382 0.60 0.369 0.62 0.355 0.64 0.343 0.66 0.332 0.68 0.325 0.70 0.321 0.72 0.323 0.74 0.332 0.76 0.348 0.78 0.374 0.80 0.412 0.82 0.463 0.84 0.528 0.86 0.606 0.88 0.694 0.90 0.785 0.92 0.872 0.94 0.942 0.96 0.987 0.98 1.000 1.00 0.977 /
\end{sparkline}
 &     &  6598.0\phantom{0} $\pm$ 2.1\phantom{0} & 6601.02 & $\pm$0.12  & $\pm$0.12  & $\pm$0.18  \\
Pr L$_1$N$_3$             & L$\gamma_3$      & \begin{sparkline}{6}
    \spark 0.00 0.409 0.02 0.412 0.04 0.411 0.06 0.405 0.08 0.397 0.10 0.385 0.12 0.372 0.14 0.359 0.16 0.346 0.18 0.335 0.20 0.326 0.22 0.322 0.24 0.322 0.26 0.329 0.28 0.343 0.30 0.366 0.32 0.401 0.34 0.449 0.36 0.510 0.38 0.586 0.40 0.672 0.42 0.763 0.44 0.852 0.46 0.927 0.48 0.979 0.50 1.000 0.52 0.985 0.54 0.937 0.56 0.862 0.58 0.768 0.60 0.667 0.62 0.568 0.64 0.477 0.66 0.398 0.68 0.332 0.70 0.279 0.72 0.236 0.74 0.202 0.76 0.175 0.78 0.154 0.80 0.137 0.82 0.122 0.84 0.111 0.86 0.101 0.88 0.092 0.90 0.085 0.92 0.079 0.94 0.073 0.96 0.068 0.98 0.064 1.00 0.060 /
\end{sparkline}
 &     &  6615.9\phantom{0} $\pm$ 2.1\phantom{0} & 6617.10 & $\pm$0.03  & $\pm$0.12  & $\pm$0.18  \\
Pr L$_1$N$_\mathrm{4,5}$  & L$\gamma_{11}$   & \begin{sparkline}{6}
    \spark 0.00 0.281 0.02 0.282 0.04 0.283 0.06 0.284 0.08 0.287 0.10 0.291 0.12 0.297 0.14 0.304 0.16 0.314 0.18 0.327 0.20 0.345 0.22 0.367 0.24 0.394 0.26 0.428 0.28 0.468 0.30 0.515 0.32 0.569 0.34 0.628 0.36 0.691 0.38 0.756 0.40 0.819 0.42 0.877 0.44 0.928 0.46 0.967 0.48 0.991 0.50 1.000 0.52 0.992 0.54 0.968 0.56 0.929 0.58 0.878 0.60 0.818 0.62 0.753 0.64 0.686 0.66 0.619 0.68 0.556 0.70 0.498 0.72 0.446 0.74 0.401 0.76 0.363 0.78 0.331 0.80 0.305 0.82 0.284 0.84 0.267 0.86 0.253 0.88 0.242 0.90 0.233 0.92 0.225 0.94 0.219 0.96 0.214 0.98 0.210 1.00 0.206 /
\end{sparkline}
 & CS  &  6718.7\phantom{0} $\pm$ 2.1\phantom{0} & 6718.30 & $\pm$0.34  & $\pm$0.12  & $\pm$0.18  \\
Pr L$_1$O$_\mathrm{2,3}$  & L$\gamma_{4,4'}$ & \begin{sparkline}{6}
    \spark 0.00 0.057 0.02 0.061 0.04 0.067 0.06 0.074 0.08 0.082 0.10 0.092 0.12 0.103 0.14 0.118 0.16 0.136 0.18 0.158 0.20 0.185 0.22 0.218 0.24 0.257 0.26 0.303 0.28 0.357 0.30 0.419 0.32 0.487 0.34 0.561 0.36 0.639 0.38 0.716 0.40 0.792 0.42 0.860 0.44 0.919 0.46 0.963 0.48 0.991 0.50 1.000 0.52 0.990 0.54 0.962 0.56 0.918 0.58 0.860 0.60 0.792 0.62 0.718 0.64 0.642 0.66 0.568 0.68 0.498 0.70 0.434 0.72 0.378 0.74 0.330 0.76 0.291 0.78 0.260 0.80 0.235 0.82 0.217 0.84 0.204 0.86 0.194 0.88 0.186 0.90 0.180 0.92 0.174 0.94 0.167 0.96 0.160 0.98 0.151 1.00 0.141 /
\end{sparkline}
 & CS  &  6815.0\phantom{0} $\pm$ 1.5\phantom{0} & 6814.04 & $\pm$0.07  & $\pm$0.12  & $\pm$0.18  \\
Pr L$_1$N$_\mathrm{6,7}$  &                  & \begin{sparkline}{6}
    \spark 0.00 0.923 0.02 0.966 0.04 0.992 0.06 1.000 0.08 0.988 0.10 0.959 0.12 0.912 0.14 0.853 0.16 0.784 0.18 0.710 0.20 0.634 0.22 0.560 0.24 0.490 0.26 0.427 0.28 0.372 0.30 0.325 0.32 0.287 0.34 0.256 0.36 0.233 0.38 0.215 0.40 0.202 0.42 0.193 0.44 0.185 0.46 0.179 0.48 0.173 0.50 0.166 0.52 0.159 0.54 0.150 0.56 0.140 0.58 0.129 0.60 0.118 0.62 0.107 0.64 0.096 0.66 0.085 0.68 0.075 0.70 0.067 0.72 0.059 0.74 0.052 0.76 0.047 0.78 0.042 0.80 0.038 0.82 0.035 0.84 0.032 0.86 0.030 0.88 0.028 0.90 0.027 0.92 0.025 0.94 0.024 0.96 0.023 0.98 0.022 1.00 0.021 /
\end{sparkline}
 & Zsc &  6830.4          & $\star$\ \emph{6829.2\phantom{0} }& $\pm$0.45  & $\pm$0.11  & $\pm$1.0\phantom{0}  \\
\end{tabular}

%% file: Tables/peak_data_Nd.tex
%
\begin{tabular}{lllllrlrr}
\multicolumn{2}{c}{Line name} & & & & Peak & \multicolumn{3}{c}{Uncertainty} \\
IUPAC & Siegbahn & Spark & &  Reference &  Estimate & Stat & Sys & Peak  \\
\hline
Nd L$_3$M$_1$             & L$\ell$          & \begin{sparkline}{6}
    \spark 0.00 0.330 0.02 0.349 0.04 0.369 0.06 0.391 0.08 0.415 0.10 0.440 0.12 0.466 0.14 0.495 0.16 0.525 0.18 0.556 0.20 0.590 0.22 0.624 0.24 0.660 0.26 0.696 0.28 0.733 0.30 0.770 0.32 0.806 0.34 0.841 0.36 0.874 0.38 0.905 0.40 0.932 0.42 0.956 0.44 0.975 0.46 0.989 0.48 0.997 0.50 1.000 0.52 0.997 0.54 0.988 0.56 0.974 0.58 0.955 0.60 0.932 0.62 0.905 0.64 0.874 0.66 0.841 0.68 0.806 0.70 0.770 0.72 0.734 0.74 0.698 0.76 0.662 0.78 0.627 0.80 0.593 0.82 0.560 0.84 0.529 0.86 0.500 0.88 0.472 0.90 0.446 0.92 0.422 0.94 0.399 0.96 0.378 0.98 0.358 1.00 0.340 /
\end{sparkline}
 &     & 4631.85 $\pm$ 0.05 & 4631.96 & $\pm$0.05  & $\pm$0.12  & $\pm$0.18  \\
Nd L$_2$M$_1$             & L$\eta$          & \begin{sparkline}{6}
    \spark 0.00 0.384 0.02 0.400 0.04 0.418 0.06 0.437 0.08 0.458 0.10 0.480 0.12 0.504 0.14 0.529 0.16 0.556 0.18 0.585 0.20 0.615 0.22 0.647 0.24 0.680 0.26 0.714 0.28 0.748 0.30 0.783 0.32 0.817 0.34 0.851 0.36 0.883 0.38 0.912 0.40 0.938 0.42 0.960 0.44 0.978 0.46 0.991 0.48 0.998 0.50 1.000 0.52 0.996 0.54 0.988 0.56 0.974 0.58 0.956 0.60 0.935 0.62 0.911 0.64 0.885 0.66 0.857 0.68 0.829 0.70 0.801 0.72 0.773 0.74 0.747 0.76 0.721 0.78 0.697 0.80 0.674 0.82 0.653 0.84 0.633 0.86 0.615 0.88 0.599 0.90 0.584 0.92 0.571 0.94 0.559 0.96 0.548 0.98 0.539 1.00 0.530 /
\end{sparkline}
 &     & 5145.25 $\pm$ 0.17 & 5145.04 & $\pm$0.11  & $\pm$0.12  & $\pm$0.18  \\
Nd L$_3$M$_4$             & L$\alpha_2$      & \begin{sparkline}{6}
    \spark 0.00 0.114 0.02 0.122 0.04 0.130 0.06 0.140 0.08 0.151 0.10 0.165 0.12 0.180 0.14 0.198 0.16 0.219 0.18 0.244 0.20 0.274 0.22 0.307 0.24 0.346 0.26 0.390 0.28 0.439 0.30 0.492 0.32 0.550 0.34 0.611 0.36 0.674 0.38 0.738 0.40 0.802 0.42 0.862 0.44 0.916 0.46 0.959 0.48 0.988 0.50 1.000 0.52 0.992 0.54 0.965 0.56 0.921 0.58 0.863 0.60 0.797 0.62 0.728 0.64 0.662 0.66 0.602 0.68 0.550 0.70 0.509 0.72 0.477 0.74 0.455 0.76 0.441 0.78 0.435 0.80 0.436 0.82 0.444 0.84 0.458 0.86 0.477 0.88 0.504 0.90 0.537 0.92 0.579 0.94 0.630 0.96 0.692 0.98 0.768 1.00 0.861 /
\end{sparkline}
 &     &  5207.7\phantom{0} $\pm$ 1.1\phantom{0} & 5207.30 & $\pm$0.02  & $\pm$0.13  & $\pm$0.18  \\
Nd L$_3$M$_5$             & L$\alpha_1$      & \begin{sparkline}{6}
    \spark 0.00 0.056 0.02 0.059 0.04 0.064 0.06 0.069 0.08 0.075 0.10 0.083 0.12 0.093 0.14 0.105 0.16 0.119 0.18 0.135 0.20 0.155 0.22 0.178 0.24 0.205 0.26 0.238 0.28 0.277 0.30 0.323 0.32 0.380 0.34 0.446 0.36 0.523 0.38 0.609 0.40 0.700 0.42 0.790 0.44 0.873 0.46 0.940 0.48 0.984 0.50 1.000 0.52 0.986 0.54 0.944 0.56 0.878 0.58 0.795 0.60 0.702 0.62 0.608 0.64 0.518 0.66 0.436 0.68 0.365 0.70 0.305 0.72 0.255 0.74 0.214 0.76 0.182 0.78 0.155 0.80 0.134 0.82 0.117 0.84 0.102 0.86 0.091 0.88 0.081 0.90 0.073 0.92 0.066 0.94 0.060 0.96 0.054 0.98 0.050 1.00 0.046 /
\end{sparkline}
 &     & 5230.24 $\pm$ 0.04 & 5229.93 & $\pm$0.004 & $\pm$0.12  & $\pm$0.18  \\
Nd L$_2$M$_4$+L$_1$M$_2$  & L$\beta_{1,4}$   & \begin{sparkline}{6}
    \spark 0.00 0.052 0.02 0.057 0.04 0.063 0.06 0.070 0.08 0.078 0.10 0.088 0.12 0.101 0.14 0.115 0.16 0.134 0.18 0.156 0.20 0.182 0.22 0.213 0.24 0.250 0.26 0.292 0.28 0.339 0.30 0.391 0.32 0.448 0.34 0.511 0.36 0.579 0.38 0.652 0.40 0.728 0.42 0.805 0.44 0.877 0.46 0.937 0.48 0.980 0.50 1.000 0.52 0.992 0.54 0.957 0.56 0.898 0.58 0.819 0.60 0.729 0.62 0.634 0.64 0.543 0.66 0.458 0.68 0.383 0.70 0.320 0.72 0.268 0.74 0.225 0.76 0.191 0.78 0.164 0.80 0.142 0.82 0.124 0.84 0.110 0.86 0.098 0.88 0.089 0.90 0.082 0.92 0.076 0.94 0.071 0.96 0.067 0.98 0.064 1.00 0.061 /
\end{sparkline}
 &     & 5721.45 $\pm$ 0.05 & 5720.96 & $\pm$0.004 & $\pm$0.12  & $\pm$0.18  \\
Nd ---                    & L$\beta$'        & \begin{sparkline}{6}
    \spark 0.00 1.000 0.02 0.881 0.04 0.760 0.06 0.646 0.08 0.542 0.10 0.453 0.12 0.378 0.14 0.317 0.16 0.267 0.18 0.227 0.20 0.196 0.22 0.170 0.24 0.150 0.26 0.133 0.28 0.120 0.30 0.109 0.32 0.101 0.34 0.094 0.36 0.088 0.38 0.083 0.40 0.080 0.42 0.076 0.44 0.074 0.46 0.071 0.48 0.068 0.50 0.066 0.52 0.063 0.54 0.059 0.56 0.056 0.58 0.052 0.60 0.049 0.62 0.045 0.64 0.042 0.66 0.038 0.68 0.036 0.70 0.033 0.72 0.031 0.74 0.028 0.76 0.027 0.78 0.025 0.80 0.024 0.82 0.022 0.84 0.021 0.86 0.020 0.88 0.019 0.90 0.018 0.92 0.017 0.94 0.017 0.96 0.016 0.98 0.015 1.00 0.015 /
\end{sparkline}
 & CS  &  5741.8          & $\star$\ \emph{5737.5\phantom{0} }& $\pm$0.03  & $\pm$0.12  & $\pm$1.0\phantom{0}  \\
Nd L$_1$M$_3$             & L$\beta_3$       & \begin{sparkline}{6}
    \spark 0.00 0.138 0.02 0.147 0.04 0.157 0.06 0.169 0.08 0.181 0.10 0.195 0.12 0.212 0.14 0.230 0.16 0.251 0.18 0.276 0.20 0.304 0.22 0.336 0.24 0.372 0.26 0.414 0.28 0.461 0.30 0.513 0.32 0.570 0.34 0.631 0.36 0.696 0.38 0.760 0.40 0.823 0.42 0.881 0.44 0.930 0.46 0.968 0.48 0.992 0.50 1.000 0.52 0.992 0.54 0.968 0.56 0.930 0.58 0.881 0.60 0.823 0.62 0.760 0.64 0.695 0.66 0.630 0.68 0.568 0.70 0.511 0.72 0.458 0.74 0.410 0.76 0.368 0.78 0.331 0.80 0.299 0.82 0.271 0.84 0.246 0.86 0.225 0.88 0.206 0.90 0.190 0.92 0.175 0.94 0.163 0.96 0.152 0.98 0.142 1.00 0.133 /
\end{sparkline}
 &     & 5827.80 $\pm$ 0.05 & 5827.89 & $\pm$0.02  & $\pm$0.11  & $\pm$0.18  \\
Nd L$_3$N$_1$             & L$\beta_6$       & \begin{sparkline}{6}
    \spark 0.00 0.223 0.02 0.229 0.04 0.236 0.06 0.244 0.08 0.253 0.10 0.265 0.12 0.278 0.14 0.293 0.16 0.311 0.18 0.332 0.20 0.356 0.22 0.384 0.24 0.417 0.26 0.454 0.28 0.497 0.30 0.545 0.32 0.597 0.34 0.654 0.36 0.713 0.38 0.773 0.40 0.832 0.42 0.886 0.44 0.933 0.46 0.969 0.48 0.992 0.50 1.000 0.52 0.993 0.54 0.970 0.56 0.935 0.58 0.888 0.60 0.832 0.62 0.772 0.64 0.710 0.66 0.648 0.68 0.588 0.70 0.533 0.72 0.482 0.74 0.436 0.76 0.395 0.78 0.359 0.80 0.328 0.82 0.301 0.84 0.277 0.86 0.256 0.88 0.237 0.90 0.221 0.92 0.207 0.94 0.195 0.96 0.184 0.98 0.174 1.00 0.165 /
\end{sparkline}
 &     & 5892.99 $\pm$ 0.25 & 5891.61 & $\pm$0.05  & $\pm$0.10  & $\pm$0.18  \\
Nd ---                    & L$\beta_{14}$    & \begin{sparkline}{6}
    \spark 0.00 0.041 0.02 0.043 0.04 0.045 0.06 0.047 0.08 0.049 0.10 0.052 0.12 0.054 0.14 0.057 0.16 0.060 0.18 0.063 0.20 0.067 0.22 0.070 0.24 0.074 0.26 0.078 0.28 0.082 0.30 0.086 0.32 0.091 0.34 0.095 0.36 0.099 0.38 0.104 0.40 0.108 0.42 0.113 0.44 0.117 0.46 0.121 0.48 0.125 0.50 0.128 0.52 0.132 0.54 0.136 0.56 0.140 0.58 0.143 0.60 0.148 0.62 0.152 0.64 0.158 0.66 0.164 0.68 0.171 0.70 0.180 0.72 0.190 0.74 0.203 0.76 0.218 0.78 0.237 0.80 0.260 0.82 0.288 0.84 0.323 0.86 0.365 0.88 0.416 0.90 0.477 0.92 0.550 0.94 0.637 0.96 0.740 0.98 0.860 1.00 1.000 /
\end{sparkline}
 & CS  &  6068.5          & $\star$\ \emph{6069.3\phantom{0} }& $\pm$0.20  & $\pm$0.12  & $\pm$1.0\phantom{0}  \\
Nd L$_3$N$_5$             & L$\beta_2$       & \begin{sparkline}{6}
    \spark 0.00 0.064 0.02 0.068 0.04 0.072 0.06 0.077 0.08 0.083 0.10 0.091 0.12 0.100 0.14 0.112 0.16 0.126 0.18 0.143 0.20 0.163 0.22 0.188 0.24 0.218 0.26 0.253 0.28 0.294 0.30 0.342 0.32 0.397 0.34 0.461 0.36 0.534 0.38 0.614 0.40 0.700 0.42 0.786 0.44 0.867 0.46 0.935 0.48 0.982 0.50 1.000 0.52 0.987 0.54 0.944 0.56 0.875 0.58 0.790 0.60 0.696 0.62 0.603 0.64 0.516 0.66 0.440 0.68 0.376 0.70 0.324 0.72 0.280 0.74 0.244 0.76 0.214 0.78 0.188 0.80 0.166 0.82 0.146 0.84 0.129 0.86 0.114 0.88 0.101 0.90 0.090 0.92 0.081 0.94 0.072 0.96 0.065 0.98 0.059 1.00 0.054 /
\end{sparkline}
 &     & 6091.25 $\pm$ 0.26 & 6090.42 & $\pm$0.009 & $\pm$0.12  & $\pm$0.18  \\
Nd L$_1$M$_4$             & L$\beta_{10}$    & \begin{sparkline}{6}
    \spark 0.00 1.000 0.02 0.939 0.04 0.883 0.06 0.833 0.08 0.787 0.10 0.746 0.12 0.708 0.14 0.674 0.16 0.642 0.18 0.613 0.20 0.586 0.22 0.561 0.24 0.539 0.26 0.518 0.28 0.499 0.30 0.482 0.32 0.467 0.34 0.455 0.36 0.447 0.38 0.443 0.40 0.445 0.42 0.452 0.44 0.463 0.46 0.473 0.48 0.478 0.50 0.475 0.52 0.460 0.54 0.437 0.56 0.409 0.58 0.380 0.60 0.354 0.62 0.333 0.64 0.317 0.66 0.306 0.68 0.297 0.70 0.290 0.72 0.285 0.74 0.281 0.76 0.278 0.78 0.275 0.80 0.273 0.82 0.272 0.84 0.272 0.86 0.272 0.88 0.273 0.90 0.276 0.92 0.280 0.94 0.285 0.96 0.293 0.98 0.303 1.00 0.317 /
\end{sparkline}
 &     & 6124.97 $\pm$ 0.41 & \emph{6125.45 }& $\pm$0.20  & $\pm$0.12  & $\pm$0.35  \\
Nd L$_1$M$_5$             & L$\beta_{9}$     & \begin{sparkline}{6}
    \spark 0.00 0.492 0.02 0.483 0.04 0.476 0.06 0.470 0.08 0.466 0.10 0.463 0.12 0.461 0.14 0.460 0.16 0.460 0.18 0.462 0.20 0.467 0.22 0.473 0.24 0.483 0.26 0.496 0.28 0.513 0.30 0.537 0.32 0.567 0.34 0.605 0.36 0.652 0.38 0.708 0.40 0.771 0.42 0.836 0.44 0.899 0.46 0.951 0.48 0.987 0.50 1.000 0.52 0.988 0.54 0.954 0.56 0.901 0.58 0.838 0.60 0.770 0.62 0.705 0.64 0.647 0.66 0.597 0.68 0.556 0.70 0.522 0.72 0.496 0.74 0.476 0.76 0.460 0.78 0.449 0.80 0.440 0.82 0.433 0.84 0.429 0.86 0.426 0.88 0.425 0.90 0.425 0.92 0.426 0.94 0.430 0.96 0.434 0.98 0.441 1.00 0.449 /
\end{sparkline}
 &     & 6148.82 $\pm$ 0.41 & 6146.97 & $\pm$0.12  & $\pm$0.12  & $\pm$0.18  \\
Nd L$_3$O$_1$             & L$\beta_7$       & \begin{sparkline}{6}
    \spark 0.00 0.667 0.02 0.647 0.04 0.632 0.06 0.621 0.08 0.613 0.10 0.608 0.12 0.605 0.14 0.604 0.16 0.605 0.18 0.609 0.20 0.614 0.22 0.622 0.24 0.633 0.26 0.646 0.28 0.663 0.30 0.684 0.32 0.710 0.34 0.740 0.36 0.775 0.38 0.814 0.40 0.855 0.42 0.897 0.44 0.936 0.46 0.969 0.48 0.991 0.50 1.000 0.52 0.994 0.54 0.975 0.56 0.943 0.58 0.902 0.60 0.855 0.62 0.807 0.64 0.760 0.66 0.716 0.68 0.676 0.70 0.641 0.72 0.611 0.74 0.584 0.76 0.561 0.78 0.541 0.80 0.523 0.82 0.507 0.84 0.493 0.86 0.480 0.88 0.469 0.90 0.458 0.92 0.448 0.94 0.438 0.96 0.430 0.98 0.421 1.00 0.414 /
\end{sparkline}
 & CS  &  6170.8\phantom{0} $\pm$ 0.9\phantom{0} & 6170.21 & $\pm$0.21  & $\pm$0.13  & $\pm$0.18  \\
Nd L$_3$N$_\mathrm{6,7}$  & Lu               & \begin{sparkline}{6}
    \spark 0.00 0.595 0.02 0.588 0.04 0.582 0.06 0.577 0.08 0.572 0.10 0.568 0.12 0.566 0.14 0.564 0.16 0.565 0.18 0.566 0.20 0.570 0.22 0.576 0.24 0.586 0.26 0.598 0.28 0.615 0.30 0.636 0.32 0.663 0.34 0.696 0.36 0.734 0.38 0.778 0.40 0.826 0.42 0.875 0.44 0.921 0.46 0.960 0.48 0.988 0.50 1.000 0.52 0.995 0.54 0.971 0.56 0.933 0.58 0.883 0.60 0.825 0.62 0.765 0.64 0.705 0.66 0.650 0.68 0.600 0.70 0.555 0.72 0.517 0.74 0.483 0.76 0.455 0.78 0.430 0.80 0.409 0.82 0.391 0.84 0.375 0.86 0.360 0.88 0.348 0.90 0.336 0.92 0.326 0.94 0.317 0.96 0.308 0.98 0.301 1.00 0.293 /
\end{sparkline}
 &     &  6202.3\phantom{0} $\pm$ 0.5\phantom{0} & 6206.82 & $\pm$0.24  & $\pm$0.12  & $\pm$0.18  \\
Nd L$_2$N$_1$             & L$\gamma_5$      & \begin{sparkline}{6}
    \spark 0.00 0.111 0.02 0.116 0.04 0.123 0.06 0.130 0.08 0.139 0.10 0.149 0.12 0.160 0.14 0.174 0.16 0.191 0.18 0.211 0.20 0.234 0.22 0.263 0.24 0.297 0.26 0.337 0.28 0.384 0.30 0.439 0.32 0.500 0.34 0.568 0.36 0.640 0.38 0.715 0.40 0.788 0.42 0.857 0.44 0.916 0.46 0.961 0.48 0.990 0.50 1.000 0.52 0.990 0.54 0.961 0.56 0.916 0.58 0.857 0.60 0.788 0.62 0.715 0.64 0.640 0.66 0.568 0.68 0.500 0.70 0.439 0.72 0.384 0.74 0.337 0.76 0.297 0.78 0.263 0.80 0.234 0.82 0.210 0.84 0.191 0.86 0.174 0.88 0.160 0.90 0.148 0.92 0.138 0.94 0.130 0.96 0.122 0.98 0.116 1.00 0.111 /
\end{sparkline}
 &     & 6405.29 $\pm$ 0.33 & 6404.97 & $\pm$0.10  & $\pm$0.31  & $\pm$0.18  \\
Nd ---                    & L$\gamma_9$      & \begin{sparkline}{6}
    \spark 0.00 0.046 0.02 0.049 0.04 0.052 0.06 0.056 0.08 0.059 0.10 0.063 0.12 0.068 0.14 0.073 0.16 0.078 0.18 0.084 0.20 0.090 0.22 0.097 0.24 0.104 0.26 0.112 0.28 0.120 0.30 0.128 0.32 0.136 0.34 0.145 0.36 0.154 0.38 0.162 0.40 0.170 0.42 0.177 0.44 0.183 0.46 0.189 0.48 0.193 0.50 0.197 0.52 0.199 0.54 0.200 0.56 0.201 0.58 0.201 0.60 0.200 0.62 0.199 0.64 0.199 0.66 0.199 0.68 0.200 0.70 0.202 0.72 0.205 0.74 0.210 0.76 0.218 0.78 0.229 0.80 0.244 0.82 0.263 0.84 0.290 0.86 0.326 0.88 0.374 0.90 0.436 0.92 0.515 0.94 0.613 0.96 0.730 0.98 0.861 1.00 1.000 /
\end{sparkline}
 & CS  &  6544.9          & \emph{6580.7\phantom{0} }& $\pm$0.09  & $\pm$0.11  & $\pm$0.5\phantom{0}  \\
Nd L$_2$N$_4$             & L$\gamma_1$      & \begin{sparkline}{6}
    \spark 0.00 0.140 0.02 0.139 0.04 0.139 0.06 0.139 0.08 0.139 0.10 0.140 0.12 0.142 0.14 0.146 0.16 0.151 0.18 0.158 0.20 0.167 0.22 0.180 0.24 0.198 0.26 0.221 0.28 0.252 0.30 0.294 0.32 0.346 0.34 0.412 0.36 0.491 0.38 0.581 0.40 0.678 0.42 0.775 0.44 0.864 0.46 0.936 0.48 0.984 0.50 1.000 0.52 0.984 0.54 0.937 0.56 0.864 0.58 0.775 0.60 0.678 0.62 0.581 0.64 0.491 0.66 0.412 0.68 0.345 0.70 0.291 0.72 0.247 0.74 0.213 0.76 0.185 0.78 0.161 0.80 0.141 0.82 0.124 0.84 0.109 0.86 0.096 0.88 0.084 0.90 0.074 0.92 0.066 0.94 0.058 0.96 0.052 0.98 0.047 1.00 0.042 /
\end{sparkline}
 &     & 6601.16 $\pm$ 0.24 & 6600.36 & $\pm$0.01  & $\pm$0.12  & $\pm$0.18  \\
Nd L$_2$O$_1$             & L$\gamma_8$      & \begin{sparkline}{6}
    \spark 0.00 0.545 0.02 0.542 0.04 0.541 0.06 0.542 0.08 0.545 0.10 0.549 0.12 0.556 0.14 0.565 0.16 0.576 0.18 0.590 0.20 0.607 0.22 0.626 0.24 0.649 0.26 0.674 0.28 0.702 0.30 0.734 0.32 0.767 0.34 0.802 0.36 0.837 0.38 0.873 0.40 0.906 0.42 0.937 0.44 0.963 0.46 0.982 0.48 0.995 0.50 1.000 0.52 0.996 0.54 0.985 0.56 0.965 0.58 0.938 0.60 0.906 0.62 0.869 0.64 0.829 0.66 0.787 0.68 0.746 0.70 0.705 0.72 0.665 0.74 0.628 0.76 0.594 0.78 0.562 0.80 0.533 0.82 0.507 0.84 0.484 0.86 0.463 0.88 0.444 0.90 0.427 0.92 0.412 0.94 0.399 0.96 0.387 0.98 0.376 1.00 0.367 /
\end{sparkline}
 & CS  &  6683.0\phantom{0} $\pm$ 1.8\phantom{0} & 6682.53 & $\pm$0.39  & $\pm$0.13  & $\pm$0.18  \\
Nd L$_2$N$_\mathrm{6,7}$  & Lv               & \begin{sparkline}{6}
    \spark 0.00 0.257 0.02 0.256 0.04 0.255 0.06 0.255 0.08 0.256 0.10 0.258 0.12 0.261 0.14 0.265 0.16 0.272 0.18 0.280 0.20 0.291 0.22 0.306 0.24 0.326 0.26 0.352 0.28 0.384 0.30 0.426 0.32 0.476 0.34 0.536 0.36 0.604 0.38 0.678 0.40 0.756 0.42 0.831 0.44 0.899 0.46 0.953 0.48 0.988 0.50 1.000 0.52 0.988 0.54 0.954 0.56 0.900 0.58 0.832 0.60 0.755 0.62 0.676 0.64 0.598 0.66 0.527 0.68 0.464 0.70 0.411 0.72 0.366 0.74 0.330 0.76 0.301 0.78 0.278 0.80 0.260 0.82 0.246 0.84 0.235 0.86 0.225 0.88 0.218 0.90 0.212 0.92 0.207 0.94 0.202 0.96 0.198 0.98 0.195 1.00 0.192 /
\end{sparkline}
 &     &  6719.0\phantom{0} $\pm$ 0.6\phantom{0} & 6721.32 & $\pm$0.20  & $\pm$0.12  & $\pm$0.18  \\
Nd ---                    & L$\gamma_{10}$   & \begin{sparkline}{6}
    \spark 0.00 0.330 0.02 0.345 0.04 0.361 0.06 0.378 0.08 0.396 0.10 0.415 0.12 0.435 0.14 0.455 0.16 0.477 0.18 0.499 0.20 0.522 0.22 0.545 0.24 0.569 0.26 0.593 0.28 0.617 0.30 0.641 0.32 0.664 0.34 0.687 0.36 0.708 0.38 0.728 0.40 0.747 0.42 0.763 0.44 0.777 0.46 0.788 0.48 0.797 0.50 0.803 0.52 0.806 0.54 0.807 0.56 0.805 0.58 0.801 0.60 0.795 0.62 0.788 0.64 0.779 0.66 0.770 0.68 0.761 0.70 0.753 0.72 0.745 0.74 0.739 0.76 0.735 0.78 0.734 0.80 0.736 0.82 0.742 0.84 0.753 0.86 0.768 0.88 0.789 0.90 0.815 0.92 0.846 0.94 0.882 0.96 0.921 0.98 0.961 1.00 1.000 /
\end{sparkline}
 & CS  &  6865.2          & 6862.90 & $\pm$0.20  & $\pm$0.11  & $\pm$0.18  \\
Nd L$_1$N$_2$             & L$\gamma_2$      & \begin{sparkline}{6}
    \spark 0.00 0.286 0.02 0.284 0.04 0.281 0.06 0.278 0.08 0.275 0.10 0.272 0.12 0.268 0.14 0.266 0.16 0.263 0.18 0.262 0.20 0.261 0.22 0.262 0.24 0.264 0.26 0.268 0.28 0.273 0.30 0.280 0.32 0.289 0.34 0.300 0.36 0.313 0.38 0.327 0.40 0.341 0.42 0.355 0.44 0.368 0.46 0.378 0.48 0.384 0.50 0.386 0.52 0.384 0.54 0.378 0.56 0.368 0.58 0.355 0.60 0.341 0.62 0.326 0.64 0.312 0.66 0.300 0.68 0.291 0.70 0.285 0.72 0.285 0.74 0.289 0.76 0.301 0.78 0.320 0.80 0.349 0.82 0.389 0.84 0.443 0.86 0.511 0.88 0.592 0.90 0.684 0.92 0.779 0.94 0.869 0.96 0.942 0.98 0.988 1.00 1.000 /
\end{sparkline}
 &     & 6884.03 $\pm$ 0.34 & 6882.82 & $\pm$0.14  & $\pm$0.10  & $\pm$0.18  \\
Nd L$_1$N$_3$             & L$\gamma_3$      & \begin{sparkline}{6}
    \spark 0.00 0.386 0.02 0.385 0.04 0.380 0.06 0.371 0.08 0.359 0.10 0.345 0.12 0.330 0.14 0.315 0.16 0.303 0.18 0.293 0.20 0.286 0.22 0.284 0.24 0.287 0.26 0.297 0.28 0.314 0.30 0.340 0.32 0.378 0.34 0.428 0.36 0.492 0.38 0.571 0.40 0.660 0.42 0.755 0.44 0.848 0.46 0.926 0.48 0.980 0.50 1.000 0.52 0.984 0.54 0.933 0.56 0.854 0.58 0.758 0.60 0.655 0.62 0.556 0.64 0.466 0.66 0.389 0.68 0.325 0.70 0.274 0.72 0.233 0.74 0.201 0.76 0.176 0.78 0.155 0.80 0.139 0.82 0.125 0.84 0.114 0.86 0.105 0.88 0.097 0.90 0.090 0.92 0.084 0.94 0.079 0.96 0.074 0.98 0.070 1.00 0.067 /
\end{sparkline}
 &     & 6900.44 $\pm$ 0.34 & 6899.93 & $\pm$0.04  & $\pm$0.10  & $\pm$0.18  \\
Nd L$_1$N$_\mathrm{4,5}$  & L$\gamma_{11}$   & \begin{sparkline}{6}
    \spark 0.00 0.458 0.02 0.452 0.04 0.447 0.06 0.442 0.08 0.438 0.10 0.434 0.12 0.430 0.14 0.427 0.16 0.426 0.18 0.426 0.20 0.429 0.22 0.434 0.24 0.444 0.26 0.460 0.28 0.483 0.30 0.514 0.32 0.553 0.34 0.602 0.36 0.659 0.38 0.722 0.40 0.788 0.42 0.853 0.44 0.911 0.46 0.958 0.48 0.989 0.50 1.000 0.52 0.991 0.54 0.961 0.56 0.915 0.58 0.854 0.60 0.786 0.62 0.713 0.64 0.642 0.66 0.576 0.68 0.518 0.70 0.468 0.72 0.427 0.74 0.395 0.76 0.370 0.78 0.352 0.80 0.338 0.82 0.328 0.84 0.320 0.86 0.314 0.88 0.310 0.90 0.306 0.92 0.304 0.94 0.301 0.96 0.299 0.98 0.297 1.00 0.296 /
\end{sparkline}
 &     & 7007.74 $\pm$ 0.36 & 7007.12 & $\pm$0.34  & $\pm$0.11  & $\pm$0.18  \\
Nd L$_1$O$_\mathrm{2,3}$  & L$\gamma_{4,4'}$ & \begin{sparkline}{6}
    \spark 0.00 0.065 0.02 0.070 0.04 0.075 0.06 0.081 0.08 0.088 0.10 0.097 0.12 0.107 0.14 0.120 0.16 0.136 0.18 0.155 0.20 0.179 0.22 0.209 0.24 0.244 0.26 0.287 0.28 0.338 0.30 0.398 0.32 0.464 0.34 0.538 0.36 0.617 0.38 0.697 0.40 0.776 0.42 0.849 0.44 0.911 0.46 0.960 0.48 0.990 0.50 1.000 0.52 0.989 0.54 0.959 0.56 0.911 0.58 0.849 0.60 0.776 0.62 0.699 0.64 0.620 0.66 0.544 0.68 0.473 0.70 0.410 0.72 0.356 0.74 0.311 0.76 0.274 0.78 0.245 0.80 0.223 0.82 0.206 0.84 0.193 0.86 0.182 0.88 0.173 0.90 0.164 0.92 0.155 0.94 0.146 0.96 0.136 0.98 0.125 1.00 0.114 /
\end{sparkline}
 & CS  &  7107.0\phantom{0} $\pm$ 1.6\phantom{0} & 7106.86 & $\pm$0.08  & $\pm$0.11  & $\pm$0.18  \\
Nd L$_1$N$_\mathrm{6,7}$  &                  & \begin{sparkline}{6}
    \spark 0.00 0.807 0.02 0.877 0.04 0.934 0.06 0.976 0.08 0.998 0.10 1.000 0.12 0.981 0.14 0.943 0.16 0.888 0.18 0.821 0.20 0.746 0.22 0.667 0.24 0.589 0.26 0.514 0.28 0.446 0.30 0.387 0.32 0.336 0.34 0.295 0.36 0.261 0.38 0.235 0.40 0.215 0.42 0.200 0.44 0.188 0.46 0.178 0.48 0.169 0.50 0.161 0.52 0.151 0.54 0.142 0.56 0.131 0.58 0.120 0.60 0.109 0.62 0.098 0.64 0.088 0.66 0.078 0.68 0.069 0.70 0.062 0.72 0.056 0.74 0.050 0.76 0.046 0.78 0.042 0.80 0.039 0.82 0.037 0.84 0.035 0.86 0.033 0.88 0.031 0.90 0.030 0.92 0.029 0.94 0.028 0.96 0.027 0.98 0.026 1.00 0.025 /
\end{sparkline}
 &     &  7122.1\phantom{0} $\pm$ 2.0\phantom{0} & $\star$\ \emph{7121.4\phantom{0} }& $\pm$0.6\phantom{0}  & $\pm$0.11  & $\pm$1.0\phantom{0}  \\
\end{tabular}

%% file: Tables/peak_data_Tb.tex
%
\begin{tabular}{lllllrlrr}
\multicolumn{2}{c}{Line name} & & & & Peak & \multicolumn{3}{c}{Uncertainty} \\
IUPAC & Siegbahn & Spark & &  Reference &  Estimate & Stat & Sys & Peak  \\
\hline
Tb L$_3$M$_1$             & L$\ell$          & \begin{sparkline}{6}
    \spark 0.00 0.314 0.02 0.334 0.04 0.356 0.06 0.379 0.08 0.404 0.10 0.431 0.12 0.460 0.14 0.490 0.16 0.523 0.18 0.556 0.20 0.591 0.22 0.627 0.24 0.664 0.26 0.702 0.28 0.739 0.30 0.777 0.32 0.813 0.34 0.847 0.36 0.880 0.38 0.909 0.40 0.936 0.42 0.958 0.44 0.976 0.46 0.989 0.48 0.997 0.50 1.000 0.52 0.997 0.54 0.989 0.56 0.976 0.58 0.958 0.60 0.936 0.62 0.909 0.64 0.880 0.66 0.847 0.68 0.813 0.70 0.777 0.72 0.740 0.74 0.702 0.76 0.665 0.78 0.628 0.80 0.592 0.82 0.557 0.84 0.523 0.86 0.491 0.88 0.461 0.90 0.432 0.92 0.405 0.94 0.380 0.96 0.357 0.98 0.335 1.00 0.315 /
\end{sparkline}
 &     &  5546.8\phantom{0} $\pm$ 0.7\phantom{0} & 5551.45 & $\pm$0.04  & $\pm$0.12  & $\pm$0.18  \\
Tb L$_3$M$_4$             & L$\alpha_2$      & \begin{sparkline}{6}
    \spark 0.00 0.095 0.02 0.102 0.04 0.109 0.06 0.117 0.08 0.126 0.10 0.137 0.12 0.151 0.14 0.166 0.16 0.184 0.18 0.205 0.20 0.230 0.22 0.258 0.24 0.289 0.26 0.325 0.28 0.365 0.30 0.410 0.32 0.460 0.34 0.518 0.36 0.582 0.38 0.654 0.40 0.730 0.42 0.808 0.44 0.880 0.46 0.941 0.48 0.983 0.50 1.000 0.52 0.989 0.54 0.951 0.56 0.890 0.58 0.812 0.60 0.726 0.62 0.639 0.64 0.558 0.66 0.486 0.68 0.425 0.70 0.376 0.72 0.338 0.74 0.309 0.76 0.287 0.78 0.272 0.80 0.262 0.82 0.257 0.84 0.254 0.86 0.255 0.88 0.258 0.90 0.264 0.92 0.272 0.94 0.282 0.96 0.295 0.98 0.311 1.00 0.330 /
\end{sparkline}
 &     &  6238.1\phantom{0} $\pm$ 0.9\phantom{0} & 6239.51 & $\pm$0.01  & $\pm$0.12  & $\pm$0.18  \\
Tb L$_3$M$_5$             & L$\alpha_1$      & \begin{sparkline}{6}
    \spark 0.00 0.069 0.02 0.077 0.04 0.086 0.06 0.098 0.08 0.111 0.10 0.127 0.12 0.145 0.14 0.164 0.16 0.186 0.18 0.209 0.20 0.233 0.22 0.258 0.24 0.285 0.26 0.316 0.28 0.350 0.30 0.391 0.32 0.439 0.34 0.495 0.36 0.560 0.38 0.634 0.40 0.713 0.42 0.794 0.44 0.871 0.46 0.935 0.48 0.980 0.50 1.000 0.52 0.990 0.54 0.951 0.56 0.887 0.58 0.803 0.60 0.709 0.62 0.611 0.64 0.518 0.66 0.433 0.68 0.360 0.70 0.299 0.72 0.248 0.74 0.208 0.76 0.176 0.78 0.151 0.80 0.130 0.82 0.114 0.84 0.101 0.86 0.090 0.88 0.081 0.90 0.074 0.92 0.067 0.94 0.062 0.96 0.057 0.98 0.053 1.00 0.050 /
\end{sparkline}
 &     &  6272.8\phantom{0} $\pm$ 0.9\phantom{0} & 6274.35 & $\pm$0.004 & $\pm$0.12  & $\pm$0.18  \\
Tb L$_2$M$_1$             & L$\eta$          & \begin{sparkline}{6}
    \spark 0.00 1.000 0.02 0.953 0.04 0.882 0.06 0.793 0.08 0.695 0.10 0.596 0.12 0.503 0.14 0.420 0.16 0.348 0.18 0.289 0.20 0.240 0.22 0.202 0.24 0.171 0.26 0.147 0.28 0.127 0.30 0.112 0.32 0.099 0.34 0.089 0.36 0.080 0.38 0.073 0.40 0.067 0.42 0.062 0.44 0.057 0.46 0.053 0.48 0.049 0.50 0.046 0.52 0.043 0.54 0.041 0.56 0.038 0.58 0.036 0.60 0.034 0.62 0.032 0.64 0.030 0.66 0.028 0.68 0.026 0.70 0.025 0.72 0.023 0.74 0.022 0.76 0.021 0.78 0.020 0.80 0.019 0.82 0.018 0.84 0.017 0.86 0.016 0.88 0.015 0.90 0.015 0.92 0.014 0.94 0.013 0.96 0.013 0.98 0.012 1.00 0.012 /
\end{sparkline}
 &     &  6283.9\phantom{0} $\pm$ 0.9\phantom{0} & $\star$\ \emph{6290.5\phantom{0} }& $\pm$0.24  & $\pm$0.12  & $\pm$2.0\phantom{0}  \\
Tb L$_1$M$_2$             & L$\beta_4$       & \begin{sparkline}{6}
    \spark 0.00 0.187 0.02 0.197 0.04 0.208 0.06 0.220 0.08 0.233 0.10 0.248 0.12 0.264 0.14 0.283 0.16 0.304 0.18 0.328 0.20 0.356 0.22 0.387 0.24 0.422 0.26 0.462 0.28 0.506 0.30 0.556 0.32 0.609 0.34 0.667 0.36 0.727 0.38 0.787 0.40 0.846 0.42 0.898 0.44 0.943 0.46 0.976 0.48 0.995 0.50 1.000 0.52 0.991 0.54 0.968 0.56 0.935 0.58 0.895 0.60 0.851 0.62 0.805 0.64 0.761 0.66 0.721 0.68 0.685 0.70 0.655 0.72 0.631 0.74 0.613 0.76 0.600 0.78 0.593 0.80 0.592 0.82 0.595 0.84 0.604 0.86 0.618 0.88 0.637 0.90 0.662 0.92 0.692 0.94 0.729 0.96 0.773 0.98 0.825 1.00 0.886 /
\end{sparkline}
 &     &  6940.3\phantom{0} $\pm$ 1.1\phantom{0} & 6942.08 & $\pm$0.04  & $\pm$0.10  & $\pm$0.18  \\
Tb L$_2$M$_4$             & L$\beta_1$       & \begin{sparkline}{6}
    \spark 0.00 0.048 0.02 0.052 0.04 0.056 0.06 0.061 0.08 0.067 0.10 0.075 0.12 0.084 0.14 0.094 0.16 0.107 0.18 0.123 0.20 0.142 0.22 0.165 0.24 0.192 0.26 0.225 0.28 0.263 0.30 0.308 0.32 0.360 0.34 0.418 0.36 0.486 0.38 0.562 0.40 0.648 0.42 0.740 0.44 0.832 0.46 0.913 0.48 0.973 0.50 1.000 0.52 0.987 0.54 0.933 0.56 0.847 0.58 0.740 0.60 0.627 0.62 0.517 0.64 0.420 0.66 0.339 0.68 0.274 0.70 0.223 0.72 0.184 0.74 0.154 0.76 0.130 0.78 0.112 0.80 0.097 0.82 0.085 0.84 0.076 0.86 0.068 0.88 0.061 0.90 0.055 0.92 0.051 0.94 0.047 0.96 0.043 0.98 0.041 1.00 0.038 /
\end{sparkline}
 &     &  6977.8\phantom{0} $\pm$ 1.7\phantom{0} & 6977.16 & $\pm$0.006 & $\pm$0.10  & $\pm$0.18  \\
Tb ---                    & L$\beta$'        & \begin{sparkline}{6}
    \spark 0.00 1.000 0.02 0.859 0.04 0.746 0.06 0.655 0.08 0.581 0.10 0.520 0.12 0.469 0.14 0.427 0.16 0.391 0.18 0.360 0.20 0.334 0.22 0.313 0.24 0.295 0.26 0.280 0.28 0.268 0.30 0.259 0.32 0.253 0.34 0.250 0.36 0.249 0.38 0.250 0.40 0.253 0.42 0.256 0.44 0.259 0.46 0.260 0.48 0.259 0.50 0.255 0.52 0.247 0.54 0.236 0.56 0.223 0.58 0.208 0.60 0.192 0.62 0.176 0.64 0.162 0.66 0.148 0.68 0.135 0.70 0.125 0.72 0.115 0.74 0.107 0.76 0.100 0.78 0.093 0.80 0.088 0.82 0.083 0.84 0.079 0.86 0.075 0.88 0.072 0.90 0.069 0.92 0.066 0.94 0.063 0.96 0.061 0.98 0.058 1.00 0.056 /
\end{sparkline}
 & CS  &  7002.7          & \emph{7003.41 }& $\pm$0.10  & $\pm$0.10  & $\pm$0.35  \\
Tb L$_1$M$_3$             & L$\beta_3$       & \begin{sparkline}{6}
    \spark 0.00 0.143 0.02 0.153 0.04 0.165 0.06 0.178 0.08 0.193 0.10 0.210 0.12 0.230 0.14 0.253 0.16 0.279 0.18 0.309 0.20 0.342 0.22 0.380 0.24 0.423 0.26 0.469 0.28 0.520 0.30 0.575 0.32 0.633 0.34 0.692 0.36 0.751 0.38 0.808 0.40 0.861 0.42 0.908 0.44 0.947 0.46 0.976 0.48 0.994 0.50 1.000 0.52 0.994 0.54 0.975 0.56 0.946 0.58 0.908 0.60 0.862 0.62 0.811 0.64 0.757 0.66 0.702 0.68 0.648 0.70 0.596 0.72 0.549 0.74 0.506 0.76 0.468 0.78 0.436 0.80 0.409 0.82 0.387 0.84 0.371 0.86 0.359 0.88 0.351 0.90 0.346 0.92 0.344 0.94 0.344 0.96 0.344 0.98 0.345 1.00 0.346 /
\end{sparkline}
 &     &  7096.1\phantom{0} $\pm$ 1.2\phantom{0} & 7097.08 & $\pm$0.04  & $\pm$0.11  & $\pm$0.18  \\
Tb L$_3$N$_1$             & L$\beta_6$       & \begin{sparkline}{6}
    \spark 0.00 1.000 0.02 0.958 0.04 0.909 0.06 0.854 0.08 0.796 0.10 0.738 0.12 0.681 0.14 0.626 0.16 0.576 0.18 0.531 0.20 0.492 0.22 0.458 0.24 0.430 0.26 0.408 0.28 0.391 0.30 0.379 0.32 0.371 0.34 0.367 0.36 0.365 0.38 0.365 0.40 0.366 0.42 0.367 0.44 0.367 0.46 0.365 0.48 0.362 0.50 0.356 0.52 0.348 0.54 0.337 0.56 0.324 0.58 0.308 0.60 0.291 0.62 0.273 0.64 0.254 0.66 0.235 0.68 0.216 0.70 0.199 0.72 0.182 0.74 0.166 0.76 0.152 0.78 0.139 0.80 0.128 0.82 0.118 0.84 0.109 0.86 0.101 0.88 0.094 0.90 0.087 0.92 0.082 0.94 0.077 0.96 0.072 0.98 0.069 1.00 0.065 /
\end{sparkline}
 &     &  7116.4\phantom{0} $\pm$ 1.2\phantom{0} & $\star$\ \emph{7117.1\phantom{0} }& $\pm$0.12  & $\pm$0.11  & $\pm$1.0\phantom{0}  \\
Tb ---                    & L$\beta_{14}$    & \begin{sparkline}{6}
    \spark 0.00 0.082 0.02 0.087 0.04 0.093 0.06 0.099 0.08 0.106 0.10 0.114 0.12 0.123 0.14 0.133 0.16 0.145 0.18 0.158 0.20 0.173 0.22 0.190 0.24 0.209 0.26 0.231 0.28 0.256 0.30 0.284 0.32 0.315 0.34 0.348 0.36 0.383 0.38 0.420 0.40 0.456 0.42 0.490 0.44 0.521 0.46 0.546 0.48 0.564 0.50 0.575 0.52 0.578 0.54 0.574 0.56 0.563 0.58 0.547 0.60 0.528 0.62 0.509 0.64 0.490 0.66 0.474 0.68 0.462 0.70 0.454 0.72 0.452 0.74 0.455 0.76 0.463 0.78 0.477 0.80 0.496 0.82 0.520 0.84 0.548 0.86 0.581 0.88 0.618 0.90 0.660 0.92 0.707 0.94 0.760 0.96 0.824 0.98 0.902 1.00 1.000 /
\end{sparkline}
 & CS  &  7339.9          & 7339.71 & $\pm$0.07  & $\pm$0.14  & $\pm$0.18  \\
Tb L$_3$N$_\mathrm{4,5}$  & L$\beta_{2,15}$  & \begin{sparkline}{6}
    \spark 0.00 0.130 0.02 0.130 0.04 0.133 0.06 0.137 0.08 0.142 0.10 0.149 0.12 0.157 0.14 0.166 0.16 0.177 0.18 0.189 0.20 0.202 0.22 0.217 0.24 0.235 0.26 0.257 0.28 0.285 0.30 0.321 0.32 0.366 0.34 0.423 0.36 0.493 0.38 0.576 0.40 0.667 0.42 0.763 0.44 0.853 0.46 0.929 0.48 0.981 0.50 1.000 0.52 0.984 0.54 0.935 0.56 0.858 0.58 0.763 0.60 0.660 0.62 0.558 0.64 0.463 0.66 0.381 0.68 0.312 0.70 0.255 0.72 0.211 0.74 0.176 0.76 0.149 0.78 0.127 0.80 0.110 0.82 0.096 0.84 0.085 0.86 0.076 0.88 0.068 0.90 0.062 0.92 0.056 0.94 0.051 0.96 0.047 0.98 0.044 1.00 0.040 /
\end{sparkline}
 &     &  7366.7\phantom{0} $\pm$ 1.3\phantom{0} & 7366.24 & $\pm$0.02  & $\pm$0.14  & $\pm$0.18  \\
Tb L$_1$M$_4$             & L$\beta_{10}$    & \begin{sparkline}{6}
    \spark 0.00 0.769 0.02 0.755 0.04 0.742 0.06 0.731 0.08 0.721 0.10 0.712 0.12 0.705 0.14 0.699 0.16 0.696 0.18 0.694 0.20 0.695 0.22 0.698 0.24 0.704 0.26 0.714 0.28 0.727 0.30 0.744 0.32 0.766 0.34 0.792 0.36 0.823 0.38 0.857 0.40 0.892 0.42 0.927 0.44 0.958 0.46 0.983 0.48 0.997 0.50 1.000 0.52 0.990 0.54 0.968 0.56 0.937 0.58 0.898 0.60 0.856 0.62 0.814 0.64 0.773 0.66 0.735 0.68 0.702 0.70 0.674 0.72 0.650 0.74 0.630 0.76 0.614 0.78 0.602 0.80 0.593 0.82 0.587 0.84 0.584 0.86 0.582 0.88 0.583 0.90 0.586 0.92 0.591 0.94 0.598 0.96 0.607 0.98 0.617 1.00 0.630 /
\end{sparkline}
 &     &  7436.1\phantom{0} $\pm$ 2.0\phantom{0} & \emph{7431.0\phantom{0} }& $\pm$0.7\phantom{0}  & $\pm$0.14  & $\pm$0.35  \\
Tb L$_1$M$_5$+L$_3$O$_1$  & L$\beta_{9,7}$   & \begin{sparkline}{6}
    \spark 0.00 0.323 0.02 0.329 0.04 0.337 0.06 0.345 0.08 0.355 0.10 0.365 0.12 0.377 0.14 0.390 0.16 0.405 0.18 0.421 0.20 0.438 0.22 0.458 0.24 0.479 0.26 0.504 0.28 0.531 0.30 0.563 0.32 0.599 0.34 0.641 0.36 0.688 0.38 0.740 0.40 0.797 0.42 0.854 0.44 0.909 0.46 0.955 0.48 0.987 0.50 1.000 0.52 0.992 0.54 0.963 0.56 0.915 0.58 0.854 0.60 0.785 0.62 0.714 0.64 0.646 0.66 0.582 0.68 0.526 0.70 0.477 0.72 0.434 0.74 0.398 0.76 0.367 0.78 0.341 0.80 0.319 0.82 0.299 0.84 0.283 0.86 0.268 0.88 0.255 0.90 0.244 0.92 0.234 0.94 0.225 0.96 0.218 0.98 0.211 1.00 0.204 /
\end{sparkline}
 & CS  &  7469.5\phantom{0} $\pm$ 2.0\phantom{0} & 7467.59 & $\pm$0.20  & $\pm$0.14  & $\pm$0.18  \\
Tb L$_3$N$_\mathrm{6,7}$  & Lu               & \begin{sparkline}{6}
    \spark 0.00 0.410 0.02 0.407 0.04 0.404 0.06 0.402 0.08 0.401 0.10 0.401 0.12 0.401 0.14 0.402 0.16 0.403 0.18 0.406 0.20 0.410 0.22 0.415 0.24 0.421 0.26 0.430 0.28 0.442 0.30 0.457 0.32 0.479 0.34 0.509 0.36 0.550 0.38 0.605 0.40 0.675 0.42 0.758 0.44 0.845 0.46 0.924 0.48 0.980 0.50 1.000 0.52 0.980 0.54 0.924 0.56 0.845 0.58 0.758 0.60 0.676 0.62 0.605 0.64 0.549 0.66 0.507 0.68 0.475 0.70 0.451 0.72 0.433 0.74 0.418 0.76 0.405 0.78 0.394 0.80 0.384 0.82 0.374 0.84 0.365 0.86 0.356 0.88 0.348 0.90 0.340 0.92 0.332 0.94 0.324 0.96 0.316 0.98 0.309 1.00 0.302 /
\end{sparkline}
 & Th  &  7504.7\phantom{0} $\pm$ 1.7\phantom{0} & 7512.35 & $\pm$0.31  & $\pm$0.15  & $\pm$0.18  \\
Tb L$_2$N$_1$             & L$\gamma_5$      & \begin{sparkline}{6}
    \spark 0.00 0.111 0.02 0.118 0.04 0.125 0.06 0.134 0.08 0.145 0.10 0.159 0.12 0.177 0.14 0.202 0.16 0.234 0.18 0.276 0.20 0.328 0.22 0.390 0.24 0.462 0.26 0.539 0.28 0.618 0.30 0.693 0.32 0.759 0.34 0.815 0.36 0.858 0.38 0.891 0.40 0.916 0.42 0.937 0.44 0.956 0.46 0.973 0.48 0.988 0.50 0.998 0.52 1.000 0.54 0.990 0.56 0.967 0.58 0.930 0.60 0.881 0.62 0.822 0.64 0.755 0.66 0.686 0.68 0.617 0.70 0.550 0.72 0.488 0.74 0.432 0.76 0.383 0.78 0.340 0.80 0.303 0.82 0.272 0.84 0.246 0.86 0.223 0.88 0.204 0.90 0.188 0.92 0.175 0.94 0.163 0.96 0.153 0.98 0.144 1.00 0.136 /
\end{sparkline}
 &     &  7853.4\phantom{0} $\pm$ 1.5\phantom{0} & 7856.84 & $\pm$0.14  & $\pm$0.20  & $\pm$0.18  \\
Tb ---                    & L$\gamma_9$      & \begin{sparkline}{6}
    \spark 0.00 0.050 0.02 0.053 0.04 0.056 0.06 0.059 0.08 0.063 0.10 0.068 0.12 0.072 0.14 0.078 0.16 0.085 0.18 0.093 0.20 0.102 0.22 0.114 0.24 0.127 0.26 0.143 0.28 0.161 0.30 0.183 0.32 0.208 0.34 0.235 0.36 0.265 0.38 0.296 0.40 0.328 0.42 0.358 0.44 0.385 0.46 0.406 0.48 0.421 0.50 0.427 0.52 0.426 0.54 0.417 0.56 0.402 0.58 0.382 0.60 0.361 0.62 0.339 0.64 0.320 0.66 0.306 0.68 0.297 0.70 0.295 0.72 0.298 0.74 0.307 0.76 0.320 0.78 0.336 0.80 0.357 0.82 0.382 0.84 0.413 0.86 0.454 0.88 0.507 0.90 0.573 0.92 0.652 0.94 0.740 0.96 0.833 0.98 0.923 1.00 1.000 /
\end{sparkline}
 & CS  &  8075.9          & 8076.40 & $\pm$0.04  & $\pm$0.22  & $\pm$0.18  \\
Tb L$_2$N$_4$             & L$\gamma_1$      & \begin{sparkline}{6}
    \spark 0.00 0.382 0.02 0.367 0.04 0.348 0.06 0.328 0.08 0.308 0.10 0.292 0.12 0.280 0.14 0.273 0.16 0.272 0.18 0.276 0.20 0.285 0.22 0.298 0.24 0.314 0.26 0.334 0.28 0.359 0.30 0.390 0.32 0.431 0.34 0.482 0.36 0.547 0.38 0.623 0.40 0.706 0.42 0.792 0.44 0.873 0.46 0.939 0.48 0.983 0.50 1.000 0.52 0.986 0.54 0.944 0.56 0.876 0.58 0.790 0.60 0.695 0.62 0.597 0.64 0.504 0.66 0.419 0.68 0.346 0.70 0.284 0.72 0.234 0.74 0.194 0.76 0.162 0.78 0.137 0.80 0.118 0.82 0.102 0.84 0.090 0.86 0.079 0.88 0.071 0.90 0.064 0.92 0.058 0.94 0.053 0.96 0.049 0.98 0.045 1.00 0.041 /
\end{sparkline}
 &     &  8101.8\phantom{0} $\pm$ 1.6\phantom{0} & 8098.57 & $\pm$0.02  & $\pm$0.22  & $\pm$0.18  \\
Tb L$_2$O$_1$             & L$\gamma_8$      & \begin{sparkline}{6}
    \spark 0.00 0.402 0.02 0.405 0.04 0.409 0.06 0.413 0.08 0.419 0.10 0.427 0.12 0.436 0.14 0.446 0.16 0.459 0.18 0.475 0.20 0.493 0.22 0.515 0.24 0.540 0.26 0.569 0.28 0.602 0.30 0.640 0.32 0.681 0.34 0.726 0.36 0.772 0.38 0.820 0.40 0.867 0.42 0.910 0.44 0.947 0.46 0.975 0.48 0.994 0.50 1.000 0.52 0.994 0.54 0.976 0.56 0.947 0.58 0.909 0.60 0.864 0.62 0.815 0.64 0.764 0.66 0.712 0.68 0.663 0.70 0.616 0.72 0.573 0.74 0.535 0.76 0.500 0.78 0.470 0.80 0.443 0.82 0.419 0.84 0.399 0.86 0.381 0.88 0.365 0.90 0.351 0.92 0.339 0.94 0.328 0.96 0.319 0.98 0.310 1.00 0.303 /
\end{sparkline}
 & CS  &  8212.0\phantom{0} $\pm$ 1.1\phantom{0} & 8204.3\phantom{0} & $\pm$0.45  & $\pm$0.24  & $\pm$0.18  \\
Tb L$_2$N$_\mathrm{6,7}$  & Lv               & \begin{sparkline}{6}
    \spark 0.00 0.226 0.02 0.224 0.04 0.222 0.06 0.220 0.08 0.219 0.10 0.218 0.12 0.217 0.14 0.217 0.16 0.218 0.18 0.219 0.20 0.222 0.22 0.227 0.24 0.234 0.26 0.246 0.28 0.264 0.30 0.290 0.32 0.328 0.34 0.381 0.36 0.450 0.38 0.535 0.40 0.633 0.42 0.738 0.44 0.839 0.46 0.923 0.48 0.980 0.50 1.000 0.52 0.980 0.54 0.924 0.56 0.839 0.58 0.738 0.60 0.632 0.62 0.532 0.64 0.446 0.66 0.375 0.68 0.321 0.70 0.282 0.72 0.254 0.74 0.235 0.76 0.222 0.78 0.213 0.80 0.207 0.82 0.203 0.84 0.200 0.86 0.197 0.88 0.196 0.90 0.195 0.92 0.194 0.94 0.193 0.96 0.192 0.98 0.192 1.00 0.192 /
\end{sparkline}
 & Th  &  8242.0\phantom{0} $\pm$ 1.5\phantom{0} & 8251.15 & $\pm$0.24  & $\pm$0.24  & $\pm$0.18  \\
Tb L$_1$N$_2$             & L$\gamma_2$      & \begin{sparkline}{6}
    \spark 0.00 0.484 0.02 0.502 0.04 0.524 0.06 0.550 0.08 0.581 0.10 0.614 0.12 0.650 0.14 0.684 0.16 0.715 0.18 0.740 0.20 0.759 0.22 0.770 0.24 0.776 0.26 0.778 0.28 0.779 0.30 0.783 0.32 0.790 0.34 0.804 0.36 0.823 0.38 0.847 0.40 0.876 0.42 0.906 0.44 0.935 0.46 0.960 0.48 0.978 0.50 0.985 0.52 0.981 0.54 0.965 0.56 0.938 0.58 0.902 0.60 0.859 0.62 0.813 0.64 0.767 0.66 0.722 0.68 0.682 0.70 0.647 0.72 0.618 0.74 0.594 0.76 0.575 0.78 0.562 0.80 0.554 0.82 0.553 0.84 0.557 0.86 0.569 0.88 0.589 0.90 0.620 0.92 0.663 0.94 0.721 0.96 0.796 0.98 0.889 1.00 1.000 /
\end{sparkline}
 &     &  8397.6\phantom{0} $\pm$ 1.7\phantom{0} & 8398.5\phantom{0} & $\pm$0.26  & $\pm$3.0\phantom{0}  & $\pm$0.18  \\
Tb L$_1$N$_3$             & L$\gamma_3$      & \begin{sparkline}{6}
    \spark 0.00 0.425 0.02 0.401 0.04 0.379 0.06 0.360 0.08 0.344 0.10 0.331 0.12 0.321 0.14 0.315 0.16 0.312 0.18 0.312 0.20 0.315 0.22 0.323 0.24 0.336 0.26 0.356 0.28 0.382 0.30 0.418 0.32 0.463 0.34 0.519 0.36 0.585 0.38 0.659 0.40 0.738 0.42 0.816 0.44 0.888 0.46 0.947 0.48 0.986 0.50 1.000 0.52 0.988 0.54 0.950 0.56 0.890 0.58 0.814 0.60 0.729 0.62 0.641 0.64 0.556 0.66 0.478 0.68 0.410 0.70 0.351 0.72 0.303 0.74 0.263 0.76 0.230 0.78 0.204 0.80 0.183 0.82 0.166 0.84 0.151 0.86 0.139 0.88 0.129 0.90 0.121 0.92 0.113 0.94 0.107 0.96 0.101 0.98 0.096 1.00 0.091 /
\end{sparkline}
 &     &  8423.9\phantom{0} $\pm$ 1.7\phantom{0} & 8425.21 & $\pm$0.06  & $\pm$0.27  & $\pm$0.18  \\
Tb L$_1$N$_\mathrm{4,5}$  & L$\gamma_{11}$   & \begin{sparkline}{6}
    \spark 0.00 0.407 0.02 0.404 0.04 0.400 0.06 0.398 0.08 0.395 0.10 0.394 0.12 0.393 0.14 0.392 0.16 0.393 0.18 0.395 0.20 0.398 0.22 0.404 0.24 0.413 0.26 0.425 0.28 0.443 0.30 0.468 0.32 0.502 0.34 0.546 0.36 0.600 0.38 0.665 0.40 0.738 0.42 0.814 0.44 0.885 0.46 0.945 0.48 0.985 0.50 1.000 0.52 0.987 0.54 0.947 0.56 0.886 0.58 0.812 0.60 0.731 0.62 0.651 0.64 0.579 0.66 0.516 0.68 0.464 0.70 0.422 0.72 0.390 0.74 0.365 0.76 0.345 0.78 0.330 0.80 0.318 0.82 0.308 0.84 0.300 0.86 0.293 0.88 0.287 0.90 0.282 0.92 0.277 0.94 0.273 0.96 0.269 0.98 0.266 1.00 0.263 /
\end{sparkline}
 &     &  8558.9\phantom{0} $\pm$ 0.6\phantom{0} & 8557.3\phantom{0} & $\pm$0.7\phantom{0}  & $\pm$0.30  & $\pm$0.18  \\
Tb L$_1$O$_\mathrm{2,3}$  & L$\gamma_{4,4'}$ & \begin{sparkline}{6}
    \spark 0.00 0.068 0.02 0.073 0.04 0.078 0.06 0.084 0.08 0.091 0.10 0.099 0.12 0.108 0.14 0.119 0.16 0.132 0.18 0.148 0.20 0.167 0.22 0.190 0.24 0.218 0.26 0.252 0.28 0.293 0.30 0.343 0.32 0.403 0.34 0.472 0.36 0.550 0.38 0.635 0.40 0.723 0.42 0.809 0.44 0.886 0.46 0.947 0.48 0.986 0.50 1.000 0.52 0.986 0.54 0.947 0.56 0.886 0.58 0.809 0.60 0.723 0.62 0.635 0.64 0.550 0.66 0.472 0.68 0.402 0.70 0.343 0.72 0.292 0.74 0.251 0.76 0.217 0.78 0.188 0.80 0.165 0.82 0.146 0.84 0.131 0.86 0.117 0.88 0.106 0.90 0.097 0.92 0.089 0.94 0.082 0.96 0.075 0.98 0.070 1.00 0.065 /
\end{sparkline}
 & CS  &  8685.0\phantom{0} $\pm$ 1.2\phantom{0} & 8684.22 & $\pm$0.13  & $\pm$0.32  & $\pm$0.18  \\
\end{tabular}

%% file: Tables/peak_data_Ho.tex
%
\begin{tabular}{lllllrlrr}
\multicolumn{2}{c}{Line name} & & & & Peak & \multicolumn{3}{c}{Uncertainty} \\
IUPAC & Siegbahn & Spark & &  Reference &  Estimate & Stat & Sys & Peak  \\
\hline
Ho L$_3$M$_1$             & L$\ell$          & \begin{sparkline}{6}
    \spark 0.00 0.280 0.02 0.298 0.04 0.317 0.06 0.338 0.08 0.360 0.10 0.385 0.12 0.411 0.14 0.439 0.16 0.469 0.18 0.501 0.20 0.536 0.22 0.572 0.24 0.610 0.26 0.649 0.28 0.689 0.30 0.730 0.32 0.771 0.34 0.811 0.36 0.850 0.38 0.886 0.40 0.918 0.42 0.946 0.44 0.969 0.46 0.986 0.48 0.997 0.50 1.000 0.52 0.997 0.54 0.986 0.56 0.969 0.58 0.946 0.60 0.918 0.62 0.886 0.64 0.850 0.66 0.811 0.68 0.771 0.70 0.730 0.72 0.690 0.74 0.649 0.76 0.610 0.78 0.572 0.80 0.536 0.82 0.501 0.84 0.469 0.86 0.439 0.88 0.411 0.90 0.385 0.92 0.360 0.94 0.338 0.96 0.317 0.98 0.298 1.00 0.280 /
\end{sparkline}
 &     & 5939.96 $\pm$ 0.07 & 5940.23 & $\pm$0.012 & $\pm$0.09  & $\pm$0.18  \\
Ho L$_3$M$_4$             & L$\alpha_2$      & \begin{sparkline}{6}
    \spark 0.00 0.081 0.02 0.086 0.04 0.091 0.06 0.098 0.08 0.105 0.10 0.112 0.12 0.121 0.14 0.132 0.16 0.144 0.18 0.158 0.20 0.176 0.22 0.196 0.24 0.221 0.26 0.251 0.28 0.287 0.30 0.332 0.32 0.387 0.34 0.452 0.36 0.527 0.38 0.612 0.40 0.701 0.42 0.791 0.44 0.873 0.46 0.940 0.48 0.984 0.50 1.000 0.52 0.985 0.54 0.942 0.56 0.875 0.58 0.791 0.60 0.700 0.62 0.608 0.64 0.523 0.66 0.447 0.68 0.384 0.70 0.333 0.72 0.293 0.74 0.262 0.76 0.239 0.78 0.223 0.80 0.211 0.82 0.203 0.84 0.197 0.86 0.195 0.88 0.194 0.90 0.194 0.92 0.197 0.94 0.200 0.96 0.206 0.98 0.212 1.00 0.220 /
\end{sparkline}
 &     & 6678.48 $\pm$ 0.05 & 6678.52 & $\pm$0.02  & $\pm$0.12  & $\pm$0.18  \\
Ho L$_3$M$_5$             & L$\alpha_1$      & \begin{sparkline}{6}
    \spark 0.00 0.062 0.02 0.068 0.04 0.076 0.06 0.086 0.08 0.098 0.10 0.112 0.12 0.129 0.14 0.149 0.16 0.173 0.18 0.201 0.20 0.233 0.22 0.268 0.24 0.307 0.26 0.349 0.28 0.394 0.30 0.440 0.32 0.489 0.34 0.541 0.36 0.598 0.38 0.660 0.40 0.728 0.42 0.798 0.44 0.868 0.46 0.930 0.48 0.977 0.50 1.000 0.52 0.994 0.54 0.958 0.56 0.894 0.58 0.807 0.60 0.708 0.62 0.605 0.64 0.505 0.66 0.416 0.68 0.340 0.70 0.277 0.72 0.228 0.74 0.189 0.76 0.159 0.78 0.136 0.80 0.118 0.82 0.103 0.84 0.091 0.86 0.081 0.88 0.073 0.90 0.066 0.92 0.059 0.94 0.054 0.96 0.049 0.98 0.045 1.00 0.041 /
\end{sparkline}
 &     & 6719.68 $\pm$ 0.06 & 6719.73 & $\pm$0.005 & $\pm$0.12  & $\pm$0.18  \\
Ho L$_2$M$_1$             & L$\eta$          & \begin{sparkline}{6}
    \spark 0.00 0.768 0.02 0.760 0.04 0.752 0.06 0.746 0.08 0.742 0.10 0.739 0.12 0.738 0.14 0.739 0.16 0.742 0.18 0.747 0.20 0.754 0.22 0.763 0.24 0.775 0.26 0.789 0.28 0.805 0.30 0.823 0.32 0.843 0.34 0.865 0.36 0.888 0.38 0.911 0.40 0.933 0.42 0.954 0.44 0.972 0.46 0.986 0.48 0.996 0.50 1.000 0.52 0.998 0.54 0.990 0.56 0.976 0.58 0.956 0.60 0.931 0.62 0.901 0.64 0.868 0.66 0.833 0.68 0.796 0.70 0.759 0.72 0.722 0.74 0.686 0.76 0.651 0.78 0.619 0.80 0.588 0.82 0.559 0.84 0.532 0.86 0.508 0.88 0.485 0.90 0.464 0.92 0.445 0.94 0.427 0.96 0.411 0.98 0.396 1.00 0.382 /
\end{sparkline}
 &     & 6786.94 $\pm$ 0.27 & 6786.52 & $\pm$0.18  & $\pm$0.12  & $\pm$0.18  \\
Ho L$_1$M$_2$             & L$\beta_4$       & \begin{sparkline}{6}
    \spark 0.00 0.186 0.02 0.195 0.04 0.205 0.06 0.215 0.08 0.227 0.10 0.241 0.12 0.256 0.14 0.273 0.16 0.292 0.18 0.314 0.20 0.339 0.22 0.367 0.24 0.400 0.26 0.436 0.28 0.478 0.30 0.525 0.32 0.577 0.34 0.634 0.36 0.694 0.38 0.757 0.40 0.819 0.42 0.878 0.44 0.928 0.46 0.968 0.48 0.992 0.50 1.000 0.52 0.991 0.54 0.965 0.56 0.926 0.58 0.878 0.60 0.822 0.62 0.765 0.64 0.708 0.66 0.654 0.68 0.604 0.70 0.559 0.72 0.520 0.74 0.486 0.76 0.457 0.78 0.432 0.80 0.411 0.82 0.393 0.84 0.378 0.86 0.366 0.88 0.357 0.90 0.349 0.92 0.343 0.94 0.338 0.96 0.335 0.98 0.334 1.00 0.333 /
\end{sparkline}
 &     &  7471.1\phantom{0} $\pm$ 1.3\phantom{0} & 7471.22 & $\pm$0.04  & $\pm$0.14  & $\pm$0.18  \\
Ho L$_2$M$_4$             & L$\beta_1$       & \begin{sparkline}{6}
    \spark 0.00 0.041 0.02 0.044 0.04 0.047 0.06 0.051 0.08 0.056 0.10 0.062 0.12 0.068 0.14 0.076 0.16 0.086 0.18 0.097 0.20 0.111 0.22 0.128 0.24 0.149 0.26 0.176 0.28 0.209 0.30 0.249 0.32 0.299 0.34 0.360 0.36 0.432 0.38 0.517 0.40 0.613 0.42 0.716 0.44 0.819 0.46 0.909 0.48 0.974 0.50 1.000 0.52 0.982 0.54 0.920 0.56 0.825 0.58 0.712 0.60 0.594 0.62 0.484 0.64 0.390 0.66 0.312 0.68 0.252 0.70 0.205 0.72 0.169 0.74 0.142 0.76 0.121 0.78 0.104 0.80 0.091 0.82 0.080 0.84 0.071 0.86 0.064 0.88 0.057 0.90 0.052 0.92 0.048 0.94 0.044 0.96 0.040 0.98 0.037 1.00 0.035 /
\end{sparkline}
 &     & 7525.67 $\pm$ 0.15 & 7525.55 & $\pm$0.005 & $\pm$0.15  & $\pm$0.18  \\
Ho ---                    & L$\beta$'        & \begin{sparkline}{6}
    \spark 0.00 1.000 0.02 0.911 0.04 0.835 0.06 0.769 0.08 0.711 0.10 0.661 0.12 0.618 0.14 0.580 0.16 0.546 0.18 0.517 0.20 0.491 0.22 0.470 0.24 0.451 0.26 0.436 0.28 0.424 0.30 0.415 0.32 0.409 0.34 0.406 0.36 0.407 0.38 0.409 0.40 0.414 0.42 0.419 0.44 0.424 0.46 0.426 0.48 0.425 0.50 0.419 0.52 0.408 0.54 0.393 0.56 0.373 0.58 0.352 0.60 0.329 0.62 0.306 0.64 0.285 0.66 0.265 0.68 0.247 0.70 0.231 0.72 0.218 0.74 0.206 0.76 0.195 0.78 0.186 0.80 0.177 0.82 0.170 0.84 0.163 0.86 0.157 0.88 0.152 0.90 0.147 0.92 0.142 0.94 0.138 0.96 0.134 0.98 0.131 1.00 0.127 /
\end{sparkline}
 & CS  &  7554.3          & \emph{7558.92 }& $\pm$0.12  & $\pm$0.15  & $\pm$0.35  \\
Ho L$_3$N$_1$             & L$\beta_6$       & \begin{sparkline}{6}
    \spark 0.00 0.097 0.02 0.101 0.04 0.105 0.06 0.110 0.08 0.116 0.10 0.122 0.12 0.129 0.14 0.137 0.16 0.147 0.18 0.158 0.20 0.170 0.22 0.185 0.24 0.201 0.26 0.220 0.28 0.241 0.30 0.265 0.32 0.292 0.34 0.321 0.36 0.353 0.38 0.386 0.40 0.420 0.42 0.453 0.44 0.486 0.46 0.517 0.48 0.545 0.50 0.570 0.52 0.593 0.54 0.613 0.56 0.634 0.58 0.654 0.60 0.677 0.62 0.703 0.64 0.733 0.66 0.768 0.68 0.807 0.70 0.848 0.72 0.889 0.74 0.928 0.76 0.961 0.78 0.986 0.80 0.999 0.82 1.000 0.84 0.987 0.86 0.959 0.88 0.920 0.90 0.870 0.92 0.812 0.94 0.750 0.96 0.685 0.98 0.621 1.00 0.560 /
\end{sparkline}
 &     &  7635.8\phantom{0} $\pm$ 1.4\phantom{0} & $\star$\ \emph{7639.5\phantom{0} }& $\pm$0.12  & $\pm$0.16  & $\pm$1.0\phantom{0}  \\
Ho L$_1$M$_3$             & L$\beta_3$       & \begin{sparkline}{6}
    \spark 0.00 0.273 0.02 0.300 0.04 0.331 0.06 0.363 0.08 0.396 0.10 0.430 0.12 0.463 0.14 0.495 0.16 0.525 0.18 0.553 0.20 0.577 0.22 0.599 0.24 0.619 0.26 0.639 0.28 0.661 0.30 0.685 0.32 0.712 0.34 0.744 0.36 0.780 0.38 0.820 0.40 0.861 0.42 0.902 0.44 0.939 0.46 0.970 0.48 0.991 0.50 1.000 0.52 0.996 0.54 0.977 0.56 0.945 0.58 0.902 0.60 0.849 0.62 0.789 0.64 0.725 0.66 0.661 0.68 0.598 0.70 0.538 0.72 0.482 0.74 0.432 0.76 0.386 0.78 0.346 0.80 0.311 0.82 0.280 0.84 0.253 0.86 0.230 0.88 0.210 0.90 0.192 0.92 0.176 0.94 0.162 0.96 0.150 0.98 0.140 1.00 0.130 /
\end{sparkline}
 &     &  7651.8\phantom{0} $\pm$ 1.4\phantom{0} & 7651.72 & $\pm$0.05  & $\pm$0.16  & $\pm$0.18  \\
Ho ---                    & L$\beta_{14}$    & \begin{sparkline}{6}
    \spark 0.00 0.041 0.02 0.044 0.04 0.047 0.06 0.049 0.08 0.053 0.10 0.056 0.12 0.060 0.14 0.065 0.16 0.070 0.18 0.076 0.20 0.082 0.22 0.089 0.24 0.098 0.26 0.107 0.28 0.118 0.30 0.130 0.32 0.144 0.34 0.159 0.36 0.175 0.38 0.193 0.40 0.210 0.42 0.228 0.44 0.245 0.46 0.260 0.48 0.273 0.50 0.284 0.52 0.292 0.54 0.299 0.56 0.305 0.58 0.312 0.60 0.321 0.62 0.332 0.64 0.346 0.66 0.361 0.68 0.378 0.70 0.398 0.72 0.421 0.74 0.448 0.76 0.479 0.78 0.512 0.80 0.547 0.82 0.581 0.84 0.612 0.86 0.642 0.88 0.671 0.90 0.703 0.92 0.741 0.94 0.787 0.96 0.845 0.98 0.916 1.00 1.000 /
\end{sparkline}
 & CS  &  7894.9          & $\star$\ \emph{7884.3\phantom{0} }& $\pm$0.07  & $\pm$0.20  & $\pm$1.0\phantom{0}  \\
Ho L$_3$N$_\mathrm{4,5}$  & L$\beta_{2,15}$  & \begin{sparkline}{6}
    \spark 0.00 0.196 0.02 0.205 0.04 0.215 0.06 0.226 0.08 0.239 0.10 0.254 0.12 0.272 0.14 0.291 0.16 0.311 0.18 0.330 0.20 0.348 0.22 0.365 0.24 0.382 0.26 0.400 0.28 0.421 0.30 0.448 0.32 0.481 0.34 0.521 0.36 0.569 0.38 0.626 0.40 0.690 0.42 0.763 0.44 0.840 0.46 0.913 0.48 0.971 0.50 1.000 0.52 0.991 0.54 0.941 0.56 0.856 0.58 0.747 0.60 0.630 0.62 0.518 0.64 0.418 0.66 0.336 0.68 0.271 0.70 0.220 0.72 0.182 0.74 0.153 0.76 0.130 0.78 0.112 0.80 0.098 0.82 0.086 0.84 0.077 0.86 0.069 0.88 0.062 0.90 0.057 0.92 0.052 0.94 0.048 0.96 0.044 0.98 0.041 1.00 0.038 /
\end{sparkline}
 &     & 7911.35 $\pm$ 0.25 & 7910.32 & $\pm$0.011 & $\pm$0.21  & $\pm$0.18  \\
Ho L$_1$M$_4$             & L$\beta_{10}$    & \begin{sparkline}{6}
    \spark 0.00 0.286 0.02 0.283 0.04 0.282 0.06 0.280 0.08 0.279 0.10 0.279 0.12 0.279 0.14 0.280 0.16 0.283 0.18 0.286 0.20 0.290 0.22 0.297 0.24 0.306 0.26 0.317 0.28 0.332 0.30 0.350 0.32 0.373 0.34 0.400 0.36 0.431 0.38 0.464 0.40 0.499 0.42 0.534 0.44 0.565 0.46 0.591 0.48 0.610 0.50 0.621 0.52 0.622 0.54 0.615 0.56 0.602 0.58 0.584 0.60 0.564 0.62 0.545 0.64 0.529 0.66 0.518 0.68 0.513 0.70 0.515 0.72 0.524 0.74 0.541 0.76 0.564 0.78 0.594 0.80 0.630 0.82 0.672 0.84 0.716 0.86 0.764 0.88 0.812 0.90 0.859 0.92 0.902 0.94 0.940 0.96 0.970 0.98 0.991 1.00 1.000 /
\end{sparkline}
 &     & 8004.08 $\pm$ 0.15 & \emph{8002.7\phantom{0} }& $\pm$0.29  & $\pm$0.29  & $\pm$0.35  \\
Ho L$_3$O$_1$             & L$\beta_7$       & \begin{sparkline}{6}
    \spark 0.00 0.622 0.02 0.620 0.04 0.611 0.06 0.595 0.08 0.576 0.10 0.557 0.12 0.538 0.14 0.524 0.16 0.515 0.18 0.513 0.20 0.517 0.22 0.529 0.24 0.549 0.26 0.575 0.28 0.608 0.30 0.646 0.32 0.689 0.34 0.735 0.36 0.783 0.38 0.831 0.40 0.877 0.42 0.918 0.44 0.953 0.46 0.979 0.48 0.995 0.50 1.000 0.52 0.993 0.54 0.974 0.56 0.946 0.58 0.908 0.60 0.864 0.62 0.816 0.64 0.765 0.66 0.715 0.68 0.666 0.70 0.621 0.72 0.580 0.74 0.545 0.76 0.517 0.78 0.497 0.80 0.486 0.82 0.484 0.84 0.493 0.86 0.514 0.88 0.547 0.90 0.592 0.92 0.645 0.94 0.704 0.96 0.763 0.98 0.818 1.00 0.860 /
\end{sparkline}
 &     &                  & 8023.0\phantom{0} & $\pm$0.18  & $\pm$1.0\phantom{0}  & $\pm$0.18  \\
Ho L$_1$M$_5$             & L$\beta_{9}$     & \begin{sparkline}{6}
    \spark 0.00 1.000 0.02 0.975 0.04 0.941 0.06 0.899 0.08 0.851 0.10 0.801 0.12 0.749 0.14 0.698 0.16 0.650 0.18 0.607 0.20 0.568 0.22 0.537 0.24 0.513 0.26 0.497 0.28 0.491 0.30 0.496 0.32 0.512 0.34 0.541 0.36 0.581 0.38 0.632 0.40 0.690 0.42 0.751 0.44 0.810 0.46 0.859 0.48 0.892 0.50 0.906 0.52 0.898 0.54 0.866 0.56 0.815 0.58 0.748 0.60 0.672 0.62 0.593 0.64 0.516 0.66 0.445 0.68 0.383 0.70 0.331 0.72 0.288 0.74 0.255 0.76 0.228 0.78 0.208 0.80 0.193 0.82 0.183 0.84 0.177 0.86 0.175 0.88 0.177 0.90 0.185 0.92 0.200 0.94 0.221 0.96 0.249 0.98 0.283 1.00 0.320 /
\end{sparkline}
 &     & 8044.61 $\pm$ 0.14 & 8044.7\phantom{0} & $\pm$0.15  & $\pm$3.0\phantom{0}  & $\pm$0.18  \\
Ho L$_3$N$_\mathrm{6,7}$  & Lu               & \begin{sparkline}{6}
    \spark 0.00 1.000 0.02 0.893 0.04 0.784 0.06 0.680 0.08 0.585 0.10 0.504 0.12 0.436 0.14 0.381 0.16 0.338 0.18 0.304 0.20 0.279 0.22 0.261 0.24 0.248 0.26 0.242 0.28 0.241 0.30 0.247 0.32 0.261 0.34 0.284 0.36 0.317 0.38 0.358 0.40 0.407 0.42 0.459 0.44 0.510 0.46 0.552 0.48 0.582 0.50 0.593 0.52 0.584 0.54 0.556 0.56 0.512 0.58 0.458 0.60 0.399 0.62 0.341 0.64 0.288 0.66 0.244 0.68 0.208 0.70 0.181 0.72 0.161 0.74 0.146 0.76 0.136 0.78 0.129 0.80 0.124 0.82 0.121 0.84 0.118 0.86 0.115 0.88 0.113 0.90 0.111 0.92 0.109 0.94 0.108 0.96 0.106 0.98 0.105 1.00 0.104 /
\end{sparkline}
 & Th  &  8062.0\phantom{0} $\pm$ 1.6\phantom{0} & 8068.18 & $\pm$0.20  & $\pm$0.23  & $\pm$0.18  \\
Ho L$_2$N$_1$             & L$\gamma_5$      & \begin{sparkline}{6}
    \spark 0.00 0.099 0.02 0.104 0.04 0.111 0.06 0.118 0.08 0.127 0.10 0.136 0.12 0.148 0.14 0.162 0.16 0.178 0.18 0.197 0.20 0.220 0.22 0.248 0.24 0.281 0.26 0.321 0.28 0.368 0.30 0.422 0.32 0.483 0.34 0.552 0.36 0.626 0.38 0.702 0.40 0.778 0.42 0.849 0.44 0.911 0.46 0.959 0.48 0.990 0.50 1.000 0.52 0.990 0.54 0.959 0.56 0.911 0.58 0.849 0.60 0.778 0.62 0.703 0.64 0.626 0.66 0.553 0.68 0.484 0.70 0.423 0.72 0.369 0.74 0.323 0.76 0.284 0.78 0.251 0.80 0.223 0.82 0.200 0.84 0.181 0.86 0.165 0.88 0.152 0.90 0.140 0.92 0.131 0.94 0.122 0.96 0.115 0.98 0.109 1.00 0.104 /
\end{sparkline}
 &     &  8481.5\phantom{0} $\pm$ 1.7\phantom{0} & 8487.00 & $\pm$0.18  & $\pm$0.29  & $\pm$0.18  \\
Ho ---                    & L$\gamma_9$      & \begin{sparkline}{6}
    \spark 0.00 0.064 0.02 0.068 0.04 0.072 0.06 0.076 0.08 0.082 0.10 0.088 0.12 0.095 0.14 0.103 0.16 0.113 0.18 0.125 0.20 0.139 0.22 0.155 0.24 0.175 0.26 0.198 0.28 0.226 0.30 0.257 0.32 0.293 0.34 0.332 0.36 0.375 0.38 0.419 0.40 0.464 0.42 0.506 0.44 0.544 0.46 0.575 0.48 0.599 0.50 0.613 0.52 0.619 0.54 0.619 0.56 0.614 0.58 0.609 0.60 0.608 0.62 0.612 0.64 0.625 0.66 0.647 0.68 0.677 0.70 0.712 0.72 0.750 0.74 0.786 0.76 0.819 0.78 0.849 0.80 0.876 0.82 0.902 0.84 0.929 0.86 0.955 0.88 0.979 0.90 0.995 0.92 1.000 0.94 0.988 0.96 0.957 0.98 0.905 1.00 0.835 /
\end{sparkline}
 & CS  &  8731.0          & \emph{8731.4\phantom{0} }& $\pm$0.07  & $\pm$0.32  & $\pm$0.5\phantom{0}  \\
Ho L$_2$N$_4$             & L$\gamma_1$      & \begin{sparkline}{6}
    \spark 0.00 0.475 0.02 0.516 0.04 0.553 0.06 0.583 0.08 0.604 0.10 0.616 0.12 0.621 0.14 0.619 0.16 0.614 0.18 0.610 0.20 0.609 0.22 0.616 0.24 0.631 0.26 0.656 0.28 0.688 0.30 0.724 0.32 0.761 0.34 0.797 0.36 0.829 0.38 0.858 0.40 0.885 0.42 0.911 0.44 0.938 0.46 0.964 0.48 0.986 0.50 1.000 0.52 1.000 0.54 0.982 0.56 0.945 0.58 0.887 0.60 0.812 0.62 0.726 0.64 0.635 0.66 0.543 0.68 0.458 0.70 0.382 0.72 0.316 0.74 0.262 0.76 0.219 0.78 0.185 0.80 0.158 0.82 0.136 0.84 0.120 0.86 0.106 0.88 0.094 0.90 0.085 0.92 0.076 0.94 0.069 0.96 0.062 0.98 0.056 1.00 0.052 /
\end{sparkline}
 &     &  8747.2\phantom{0} $\pm$ 1.8\phantom{0} & 8750.84 & $\pm$0.04  & $\pm$0.32  & $\pm$0.18  \\
Ho L$_2$O$_1$             & L$\gamma_8$      & \begin{sparkline}{6}
    \spark 0.00 0.447 0.02 0.454 0.04 0.461 0.06 0.470 0.08 0.480 0.10 0.492 0.12 0.506 0.14 0.521 0.16 0.539 0.18 0.559 0.20 0.581 0.22 0.606 0.24 0.633 0.26 0.663 0.28 0.696 0.30 0.730 0.32 0.766 0.34 0.803 0.36 0.840 0.38 0.876 0.40 0.910 0.42 0.940 0.44 0.965 0.46 0.984 0.48 0.996 0.50 1.000 0.52 0.996 0.54 0.984 0.56 0.965 0.58 0.939 0.60 0.908 0.62 0.872 0.64 0.833 0.66 0.793 0.68 0.752 0.70 0.711 0.72 0.672 0.74 0.635 0.76 0.599 0.78 0.567 0.80 0.537 0.82 0.509 0.84 0.484 0.86 0.461 0.88 0.440 0.90 0.421 0.92 0.405 0.94 0.389 0.96 0.375 0.98 0.363 1.00 0.352 /
\end{sparkline}
 & CS  &  8867.0\phantom{0} $\pm$ 1.3\phantom{0} & 8866.1\phantom{0} & $\pm$0.7\phantom{0}  & $\pm$0.5\phantom{0}  & $\pm$0.18  \\
Ho L$_2$N$_\mathrm{6,7}$  & Lv               & \begin{sparkline}{6}
    \spark 0.00 0.263 0.02 0.258 0.04 0.253 0.06 0.249 0.08 0.245 0.10 0.242 0.12 0.239 0.14 0.237 0.16 0.236 0.18 0.236 0.20 0.237 0.22 0.239 0.24 0.245 0.26 0.255 0.28 0.272 0.30 0.298 0.32 0.336 0.34 0.389 0.36 0.458 0.38 0.544 0.40 0.642 0.42 0.745 0.44 0.843 0.46 0.926 0.48 0.980 0.50 1.000 0.52 0.981 0.54 0.926 0.56 0.844 0.58 0.744 0.60 0.638 0.62 0.538 0.64 0.450 0.66 0.377 0.68 0.321 0.70 0.280 0.72 0.251 0.74 0.231 0.76 0.218 0.78 0.209 0.80 0.204 0.82 0.200 0.84 0.197 0.86 0.195 0.88 0.194 0.90 0.193 0.92 0.192 0.94 0.191 0.96 0.191 0.98 0.191 1.00 0.191 /
\end{sparkline}
 & Th  &  8908.4\phantom{0} $\pm$ 0.7\phantom{0} & 8915.5\phantom{0} & $\pm$0.38  & $\pm$1.2\phantom{0}  & $\pm$0.18  \\
Ho L$_1$N$_2$             & L$\gamma_2$      & \begin{sparkline}{6}
    \spark 0.00 0.415 0.02 0.449 0.04 0.482 0.06 0.514 0.08 0.543 0.10 0.569 0.12 0.591 0.14 0.610 0.16 0.627 0.18 0.643 0.20 0.660 0.22 0.677 0.24 0.697 0.26 0.720 0.28 0.746 0.30 0.774 0.32 0.804 0.34 0.836 0.36 0.868 0.38 0.898 0.40 0.927 0.42 0.952 0.44 0.972 0.46 0.987 0.48 0.997 0.50 1.000 0.52 0.997 0.54 0.987 0.56 0.972 0.58 0.952 0.60 0.927 0.62 0.899 0.64 0.869 0.66 0.837 0.68 0.806 0.70 0.775 0.72 0.747 0.74 0.721 0.76 0.699 0.78 0.681 0.80 0.667 0.82 0.658 0.84 0.653 0.86 0.651 0.88 0.652 0.90 0.654 0.92 0.656 0.94 0.656 0.96 0.654 0.98 0.649 1.00 0.641 /
\end{sparkline}
 &     &  9051.1\phantom{0} $\pm$ 2.0\phantom{0} & 9044.3\phantom{0} & $\pm$0.25  & $\pm$0.37  & $\pm$0.18  \\
Ho L$_1$N$_3$             & L$\gamma_3$      & \begin{sparkline}{6}
    \spark 0.00 0.267 0.02 0.267 0.04 0.265 0.06 0.262 0.08 0.258 0.10 0.254 0.12 0.249 0.14 0.246 0.16 0.244 0.18 0.245 0.20 0.250 0.22 0.260 0.24 0.277 0.26 0.301 0.28 0.335 0.30 0.378 0.32 0.432 0.34 0.496 0.36 0.570 0.38 0.651 0.40 0.735 0.42 0.817 0.44 0.890 0.46 0.949 0.48 0.987 0.50 1.000 0.52 0.987 0.54 0.950 0.56 0.891 0.58 0.815 0.60 0.730 0.62 0.641 0.64 0.554 0.66 0.473 0.68 0.401 0.70 0.338 0.72 0.285 0.74 0.242 0.76 0.206 0.78 0.177 0.80 0.154 0.82 0.136 0.84 0.120 0.86 0.108 0.88 0.097 0.90 0.089 0.92 0.081 0.94 0.075 0.96 0.070 0.98 0.065 1.00 0.060 /
\end{sparkline}
 &     &  9087.6\phantom{0} $\pm$ 2.0\phantom{0} & 9086.62 & $\pm$0.07  & $\pm$0.37  & $\pm$0.18  \\
Ho L$_1$N$_\mathrm{4,5}$  & L$\gamma_{11}$   & \begin{sparkline}{6}
    \spark 0.00 0.266 0.02 0.265 0.04 0.264 0.06 0.264 0.08 0.264 0.10 0.264 0.12 0.264 0.14 0.265 0.16 0.267 0.18 0.269 0.20 0.273 0.22 0.277 0.24 0.284 0.26 0.294 0.28 0.309 0.30 0.330 0.32 0.360 0.34 0.403 0.36 0.461 0.38 0.536 0.40 0.626 0.42 0.728 0.44 0.829 0.46 0.918 0.48 0.978 0.50 1.000 0.52 0.979 0.54 0.918 0.56 0.829 0.58 0.726 0.60 0.622 0.62 0.527 0.64 0.449 0.66 0.387 0.68 0.341 0.70 0.308 0.72 0.284 0.74 0.266 0.76 0.254 0.78 0.244 0.80 0.237 0.82 0.231 0.84 0.226 0.86 0.222 0.88 0.218 0.90 0.215 0.92 0.212 0.94 0.210 0.96 0.208 0.98 0.206 1.00 0.204 /
\end{sparkline}
 & CS  &  9232.1\phantom{0} $\pm$ 1.5\phantom{0} & 9229.6\phantom{0} & $\pm$0.8\phantom{0}  & $\pm$0.40  & $\pm$0.18  \\
Ho L$_1$O$_\mathrm{2,3}$  & L$\gamma_{4,4'}$ & \begin{sparkline}{6}
    \spark 0.00 0.055 0.02 0.059 0.04 0.064 0.06 0.069 0.08 0.074 0.10 0.081 0.12 0.089 0.14 0.098 0.16 0.109 0.18 0.122 0.20 0.138 0.22 0.157 0.24 0.181 0.26 0.211 0.28 0.247 0.30 0.292 0.32 0.347 0.34 0.414 0.36 0.493 0.38 0.581 0.40 0.677 0.42 0.773 0.44 0.863 0.46 0.935 0.48 0.983 0.50 1.000 0.52 0.983 0.54 0.936 0.56 0.863 0.58 0.773 0.60 0.677 0.62 0.581 0.64 0.492 0.66 0.414 0.68 0.347 0.70 0.292 0.72 0.246 0.74 0.210 0.76 0.181 0.78 0.157 0.80 0.137 0.82 0.121 0.84 0.108 0.86 0.097 0.88 0.088 0.90 0.080 0.92 0.073 0.94 0.067 0.96 0.062 0.98 0.058 1.00 0.054 /
\end{sparkline}
 & CS  &  9374.0\phantom{0} $\pm$ 1.4\phantom{0} & 9368.6\phantom{0} & $\pm$0.10  & $\pm$0.42  & $\pm$0.18  \\
\end{tabular}

%% file: Tables/roi_table.tex
\begin{tabular}{rrrrrrrrr}
Element & $E$ range & N$_\mathrm{V}$ & $\sigma$ & $\sigma_0$ & BG low & BG high & $\chi^2$/dof & $p$-value \\
%
\hline
Pr A & 4350--4550 & 2 & 3.42 & 3.06 &   50.1 &   83.9 &  170.3/191 & 0.8562 \\
Pr B & 4850--5130 & 7 & 2.21 & 1.55 &  161.8 &  157.1 &  511.8/256 & 0.0000 \\
Pr C & 5400--5525 & 4 & 1.87 & 0.96 &  379.7 &   17.4 &  418.2/110 & 0.0000 \\
Pr D & 5560--5700 & 4 & 2.11 & 1.35 &  300.9 &  221.9 &  199.3/125 & 0.0000 \\
Pr E & 5800--5870 & 4 & 2.27 & 1.58 &  382.2 &   78.7 &   53.1/ 55 & 0.5490 \\
Pr F & 5876--5964 & 4 & 1.64 & 0.00 &    0.0 &  313.2 &   91.9/ 73 & 0.0671 \\
Pr G & 6106--6449 & 9 & 2.84 & 2.30 &   96.7 &   72.1 &  385.5/313 & 0.0032 \\
Pr H & 6517--6700 & 4 & 2.13 & 1.29 &   25.7 &    2.6 &  181.5/168 & 0.2252 \\
Pr I & 6680--6900 & 3 & 3.83 & 3.43 &   52.8 &   66.6 &  251.8/208 & 0.0205 \\
%
\hline
Nd A & 4550--4720 & 2 & 1.96 & 1.21 &   70.5 &  104.4 &  179.4/161 & 0.1523 \\
Nd B & 5050--5320 & 7 & 1.94 & 1.12 &  145.2 &  160.0 &  369.5/246 & 0.0000 \\
Nd C & 5601--5770 & 5 & 2.07 & 1.29 &  277.4 &  277.4 &  185.5/151 & 0.0293 \\
Nd D & 5805--5950 & 4 & 2.29 & 1.59 &   99.7 &  378.4 &  183.4/130 & 0.0014 \\
Nd E & 6012--6104 & 4 & 2.47 & 1.84 &  346.4 &  235.8 &  101.6/ 77 & 0.0318 \\
Nd F & 6120--6213 & 5 & 1.79 & 0.67 &  179.5 &   17.9 &   99.8/ 75 & 0.0292 \\
Nd G & 6331--6732 & 9 & 2.93 & 2.39 &   73.9 &   47.1 &  464.0/371 & 0.0007 \\
Nd H & 6790--6960 & 4 & 2.15 & 1.29 &   24.1 &    2.4 &  189.6/155 & 0.0306 \\
Nd I & 6955--7250 & 4 & 3.68 & 3.25 &   52.9 &   68.9 &  323.1/280 & 0.0389 \\
%
\hline
Tb A & 5430--5650 & 2 & 3.69 & 3.32 &  262.9 &  393.9 &  211.5/211 & 0.4782 \\
Tb B & 6175--6350 & 7 & 2.23 & 1.48 &  175.4 &  143.0 &  199.1/151 & 0.0053 \\
Tb C & 6875--7044 & 6 & 2.16 & 1.30 &  191.7 &  143.2 &  209.7/148 & 0.0006 \\
Tb D & 7070--7150 & 2 & 3.81 & 3.39 &  282.0 &  149.4 &  104.2/ 71 & 0.0063 \\
Tb E & 7300--7420 & 3 & 2.59 & 1.91 &  319.8 &   75.8 &  339.1/108 & 0.0000 \\
Tb F & 7420--7560 & 5 & 2.04 & 1.03 &   82.3 &   41.1 &  231.7/122 & 0.0000 \\
Tb G & 7750--8260 & 9 & 3.26 & 2.71 &   96.7 &   75.2 &  593.5/480 & 0.0003 \\
Tb H & 8300--8490 & 6 & 3.28 & 2.72 &   57.4 &    5.7 &  168.4/169 & 0.4986 \\
Tb I & 8500--8780 & 2 & 3.18 & 2.59 &   57.8 &   43.0 &  297.0/271 & 0.1329 \\
%
\hline
Ho A & 5850--6028 & 2 & 2.51 & 1.90 &  431.9 &  345.5 &  208.3/169 & 0.0214 \\
Ho B & 6620--6850 & 7 & 2.63 & 2.00 &  312.3 &  153.9 &  480.4/206 & 0.0000 \\
Ho C & 7300--7567 & 9 & 2.07 & 1.09 &  158.9 &   14.5 &  283.7/237 & 0.0203 \\
Ho D & 7600--7699 & 3 & 3.08 & 2.52 &  236.3 &  219.4 &  146.9/ 87 & 0.0001 \\
Ho E & 7820--7950 & 5 & 2.21 & 1.29 &  401.1 &  121.0 &  212.0/112 & 0.0000 \\
Ho F & 7972--8095 & 4 & 3.52 & 3.02 &  192.2 &  112.5 &  116.8/108 & 0.2654 \\
Ho G & 8460--8925 & 10 & 3.71 & 3.21 &   86.2 &   49.5 &  542.2/432 & 0.0002 \\
Ho H & 8950--9150 & 4 & 3.86 & 3.37 &   41.1 &   30.8 &  210.8/185 & 0.0935 \\
Ho I & 9170--9500 & 2 & 3.03 & 2.36 &   40.1 &   27.7 &  331.0/321 & 0.3377 \\
\end{tabular}

%% file: Tables/shape_data_Pr.tex
%
\begin{tabular}{llrrr}
    & $\sigma_0$ & $E_i$ & $\Gamma_i$ & $I_i$ \\
\hline
Pr A & 3.06 & 4457.56 &  6.08 &    285\,200 \\
     &      & 4480.87 & 17.87 &     35\,800 \\
\hline
Pr B & 1.55 & 4933.39 &  6.70 &    139\,400 \\
     &      & 4948.01 & 13.78 &     55\,200 \\
     &      & 5011.95 &  2.60 & 1\,411\,700 \\
     &      & 5026.93 &  1.02 &    289\,700 \\
     &      & 5032.71 &  0.81 & 3\,838\,900 \\
     &      & 5033.11 &  2.79 & 7\,798\,900 \\
     &      & 5055.60 &  2.54 &     12\,100 \\
\hline
Pr C & 0.96 & 5479.75 &  0.60 &     17\,800 \\
     &      & 5484.71 &  1.99 & 1\,265\,000 \\
     &      & 5488.58 &  2.37 & 6\,821\,300 \\
     &      & 5496.85 &  5.42 & 1\,120\,700 \\
\hline
Pr D & 1.35 & 5593.04 &  4.44 & 1\,446\,200 \\
     &      & 5617.23 &  9.49 &     24\,300 \\
     &      & 5660.20 &  3.70 &    172\,700 \\
     &      & 5678.97 & 23.93 &     38\,900 \\
\hline
Pr E & 1.58 & 5830.10 &  9.21 &    164\,900 \\
     &      & 5845.23 &  2.09 &    556\,600 \\
     &      & 5849.69 &  1.49 & 2\,295\,000 \\
     &      & 5854.65 &  4.33 &    841\,500 \\
\hline
Pr F & 0.00 & 5883.48 &  2.61 &     18\,100 \\
     &      & 5902.71 &  4.73 &     69\,800 \\
     &      & 5925.16 &  6.57 &     58\,200 \\
     &      & 5960.39 &  0.81 &      2\,900 \\
\hline
Pr G & 2.30 & 6136.23 &  3.28 &     42\,300 \\
     &      & 6159.00 & 50.00 &      7\,800 \\
     &      & 6284.32 & 22.59 &     17\,900 \\
     &      & 6305.43 &  7.46 &    216\,900 \\
     &      & 6322.18 &  1.58 &    950\,200 \\
     &      & 6329.29 &  3.33 &    123\,000 \\
     &      & 6382.00 & 10.63 &      9\,100 \\
     &      & 6402.00 &  3.23 &     11\,300 \\
     &      & 6437.51 &  1.88 &      8\,000 \\
\hline
Pr H & 1.29 & 6579.26 & 11.84 &    171\,500 \\
     &      & 6601.13 &  6.19 &    115\,900 \\
     &      & 6617.06 &  2.63 &    184\,800 \\
\hline
Pr I & 3.43 & 6718.32 &  1.85 &      4\,900 \\
     &      & 6814.04 &  2.43 &     67\,700 \\
     &      & 6829.15 &  1.58 &      5\,700 \\
\hline
\end{tabular}

%% file: Tables/shape_data_Nd.tex
%
\begin{tabular}{llrrr}
    & $\sigma_0$ & $E_i$ & $\Gamma_i$ & $I_i$ \\
\hline
Nd A & 1.21 & 4631.95 &  7.50 &    302\,100 \\
     &      & 4660.59 & 12.19 &     20\,800 \\
\hline
Nd B & 1.12 & 5144.88 &  7.58 &    167\,600 \\
     &      & 5162.00 & 19.23 &     38\,500 \\
     &      & 5204.05 &  2.90 &    423\,700 \\
     &      & 5207.64 &  2.23 &    885\,500 \\
     &      & 5223.79 &  1.26 &    230\,300 \\
     &      & 5229.90 &  2.28 & 11\,291\,000 \\
     &      & 5254.05 &  0.60 &      3\,600 \\
\hline
Nd C & 1.29 & 5630.61 & 18.53 &     28\,800 \\
     &      & 5685.02 & 18.04 &     51\,400 \\
     &      & 5716.07 &  2.12 & 1\,141\,000 \\
     &      & 5720.94 &  2.45 & 7\,402\,600 \\
     &      & 5737.47 &  3.43 &    175\,100 \\
\hline
Nd D & 1.59 & 5823.00 & 20.82 &    139\,200 \\
     &      & 5827.89 &  4.45 & 1\,296\,300 \\
     &      & 5849.98 &  9.62 &     22\,500 \\
     &      & 5891.62 &  4.52 &    191\,400 \\
\hline
Nd E & 1.84 & 6069.28 &  9.14 &    212\,000 \\
     &      & 6087.45 &  3.09 & 1\,302\,200 \\
     &      & 6090.59 &  0.60 & 1\,307\,100 \\
     &      & 6094.59 &  4.05 &    867\,300 \\
\hline
Nd F & 0.67 & 6125.45 &  0.60 &      6\,200 \\
     &      & 6146.96 &  2.79 &     36\,900 \\
     &      & 6170.15 & 29.16 &     98\,700 \\
     &      & 6170.19 &  3.76 &     25\,400 \\
     &      & 6206.79 &  3.84 &     19\,600 \\
\hline
Nd G & 2.39 & 6404.97 &  3.29 &     36\,800 \\
     &      & 6495.13 &  0.81 &         700 \\
     &      & 6580.74 &  7.93 &    242\,700 \\
     &      & 6596.00 &  5.82 &    178\,900 \\
     &      & 6600.40 &  1.01 &    654\,800 \\
     &      & 6607.27 &  2.94 &    100\,300 \\
     &      & 6656.00 &  5.04 &      2\,500 \\
     &      & 6682.65 &  6.28 &     12\,200 \\
     &      & 6721.33 &  2.25 &      9\,700 \\
\hline
Nd H & 1.29 & 6862.90 & 13.83 &    164\,700 \\
     &      & 6882.94 &  5.44 &     73\,500 \\
     &      & 6899.90 &  2.77 &    162\,100 \\
     &      & 6952.00 & 35.14 &     23\,900 \\
\hline
Nd I & 3.25 & 7007.14 &  0.60 &      3\,200 \\
     &      & 7057.00 & 31.85 &      4\,300 \\
     &      & 7106.86 &  2.69 &     58\,700 \\
     &      & 7121.41 &  1.35 &      3\,500 \\
\hline
\end{tabular}

%% file: Tables/shape_data_Tb.tex
%
\begin{tabular}{llrrr}
    & $\sigma_0$ & $E_i$ & $\Gamma_i$ & $I_i$ \\
\hline
Tb A & 3.32 & 5551.45 &  7.76 &    452\,600 \\
     &      & 5591.01 &  7.40 &      6\,700 \\
\hline
Tb B & 1.48 & 6207.43 & 49.99 &    249\,500 \\
     &      & 6234.04 &  3.34 &    263\,900 \\
     &      & 6239.52 &  2.43 & 1\,289\,300 \\
     &      & 6265.50 &  2.76 & 1\,246\,400 \\
     &      & 6270.30 &  2.39 & 1\,238\,800 \\
     &      & 6274.40 &  2.55 & 10\,941\,800 \\
     &      & 6290.54 &  6.34 &    143\,100 \\
\hline
Tb C & 1.30 & 6910.60 & 23.86 &     44\,200 \\
     &      & 6942.00 &  5.72 &    589\,600 \\
     &      & 6972.88 &  3.67 & 2\,125\,700 \\
     &      & 6977.12 &  1.77 & 3\,930\,500 \\
     &      & 6978.02 &  3.56 & 1\,755\,400 \\
     &      & 7003.41 &  4.10 &    171\,600 \\
\hline
Tb D & 3.39 & 7097.05 &  5.17 &    813\,000 \\
     &      & 7117.14 &  4.93 &    191\,600 \\
\hline
Tb E & 1.91 & 7339.71 &  5.65 &    591\,500 \\
     &      & 7356.48 &  6.53 &    453\,100 \\
     &      & 7366.23 &  2.65 & 2\,549\,400 \\
\hline
Tb F & 1.03 & 7430.95 &  5.17 &     22\,400 \\
     &      & 7461.73 & 13.09 &     41\,900 \\
     &      & 7467.66 &  4.42 &     50\,300 \\
     &      & 7512.35 &  1.91 &     10\,500 \\
     &      & 7517.00 & 21.32 &     27\,700 \\
\hline
Tb G & 2.71 & 7848.57 &  0.60 &      6\,800 \\
     &      & 7856.52 &  4.73 &     34\,600 \\
     &      & 8022.00 & 50.00 &     45\,200 \\
     &      & 8057.98 & 15.35 &     27\,800 \\
     &      & 8076.41 &  4.61 &    383\,100 \\
     &      & 8088.41 &  0.61 &     47\,400 \\
     &      & 8098.54 &  2.81 &    790\,300 \\
     &      & 8204.38 &  6.30 &     12\,600 \\
     &      & 8251.15 &  1.10 &      7\,200 \\
\hline
Tb H & 2.72 & 8366.91 &  7.30 &     16\,100 \\
     &      & 8385.01 &  0.90 &      6\,500 \\
     &      & 8393.08 & 19.87 &    164\,900 \\
     &      & 8398.73 &  3.29 &     25\,200 \\
     &      & 8425.25 &  3.42 &    110\,600 \\
     &      & 8482.00 & 15.40 &     12\,900 \\
\hline
Tb I & 2.59 & 8557.27 &  2.47 &      2\,600 \\
     &      & 8684.22 &  4.31 &     41\,900 \\
\hline
\end{tabular}

%% file: Tables/shape_data_Ho.tex
%
\begin{tabular}{llrrr}
    & $\sigma_0$ & $E_i$ & $\Gamma_i$ & $I_i$ \\
\hline
Ho A & 1.90 & 5940.23 &  8.53 &    527\,700 \\
     &      & 5984.31 &  1.42 &      2\,600 \\
\hline
Ho B & 2.00 & 6673.05 & 31.43 &    161\,300 \\
     &      & 6675.99 &  5.66 &    394\,400 \\
     &      & 6678.60 &  1.98 & 1\,263\,900 \\
     &      & 6713.86 &  3.36 & 4\,922\,500 \\
     &      & 6719.74 &  1.59 & 8\,612\,000 \\
     &      & 6725.00 &  6.19 &    950\,900 \\
     &      & 6786.95 &  7.74 &    124\,800 \\
\hline
Ho C & 1.09 & 7386.14 &  9.22 &     19\,100 \\
     &      & 7386.93 & 29.67 &     14\,300 \\
     &      & 7419.72 & 31.75 &     15\,800 \\
     &      & 7471.18 &  6.08 &    483\,000 \\
     &      & 7489.55 & 45.93 &    292\,200 \\
     &      & 7521.90 &  3.68 & 1\,006\,900 \\
     &      & 7525.64 &  2.65 & 4\,571\,300 \\
     &      & 7546.96 & 50.00 &     13\,000 \\
     &      & 7558.92 &  4.77 &     82\,700 \\
\hline
Ho D & 2.52 & 7614.79 &  2.41 &      2\,200 \\
     &      & 7639.52 &  5.27 &    231\,000 \\
     &      & 7651.85 &  5.74 &    683\,200 \\
\hline
Ho E & 1.29 & 7884.30 &  6.20 &    455\,900 \\
     &      & 7890.86 &  0.60 &     22\,800 \\
     &      & 7898.22 &  5.41 &    883\,200 \\
     &      & 7904.89 &  2.54 &    410\,300 \\
     &      & 7910.20 &  2.68 & 2\,057\,700 \\
\hline
Ho F & 3.02 & 8002.68 &  2.74 &     16\,400 \\
     &      & 8023.02 &  6.63 &     68\,000 \\
     &      & 8044.74 &  2.05 &     29\,500 \\
     &      & 8068.17 &  0.60 &     10\,100 \\
\hline
Ho G & 3.21 & 8487.00 &  4.42 &     30\,600 \\
     &      & 8554.39 &  0.62 &         500 \\
     &      & 8635.71 & 24.83 &     11\,900 \\
     &      & 8701.99 & 30.08 &     48\,000 \\
     &      & 8731.45 &  4.92 &    344\,100 \\
     &      & 8742.64 &  1.15 &    143\,700 \\
     &      & 8750.88 &  2.63 &    389\,300 \\
     &      & 8764.02 &  2.08 &      5\,300 \\
     &      & 8866.25 &  9.48 &     11\,400 \\
     &      & 8915.45 &  0.86 &      4\,700 \\
\hline
Ho H & 3.37 & 9026.17 &  3.47 &      8\,600 \\
     &      & 9044.13 & 12.50 &    102\,200 \\
     &      & 9065.81 &  4.35 &     12\,900 \\
     &      & 9086.65 &  3.58 &    107\,600 \\
\hline
Ho I & 2.36 & 9229.56 &  2.01 &      1\,600 \\
     &      & 9368.57 &  4.34 &     34\,700 \\
\hline
\end{tabular}

%% file: Tables/peak_data_supplement.tex
%
\begin{tabular}{lllrrrr}
\multicolumn{2}{c}{Line name} & &  \multicolumn{2}{c}{Peaks} & & Peak \\
IUPAC & Siegbahn & & 6\,eV & Raw & Shift & Corrected  \\
\hline
Pr L$_3$M$_4$             & L$\alpha_2$      & \begin{sparkline}{6}
    \spark 0.00 0.095 0.02 0.101 0.04 0.108 0.06 0.115 0.08 0.124 0.10 0.134 0.12 0.145 0.14 0.159 0.16 0.175 0.18 0.193 0.20 0.215 0.22 0.241 0.24 0.272 0.26 0.308 0.28 0.350 0.30 0.398 0.32 0.451 0.34 0.509 0.36 0.570 0.38 0.632 0.40 0.691 0.42 0.745 0.44 0.790 0.46 0.822 0.48 0.840 0.50 0.843 0.52 0.831 0.54 0.806 0.56 0.769 0.58 0.725 0.60 0.677 0.62 0.628 0.64 0.580 0.66 0.538 0.68 0.501 0.70 0.471 0.72 0.448 0.74 0.433 0.76 0.424 0.78 0.423 0.80 0.428 0.82 0.440 0.84 0.460 0.86 0.487 0.88 0.524 0.90 0.570 0.92 0.627 0.94 0.697 0.96 0.781 0.98 0.882 1.00 1.000 /
\end{sparkline}
 & 5012.28 & 5012.13 & $-0.25$ & 5011.88 \\
Pr L$_2$M$_4$+L$_1$M$_2$  & L$\beta_{1,4}$   & \begin{sparkline}{6}
    \spark 0.00 0.052 0.02 0.057 0.04 0.063 0.06 0.069 0.08 0.077 0.10 0.086 0.12 0.097 0.14 0.110 0.16 0.126 0.18 0.145 0.20 0.169 0.22 0.197 0.24 0.231 0.26 0.272 0.28 0.320 0.30 0.375 0.32 0.437 0.34 0.505 0.36 0.578 0.38 0.655 0.40 0.733 0.42 0.810 0.44 0.882 0.46 0.941 0.48 0.982 0.50 1.000 0.52 0.991 0.54 0.955 0.56 0.895 0.58 0.818 0.60 0.731 0.62 0.641 0.64 0.556 0.66 0.479 0.68 0.413 0.70 0.358 0.72 0.314 0.74 0.278 0.76 0.250 0.78 0.227 0.80 0.208 0.82 0.191 0.84 0.177 0.86 0.164 0.88 0.152 0.90 0.140 0.92 0.130 0.94 0.119 0.96 0.110 0.98 0.101 1.00 0.092 /
\end{sparkline}
 & 5488.22 & 5488.31 & $+0.16$ & 5488.48 \\
Pr L$_1$M$_5$             & L$\beta_{9}$     & \begin{sparkline}{6}
    \spark 0.00 0.612 0.02 0.585 0.04 0.563 0.06 0.546 0.08 0.533 0.10 0.524 0.12 0.519 0.14 0.517 0.16 0.519 0.18 0.523 0.20 0.531 0.22 0.543 0.24 0.558 0.26 0.578 0.28 0.603 0.30 0.632 0.32 0.667 0.34 0.707 0.36 0.751 0.38 0.798 0.40 0.846 0.42 0.893 0.44 0.935 0.46 0.968 0.48 0.991 0.50 1.000 0.52 0.994 0.54 0.974 0.56 0.940 0.58 0.897 0.60 0.847 0.62 0.794 0.64 0.740 0.66 0.688 0.68 0.640 0.70 0.597 0.72 0.559 0.74 0.526 0.76 0.497 0.78 0.473 0.80 0.454 0.82 0.438 0.84 0.426 0.86 0.416 0.88 0.410 0.90 0.406 0.92 0.405 0.94 0.406 0.96 0.410 0.98 0.417 1.00 0.425 /
\end{sparkline}
 & 5902.53 & 5902.60 & $+0.11$ & 5902.71 \\
Pr L$_3$O$_1$             & L$\beta_7$       & \begin{sparkline}{6}
    \spark 0.00 0.837 0.02 0.794 0.04 0.758 0.06 0.728 0.08 0.705 0.10 0.686 0.12 0.673 0.14 0.665 0.16 0.660 0.18 0.661 0.20 0.665 0.22 0.673 0.24 0.685 0.26 0.701 0.28 0.721 0.30 0.745 0.32 0.772 0.34 0.801 0.36 0.833 0.38 0.866 0.40 0.899 0.42 0.930 0.44 0.957 0.46 0.979 0.48 0.993 0.50 1.000 0.52 0.998 0.54 0.986 0.56 0.965 0.58 0.937 0.60 0.901 0.62 0.861 0.64 0.818 0.66 0.773 0.68 0.728 0.70 0.683 0.72 0.641 0.74 0.601 0.76 0.564 0.78 0.529 0.80 0.497 0.82 0.468 0.84 0.442 0.86 0.418 0.88 0.396 0.90 0.376 0.92 0.358 0.94 0.342 0.96 0.326 0.98 0.313 1.00 0.300 /
\end{sparkline}
 & 5924.67 & 5924.81 & $+0.25$ & 5925.06 \\
Pr L$_3$N$_\mathrm{6,7}$  & Lu               & \begin{sparkline}{6}
    \spark 0.00 0.788 0.02 0.767 0.04 0.748 0.06 0.730 0.08 0.713 0.10 0.697 0.12 0.682 0.14 0.667 0.16 0.654 0.18 0.642 0.20 0.630 0.22 0.620 0.24 0.610 0.26 0.602 0.28 0.595 0.30 0.591 0.32 0.590 0.34 0.594 0.36 0.607 0.38 0.634 0.40 0.677 0.42 0.739 0.44 0.817 0.46 0.898 0.48 0.966 0.50 1.000 0.52 0.990 0.54 0.937 0.56 0.853 0.58 0.757 0.60 0.667 0.62 0.594 0.64 0.539 0.66 0.500 0.68 0.474 0.70 0.456 0.72 0.443 0.74 0.432 0.76 0.424 0.78 0.416 0.80 0.410 0.82 0.404 0.84 0.398 0.86 0.393 0.88 0.388 0.90 0.383 0.92 0.379 0.94 0.374 0.96 0.370 0.98 0.366 1.00 0.362 /
\end{sparkline}
 & 5960.13 & 5960.30 & $+0.31$ & 5960.61 \\
Pr L$_2$O$_1$             & L$\gamma_8$      & \begin{sparkline}{6}
    \spark 0.00 0.554 0.02 0.547 0.04 0.540 0.06 0.533 0.08 0.526 0.10 0.521 0.12 0.516 0.14 0.513 0.16 0.513 0.18 0.514 0.20 0.519 0.22 0.527 0.24 0.540 0.26 0.558 0.28 0.581 0.30 0.611 0.32 0.646 0.34 0.688 0.36 0.735 0.38 0.785 0.40 0.837 0.42 0.886 0.44 0.931 0.46 0.966 0.48 0.990 0.50 1.000 0.52 0.994 0.54 0.973 0.56 0.938 0.58 0.891 0.60 0.836 0.62 0.776 0.64 0.714 0.66 0.653 0.68 0.595 0.70 0.543 0.72 0.496 0.74 0.456 0.76 0.421 0.78 0.391 0.80 0.366 0.82 0.345 0.84 0.327 0.86 0.312 0.88 0.299 0.90 0.288 0.92 0.278 0.94 0.270 0.96 0.263 0.98 0.257 1.00 0.252 /
\end{sparkline}
 & 6401.71 & 6401.80 & $+0.17$ & 6401.96 \\
Pr L$_1$N$_2$             & L$\gamma_2$      & \begin{sparkline}{6}
    \spark 0.00 0.273 0.02 0.268 0.04 0.263 0.06 0.259 0.08 0.255 0.10 0.252 0.12 0.250 0.14 0.248 0.16 0.248 0.18 0.250 0.20 0.253 0.22 0.257 0.24 0.263 0.26 0.271 0.28 0.281 0.30 0.293 0.32 0.306 0.34 0.321 0.36 0.337 0.38 0.353 0.40 0.369 0.42 0.383 0.44 0.395 0.46 0.405 0.48 0.410 0.50 0.412 0.52 0.410 0.54 0.403 0.56 0.394 0.58 0.382 0.60 0.369 0.62 0.355 0.64 0.343 0.66 0.332 0.68 0.325 0.70 0.321 0.72 0.323 0.74 0.332 0.76 0.348 0.78 0.374 0.80 0.412 0.82 0.463 0.84 0.528 0.86 0.606 0.88 0.694 0.90 0.785 0.92 0.872 0.94 0.942 0.96 0.987 0.98 1.000 1.00 0.977 /
\end{sparkline}
 & 6601.23 & 6601.16 & $-0.14$ & 6601.02 \\
Pr L$_1$N$_3$             & L$\gamma_3$      & \begin{sparkline}{6}
    \spark 0.00 0.409 0.02 0.412 0.04 0.411 0.06 0.405 0.08 0.397 0.10 0.385 0.12 0.372 0.14 0.359 0.16 0.346 0.18 0.335 0.20 0.326 0.22 0.322 0.24 0.322 0.26 0.329 0.28 0.343 0.30 0.366 0.32 0.401 0.34 0.449 0.36 0.510 0.38 0.586 0.40 0.672 0.42 0.763 0.44 0.852 0.46 0.927 0.48 0.979 0.50 1.000 0.52 0.985 0.54 0.937 0.56 0.862 0.58 0.768 0.60 0.667 0.62 0.568 0.64 0.477 0.66 0.398 0.68 0.332 0.70 0.279 0.72 0.236 0.74 0.202 0.76 0.175 0.78 0.154 0.80 0.137 0.82 0.122 0.84 0.111 0.86 0.101 0.88 0.092 0.90 0.085 0.92 0.079 0.94 0.073 0.96 0.068 0.98 0.064 1.00 0.060 /
\end{sparkline}
 & 6616.92 & 6616.98 & $+0.12$ & 6617.10 \\
%
\hline
Nd L$_2$M$_1$             & L$\eta$          & \begin{sparkline}{6}
    \spark 0.00 0.384 0.02 0.400 0.04 0.418 0.06 0.437 0.08 0.458 0.10 0.480 0.12 0.504 0.14 0.529 0.16 0.556 0.18 0.585 0.20 0.615 0.22 0.647 0.24 0.680 0.26 0.714 0.28 0.748 0.30 0.783 0.32 0.817 0.34 0.851 0.36 0.883 0.38 0.912 0.40 0.938 0.42 0.960 0.44 0.978 0.46 0.991 0.48 0.998 0.50 1.000 0.52 0.996 0.54 0.988 0.56 0.974 0.58 0.956 0.60 0.935 0.62 0.911 0.64 0.885 0.66 0.857 0.68 0.829 0.70 0.801 0.72 0.773 0.74 0.747 0.76 0.721 0.78 0.697 0.80 0.674 0.82 0.653 0.84 0.633 0.86 0.615 0.88 0.599 0.90 0.584 0.92 0.571 0.94 0.559 0.96 0.548 0.98 0.539 1.00 0.530 /
\end{sparkline}
 & 5145.27 & 5145.18 & $-0.14$ & 5145.04 \\
Nd L$_2$M$_4$+L$_1$M$_2$  & L$\beta_{1,4}$   & \begin{sparkline}{6}
    \spark 0.00 0.052 0.02 0.057 0.04 0.063 0.06 0.070 0.08 0.078 0.10 0.088 0.12 0.101 0.14 0.115 0.16 0.134 0.18 0.156 0.20 0.182 0.22 0.213 0.24 0.250 0.26 0.292 0.28 0.339 0.30 0.391 0.32 0.448 0.34 0.511 0.36 0.579 0.38 0.652 0.40 0.728 0.42 0.805 0.44 0.877 0.46 0.937 0.48 0.980 0.50 1.000 0.52 0.992 0.54 0.957 0.56 0.898 0.58 0.819 0.60 0.729 0.62 0.634 0.64 0.543 0.66 0.458 0.68 0.383 0.70 0.320 0.72 0.268 0.74 0.225 0.76 0.191 0.78 0.164 0.80 0.142 0.82 0.124 0.84 0.110 0.86 0.098 0.88 0.089 0.90 0.082 0.92 0.076 0.94 0.071 0.96 0.067 0.98 0.064 1.00 0.061 /
\end{sparkline}
 & 5720.59 & 5720.72 & $+0.23$ & 5720.96 \\
Nd L$_3$N$_5$             & L$\beta_2$       & \begin{sparkline}{6}
    \spark 0.00 0.064 0.02 0.068 0.04 0.072 0.06 0.077 0.08 0.083 0.10 0.091 0.12 0.100 0.14 0.112 0.16 0.126 0.18 0.143 0.20 0.163 0.22 0.188 0.24 0.218 0.26 0.253 0.28 0.294 0.30 0.342 0.32 0.397 0.34 0.461 0.36 0.534 0.38 0.614 0.40 0.700 0.42 0.786 0.44 0.867 0.46 0.935 0.48 0.982 0.50 1.000 0.52 0.987 0.54 0.944 0.56 0.875 0.58 0.790 0.60 0.696 0.62 0.603 0.64 0.516 0.66 0.440 0.68 0.376 0.70 0.324 0.72 0.280 0.74 0.244 0.76 0.214 0.78 0.188 0.80 0.166 0.82 0.146 0.84 0.129 0.86 0.114 0.88 0.101 0.90 0.090 0.92 0.081 0.94 0.072 0.96 0.065 0.98 0.059 1.00 0.054 /
\end{sparkline}
 & 6090.26 & 6090.31 & $+0.10$ & 6090.42 \\
Nd L$_3$O$_1$             & L$\beta_7$       & \begin{sparkline}{6}
    \spark 0.00 0.667 0.02 0.647 0.04 0.632 0.06 0.621 0.08 0.613 0.10 0.608 0.12 0.605 0.14 0.604 0.16 0.605 0.18 0.609 0.20 0.614 0.22 0.622 0.24 0.633 0.26 0.646 0.28 0.663 0.30 0.684 0.32 0.710 0.34 0.740 0.36 0.775 0.38 0.814 0.40 0.855 0.42 0.897 0.44 0.936 0.46 0.969 0.48 0.991 0.50 1.000 0.52 0.994 0.54 0.975 0.56 0.943 0.58 0.902 0.60 0.855 0.62 0.807 0.64 0.760 0.66 0.716 0.68 0.676 0.70 0.641 0.72 0.611 0.74 0.584 0.76 0.561 0.78 0.541 0.80 0.523 0.82 0.507 0.84 0.493 0.86 0.480 0.88 0.469 0.90 0.458 0.92 0.448 0.94 0.438 0.96 0.430 0.98 0.421 1.00 0.414 /
\end{sparkline}
 & 6169.99 & 6170.07 & $+0.14$ & 6170.21 \\
Nd L$_3$N$_\mathrm{6,7}$  & Lu               & \begin{sparkline}{6}
    \spark 0.00 0.595 0.02 0.588 0.04 0.582 0.06 0.577 0.08 0.572 0.10 0.568 0.12 0.566 0.14 0.564 0.16 0.565 0.18 0.566 0.20 0.570 0.22 0.576 0.24 0.586 0.26 0.598 0.28 0.615 0.30 0.636 0.32 0.663 0.34 0.696 0.36 0.734 0.38 0.778 0.40 0.826 0.42 0.875 0.44 0.921 0.46 0.960 0.48 0.988 0.50 1.000 0.52 0.995 0.54 0.971 0.56 0.933 0.58 0.883 0.60 0.825 0.62 0.765 0.64 0.705 0.66 0.650 0.68 0.600 0.70 0.555 0.72 0.517 0.74 0.483 0.76 0.455 0.78 0.430 0.80 0.409 0.82 0.391 0.84 0.375 0.86 0.360 0.88 0.348 0.90 0.336 0.92 0.326 0.94 0.317 0.96 0.308 0.98 0.301 1.00 0.293 /
\end{sparkline}
 & 6206.47 & 6206.59 & $+0.23$ & 6206.82 \\
Nd L$_2$O$_1$             & L$\gamma_8$      & \begin{sparkline}{6}
    \spark 0.00 0.545 0.02 0.542 0.04 0.541 0.06 0.542 0.08 0.545 0.10 0.549 0.12 0.556 0.14 0.565 0.16 0.576 0.18 0.590 0.20 0.607 0.22 0.626 0.24 0.649 0.26 0.674 0.28 0.702 0.30 0.734 0.32 0.767 0.34 0.802 0.36 0.837 0.38 0.873 0.40 0.906 0.42 0.937 0.44 0.963 0.46 0.982 0.48 0.995 0.50 1.000 0.52 0.996 0.54 0.985 0.56 0.965 0.58 0.938 0.60 0.906 0.62 0.869 0.64 0.829 0.66 0.787 0.68 0.746 0.70 0.705 0.72 0.665 0.74 0.628 0.76 0.594 0.78 0.562 0.80 0.533 0.82 0.507 0.84 0.484 0.86 0.463 0.88 0.444 0.90 0.427 0.92 0.412 0.94 0.399 0.96 0.387 0.98 0.376 1.00 0.367 /
\end{sparkline}
 & 6682.37 & 6682.42 & $+0.11$ & 6682.53 \\
%
\hline
Tb L$_3$M$_4$             & L$\alpha_2$      & \begin{sparkline}{6}
    \spark 0.00 0.095 0.02 0.102 0.04 0.109 0.06 0.117 0.08 0.126 0.10 0.137 0.12 0.151 0.14 0.166 0.16 0.184 0.18 0.205 0.20 0.230 0.22 0.258 0.24 0.289 0.26 0.325 0.28 0.365 0.30 0.410 0.32 0.460 0.34 0.518 0.36 0.582 0.38 0.654 0.40 0.730 0.42 0.808 0.44 0.880 0.46 0.941 0.48 0.983 0.50 1.000 0.52 0.989 0.54 0.951 0.56 0.890 0.58 0.812 0.60 0.726 0.62 0.639 0.64 0.558 0.66 0.486 0.68 0.425 0.70 0.376 0.72 0.338 0.74 0.309 0.76 0.287 0.78 0.272 0.80 0.262 0.82 0.257 0.84 0.254 0.86 0.255 0.88 0.258 0.90 0.264 0.92 0.272 0.94 0.282 0.96 0.295 0.98 0.311 1.00 0.330 /
\end{sparkline}
 & 6239.30 & 6239.37 & $+0.14$ & 6239.51 \\
Tb L$_3$M$_5$             & L$\alpha_1$      & \begin{sparkline}{6}
    \spark 0.00 0.069 0.02 0.077 0.04 0.086 0.06 0.098 0.08 0.111 0.10 0.127 0.12 0.145 0.14 0.164 0.16 0.186 0.18 0.209 0.20 0.233 0.22 0.258 0.24 0.285 0.26 0.316 0.28 0.350 0.30 0.391 0.32 0.439 0.34 0.495 0.36 0.560 0.38 0.634 0.40 0.713 0.42 0.794 0.44 0.871 0.46 0.935 0.48 0.980 0.50 1.000 0.52 0.990 0.54 0.951 0.56 0.887 0.58 0.803 0.60 0.709 0.62 0.611 0.64 0.518 0.66 0.433 0.68 0.360 0.70 0.299 0.72 0.248 0.74 0.208 0.76 0.176 0.78 0.151 0.80 0.130 0.82 0.114 0.84 0.101 0.86 0.090 0.88 0.081 0.90 0.074 0.92 0.067 0.94 0.062 0.96 0.057 0.98 0.053 1.00 0.050 /
\end{sparkline}
 & 6274.03 & 6274.14 & $+0.21$ & 6274.35 \\
Tb L$_1$M$_2$             & L$\beta_4$       & \begin{sparkline}{6}
    \spark 0.00 0.187 0.02 0.197 0.04 0.208 0.06 0.220 0.08 0.233 0.10 0.248 0.12 0.264 0.14 0.283 0.16 0.304 0.18 0.328 0.20 0.356 0.22 0.387 0.24 0.422 0.26 0.462 0.28 0.506 0.30 0.556 0.32 0.609 0.34 0.667 0.36 0.727 0.38 0.787 0.40 0.846 0.42 0.898 0.44 0.943 0.46 0.976 0.48 0.995 0.50 1.000 0.52 0.991 0.54 0.968 0.56 0.935 0.58 0.895 0.60 0.851 0.62 0.805 0.64 0.761 0.66 0.721 0.68 0.685 0.70 0.655 0.72 0.631 0.74 0.613 0.76 0.600 0.78 0.593 0.80 0.592 0.82 0.595 0.84 0.604 0.86 0.618 0.88 0.637 0.90 0.662 0.92 0.692 0.94 0.729 0.96 0.773 0.98 0.825 1.00 0.886 /
\end{sparkline}
 & 6942.44 & 6942.32 & $-0.24$ & 6942.08 \\
Tb L$_2$M$_4$             & L$\beta_1$       & \begin{sparkline}{6}
    \spark 0.00 0.048 0.02 0.052 0.04 0.056 0.06 0.061 0.08 0.067 0.10 0.075 0.12 0.084 0.14 0.094 0.16 0.107 0.18 0.123 0.20 0.142 0.22 0.165 0.24 0.192 0.26 0.225 0.28 0.263 0.30 0.308 0.32 0.360 0.34 0.418 0.36 0.486 0.38 0.562 0.40 0.648 0.42 0.740 0.44 0.832 0.46 0.913 0.48 0.973 0.50 1.000 0.52 0.987 0.54 0.933 0.56 0.847 0.58 0.740 0.60 0.627 0.62 0.517 0.64 0.420 0.66 0.339 0.68 0.274 0.70 0.223 0.72 0.184 0.74 0.154 0.76 0.130 0.78 0.112 0.80 0.097 0.82 0.085 0.84 0.076 0.86 0.068 0.88 0.061 0.90 0.055 0.92 0.051 0.94 0.047 0.96 0.043 0.98 0.041 1.00 0.038 /
\end{sparkline}
 & 6976.78 & 6976.90 & $+0.26$ & 6977.16 \\
Tb L$_1$M$_5$+L$_3$O$_1$  & L$\beta_{9,7}$   & \begin{sparkline}{6}
    \spark 0.00 0.323 0.02 0.329 0.04 0.337 0.06 0.345 0.08 0.355 0.10 0.365 0.12 0.377 0.14 0.390 0.16 0.405 0.18 0.421 0.20 0.438 0.22 0.458 0.24 0.479 0.26 0.504 0.28 0.531 0.30 0.563 0.32 0.599 0.34 0.641 0.36 0.688 0.38 0.740 0.40 0.797 0.42 0.854 0.44 0.909 0.46 0.955 0.48 0.987 0.50 1.000 0.52 0.992 0.54 0.963 0.56 0.915 0.58 0.854 0.60 0.785 0.62 0.714 0.64 0.646 0.66 0.582 0.68 0.526 0.70 0.477 0.72 0.434 0.74 0.398 0.76 0.367 0.78 0.341 0.80 0.319 0.82 0.299 0.84 0.283 0.86 0.268 0.88 0.255 0.90 0.244 0.92 0.234 0.94 0.225 0.96 0.218 0.98 0.211 1.00 0.204 /
\end{sparkline}
 & 7467.27 & 7467.37 & $+0.22$ & 7467.59 \\
Tb L$_2$N$_1$             & L$\gamma_5$      & \begin{sparkline}{6}
    \spark 0.00 0.111 0.02 0.118 0.04 0.125 0.06 0.134 0.08 0.145 0.10 0.159 0.12 0.177 0.14 0.202 0.16 0.234 0.18 0.276 0.20 0.328 0.22 0.390 0.24 0.462 0.26 0.539 0.28 0.618 0.30 0.693 0.32 0.759 0.34 0.815 0.36 0.858 0.38 0.891 0.40 0.916 0.42 0.937 0.44 0.956 0.46 0.973 0.48 0.988 0.50 0.998 0.52 1.000 0.54 0.990 0.56 0.967 0.58 0.930 0.60 0.881 0.62 0.822 0.64 0.755 0.66 0.686 0.68 0.617 0.70 0.550 0.72 0.488 0.74 0.432 0.76 0.383 0.78 0.340 0.80 0.303 0.82 0.272 0.84 0.246 0.86 0.223 0.88 0.204 0.90 0.188 0.92 0.175 0.94 0.163 0.96 0.153 0.98 0.144 1.00 0.136 /
\end{sparkline}
 & 7855.08 & 7855.60 & $+1.23$ & 7856.84 \\
Tb L$_2$N$_4$             & L$\gamma_1$      & \begin{sparkline}{6}
    \spark 0.00 0.382 0.02 0.367 0.04 0.348 0.06 0.328 0.08 0.308 0.10 0.292 0.12 0.280 0.14 0.273 0.16 0.272 0.18 0.276 0.20 0.285 0.22 0.298 0.24 0.314 0.26 0.334 0.28 0.359 0.30 0.390 0.32 0.431 0.34 0.482 0.36 0.547 0.38 0.623 0.40 0.706 0.42 0.792 0.44 0.873 0.46 0.939 0.48 0.983 0.50 1.000 0.52 0.986 0.54 0.944 0.56 0.876 0.58 0.790 0.60 0.695 0.62 0.597 0.64 0.504 0.66 0.419 0.68 0.346 0.70 0.284 0.72 0.234 0.74 0.194 0.76 0.162 0.78 0.137 0.80 0.118 0.82 0.102 0.84 0.090 0.86 0.079 0.88 0.071 0.90 0.064 0.92 0.058 0.94 0.053 0.96 0.049 0.98 0.045 1.00 0.041 /
\end{sparkline}
 & 8098.42 & 8098.47 & $+0.11$ & 8098.57 \\
Tb L$_1$N$_2$             & L$\gamma_2$      & \begin{sparkline}{6}
    \spark 0.00 0.484 0.02 0.502 0.04 0.524 0.06 0.550 0.08 0.581 0.10 0.614 0.12 0.650 0.14 0.684 0.16 0.715 0.18 0.740 0.20 0.759 0.22 0.770 0.24 0.776 0.26 0.778 0.28 0.779 0.30 0.783 0.32 0.790 0.34 0.804 0.36 0.823 0.38 0.847 0.40 0.876 0.42 0.906 0.44 0.935 0.46 0.960 0.48 0.978 0.50 0.985 0.52 0.981 0.54 0.965 0.56 0.938 0.58 0.902 0.60 0.859 0.62 0.813 0.64 0.767 0.66 0.722 0.68 0.682 0.70 0.647 0.72 0.618 0.74 0.594 0.76 0.575 0.78 0.562 0.80 0.554 0.82 0.553 0.84 0.557 0.86 0.569 0.88 0.589 0.90 0.620 0.92 0.663 0.94 0.721 0.96 0.796 0.98 0.889 1.00 1.000 /
\end{sparkline}
 & 8397.94 & 8398.08 & $+0.38$ & 8398.46 \\
%
\hline
Ho L$_3$M$_5$             & L$\alpha_1$      & \begin{sparkline}{6}
    \spark 0.00 0.062 0.02 0.068 0.04 0.076 0.06 0.086 0.08 0.098 0.10 0.112 0.12 0.129 0.14 0.149 0.16 0.173 0.18 0.201 0.20 0.233 0.22 0.268 0.24 0.307 0.26 0.349 0.28 0.394 0.30 0.440 0.32 0.489 0.34 0.541 0.36 0.598 0.38 0.660 0.40 0.728 0.42 0.798 0.44 0.868 0.46 0.930 0.48 0.977 0.50 1.000 0.52 0.994 0.54 0.958 0.56 0.894 0.58 0.807 0.60 0.708 0.62 0.605 0.64 0.505 0.66 0.416 0.68 0.340 0.70 0.277 0.72 0.228 0.74 0.189 0.76 0.159 0.78 0.136 0.80 0.118 0.82 0.103 0.84 0.091 0.86 0.081 0.88 0.073 0.90 0.066 0.92 0.059 0.94 0.054 0.96 0.049 0.98 0.045 1.00 0.041 /
\end{sparkline}
 & 6719.09 & 6719.30 & $+0.43$ & 6719.73 \\
Ho L$_2$M$_1$             & L$\eta$          & \begin{sparkline}{6}
    \spark 0.00 0.768 0.02 0.760 0.04 0.752 0.06 0.746 0.08 0.742 0.10 0.739 0.12 0.738 0.14 0.739 0.16 0.742 0.18 0.747 0.20 0.754 0.22 0.763 0.24 0.775 0.26 0.789 0.28 0.805 0.30 0.823 0.32 0.843 0.34 0.865 0.36 0.888 0.38 0.911 0.40 0.933 0.42 0.954 0.44 0.972 0.46 0.986 0.48 0.996 0.50 1.000 0.52 0.998 0.54 0.990 0.56 0.976 0.58 0.956 0.60 0.931 0.62 0.901 0.64 0.868 0.66 0.833 0.68 0.796 0.70 0.759 0.72 0.722 0.74 0.686 0.76 0.651 0.78 0.619 0.80 0.588 0.82 0.559 0.84 0.532 0.86 0.508 0.88 0.485 0.90 0.464 0.92 0.445 0.94 0.427 0.96 0.411 0.98 0.396 1.00 0.382 /
\end{sparkline}
 & 6786.05 & 6786.20 & $+0.32$ & 6786.52 \\
Ho L$_2$M$_4$             & L$\beta_1$       & \begin{sparkline}{6}
    \spark 0.00 0.041 0.02 0.044 0.04 0.047 0.06 0.051 0.08 0.056 0.10 0.062 0.12 0.068 0.14 0.076 0.16 0.086 0.18 0.097 0.20 0.111 0.22 0.128 0.24 0.149 0.26 0.176 0.28 0.209 0.30 0.249 0.32 0.299 0.34 0.360 0.36 0.432 0.38 0.517 0.40 0.613 0.42 0.716 0.44 0.819 0.46 0.909 0.48 0.974 0.50 1.000 0.52 0.982 0.54 0.920 0.56 0.825 0.58 0.712 0.60 0.594 0.62 0.484 0.64 0.390 0.66 0.312 0.68 0.252 0.70 0.205 0.72 0.169 0.74 0.142 0.76 0.121 0.78 0.104 0.80 0.091 0.82 0.080 0.84 0.071 0.86 0.064 0.88 0.057 0.90 0.052 0.92 0.048 0.94 0.044 0.96 0.040 0.98 0.037 1.00 0.035 /
\end{sparkline}
 & 7525.31 & 7525.38 & $+0.17$ & 7525.55 \\
Ho L$_1$M$_3$             & L$\beta_3$       & \begin{sparkline}{6}
    \spark 0.00 0.273 0.02 0.300 0.04 0.331 0.06 0.363 0.08 0.396 0.10 0.430 0.12 0.463 0.14 0.495 0.16 0.525 0.18 0.553 0.20 0.577 0.22 0.599 0.24 0.619 0.26 0.639 0.28 0.661 0.30 0.685 0.32 0.712 0.34 0.744 0.36 0.780 0.38 0.820 0.40 0.861 0.42 0.902 0.44 0.939 0.46 0.970 0.48 0.991 0.50 1.000 0.52 0.996 0.54 0.977 0.56 0.945 0.58 0.902 0.60 0.849 0.62 0.789 0.64 0.725 0.66 0.661 0.68 0.598 0.70 0.538 0.72 0.482 0.74 0.432 0.76 0.386 0.78 0.346 0.80 0.311 0.82 0.280 0.84 0.253 0.86 0.230 0.88 0.210 0.90 0.192 0.92 0.176 0.94 0.162 0.96 0.150 0.98 0.140 1.00 0.130 /
\end{sparkline}
 & 7651.25 & 7651.40 & $+0.33$ & 7651.72 \\
Ho L$_3$N$_\mathrm{4,5}$  & L$\beta_{2,15}$  & \begin{sparkline}{6}
    \spark 0.00 0.196 0.02 0.205 0.04 0.215 0.06 0.226 0.08 0.239 0.10 0.254 0.12 0.272 0.14 0.291 0.16 0.311 0.18 0.330 0.20 0.348 0.22 0.365 0.24 0.382 0.26 0.400 0.28 0.421 0.30 0.448 0.32 0.481 0.34 0.521 0.36 0.569 0.38 0.626 0.40 0.690 0.42 0.763 0.44 0.840 0.46 0.913 0.48 0.971 0.50 1.000 0.52 0.991 0.54 0.941 0.56 0.856 0.58 0.747 0.60 0.630 0.62 0.518 0.64 0.418 0.66 0.336 0.68 0.271 0.70 0.220 0.72 0.182 0.74 0.153 0.76 0.130 0.78 0.112 0.80 0.098 0.82 0.086 0.84 0.077 0.86 0.069 0.88 0.062 0.90 0.057 0.92 0.052 0.94 0.048 0.96 0.044 0.98 0.041 1.00 0.038 /
\end{sparkline}
 & 7909.57 & 7909.79 & $+0.53$ & 7910.32 \\
Ho L$_1$M$_5$             & L$\beta_{9}$     & \begin{sparkline}{6}
    \spark 0.00 1.000 0.02 0.975 0.04 0.941 0.06 0.899 0.08 0.851 0.10 0.801 0.12 0.749 0.14 0.698 0.16 0.650 0.18 0.607 0.20 0.568 0.22 0.537 0.24 0.513 0.26 0.497 0.28 0.491 0.30 0.496 0.32 0.512 0.34 0.541 0.36 0.581 0.38 0.632 0.40 0.690 0.42 0.751 0.44 0.810 0.46 0.859 0.48 0.892 0.50 0.906 0.52 0.898 0.54 0.866 0.56 0.815 0.58 0.748 0.60 0.672 0.62 0.593 0.64 0.516 0.66 0.445 0.68 0.383 0.70 0.331 0.72 0.288 0.74 0.255 0.76 0.228 0.78 0.208 0.80 0.193 0.82 0.183 0.84 0.177 0.86 0.175 0.88 0.177 0.90 0.185 0.92 0.200 0.94 0.221 0.96 0.249 0.98 0.283 1.00 0.320 /
\end{sparkline}
 & 8044.37 & 8044.47 & $+0.24$ & 8044.70 \\
Ho L$_3$N$_\mathrm{6,7}$  & Lu               & \begin{sparkline}{6}
    \spark 0.00 1.000 0.02 0.893 0.04 0.784 0.06 0.680 0.08 0.585 0.10 0.504 0.12 0.436 0.14 0.381 0.16 0.338 0.18 0.304 0.20 0.279 0.22 0.261 0.24 0.248 0.26 0.242 0.28 0.241 0.30 0.247 0.32 0.261 0.34 0.284 0.36 0.317 0.38 0.358 0.40 0.407 0.42 0.459 0.44 0.510 0.46 0.552 0.48 0.582 0.50 0.593 0.52 0.584 0.54 0.556 0.56 0.512 0.58 0.458 0.60 0.399 0.62 0.341 0.64 0.288 0.66 0.244 0.68 0.208 0.70 0.181 0.72 0.161 0.74 0.146 0.76 0.136 0.78 0.129 0.80 0.124 0.82 0.121 0.84 0.118 0.86 0.115 0.88 0.113 0.90 0.111 0.92 0.109 0.94 0.108 0.96 0.106 0.98 0.105 1.00 0.104 /
\end{sparkline}
 & 8068.00 & 8068.05 & $+0.12$ & 8068.18 \\
Ho L$_2$N$_4$             & L$\gamma_1$      & \begin{sparkline}{6}
    \spark 0.00 0.475 0.02 0.516 0.04 0.553 0.06 0.583 0.08 0.604 0.10 0.616 0.12 0.621 0.14 0.619 0.16 0.614 0.18 0.610 0.20 0.609 0.22 0.616 0.24 0.631 0.26 0.656 0.28 0.688 0.30 0.724 0.32 0.761 0.34 0.797 0.36 0.829 0.38 0.858 0.40 0.885 0.42 0.911 0.44 0.938 0.46 0.964 0.48 0.986 0.50 1.000 0.52 1.000 0.54 0.982 0.56 0.945 0.58 0.887 0.60 0.812 0.62 0.726 0.64 0.635 0.66 0.543 0.68 0.458 0.70 0.382 0.72 0.316 0.74 0.262 0.76 0.219 0.78 0.185 0.80 0.158 0.82 0.136 0.84 0.120 0.86 0.106 0.88 0.094 0.90 0.085 0.92 0.076 0.94 0.069 0.96 0.062 0.98 0.056 1.00 0.052 /
\end{sparkline}
 & 8749.15 & 8749.61 & $+1.24$ & 8750.84 \\
\end{tabular}